\documentclass[aps,floats,amssymb,amsmath,prb,nofootinbib,twocolumn,superscriptaddress]{revtex4-1}

\usepackage{booktabs}
\usepackage{calc}
\usepackage{psfrag}
\usepackage{graphicx}
\usepackage{color}
\usepackage{units}
\usepackage[dvipsnames]{xcolor}
\usepackage[unicode=true,pdfusetitle,
 bookmarks=true,bookmarksnumbered=false,bookmarksopen=false,
 breaklinks=false,pdfborder={0 0 0},backref=false,colorlinks=true]
 {hyperref}
\usepackage{geometry}
\pdfpageattr {/Group << /S /Transparency /I true /CS /DeviceRGB>>} 

\newcommand{\up}{\uparrow}
\newcommand{\dn}{\downarrow}

\newcommand{\imp}{\mathrm{imp}}

\usepackage{pdfpages}

\makeatletter
\AtBeginDocument{\let\LS@rot\@undefined}
\makeatother

\begin{document}

\title{Electrical conductivity in the Hubbard model: orbital effects of magnetic field}
\author{J. Vu\v ci\v cevi\'c}
\affiliation{Scientific Computing Laboratory, Center for the Study of Complex Systems, Institute of Physics Belgrade,
University of Belgrade, Pregrevica 118, 11080 Belgrade, Serbia}
\author{R. \v{Z}itko}

\affiliation{Jo\v{z}ef Stefan Institute, Jamova 39, SI-1000 Ljubljana,
Slovenia}
\affiliation{Faculty  of Mathematics and Physics, University of
Ljubljana, Jadranska 19, SI-1000 Ljubljana, Slovenia}

\begin{abstract}
Calculation of conductivity in the Hubbard model is a challenging task. 
Recent years have seen much progress in this respect and numerically exact solutions are now possible in certain regimes.
In this paper we discuss the calculation of conductivity for the
square lattice Hubbard model in the presence of a perpendicular
magnetic field, focusing on orbital effects.
We present the relevant formalism in all detail and in full generality, and then discuss the simplifications that arise at the level of the dynamical mean field theory (DMFT).
We prove that the Kubo bubble preserves gauge and translational
invariance, and that in the DMFT the vertex corrections cancel regardless of the magnetic field.
We present the DMFT results for the spectral function and both the longitudinal and Hall conductivity in several regimes of parameters.
We analyze thoroughly the quantum oscillations of the longitudinal conductivity and identify a high-frequency oscillation component, arising as a combined effect of scattering and temperature, in line with recent experimental observations in moir\' e systems.
\end{abstract}

\maketitle

\section{Introduction}

Strong correlations in electronic systems have a profound effect on conductivity, and lead to a range of unconventional behaviors which are at the center of interest in condensed matter theory.
One such behavior is the linear temperature dependence of resistivity, observed in the cuprate superconductors\cite{takagi1992,Keimer2015,Legros2018}.
The linear resistivity has been reproduced by numerical simulation of the Hubbard model\cite{Pruschke1993,Jarrel1994,Bergeron2011,deng,vucicevic2015prl,perepelitsky2016,kokalj2017,Vucicevic2019,Huang2019,Vranic2020}, as well as with ultra cold atom simulations\cite{Brown2018}.
It is viewed as an effect of proximity to the Mott
transition\cite{Terletska2011,vucicevic2013,vucicevic2015prl} or quantum critical points\cite{Bergeron2011,Krempa2012,Rameau2014,Hartnoll2014}, as well as a generic high-temperature feature of correlated materials which are well approximated by a single-band model \cite{perepelitsky2016}.

External magnetic fields are also known to affect the transport
properties of electronic systems \cite{thomson}, sometimes
drastically: in the context of the two-dimensional electron gas,
magnetic field leads to the well-known quantum Hall effect (QHE), where
conductivity displays intricate dependence on the magnetic field
\cite{klitzing1980,Thouless1982,niu1985,ezawa}.
The effect of the Coulomb interaction is here essential for the
understanding of the fractional QHE\cite{tsui1982,laughlin1983}.
In conventional metals, resistivity is an oscillatory function of
the magnetic field, which is the well known Shubnikov-de Haas effect
(SdH)\cite{SdH1,shoenberg}.
However, the SdH effect is often observed even in states which are
assumed to be correlated and are not yet fully understood\cite{LeBoeuf2007,Moetakef2012,Sebastian2015}.
In such cases, an effective Fermi liquid description of the material
is often invoked to analyze the experimental data, and to map out the geometry of the Fermi surface.
It is, therefore, of great importance to understand the interplay of strong coupling and magnetic fields in lattice systems.
The study of magnetoresistance in correlated lattice models has been so far limited to perturbative approaches, either for weak fields\cite{Voruganti1992,Ding2017,wang2021arxiv}, or weak interactions\cite{Thouless1982,Auerbach2019,Wijesinghe2019}. 
To the best of our knowledge, 
the only non-perturbative calculations of magnetotransport
were limited to the transversal conductivity\cite{Assad1995,Markov2019,Wang2020}.
Non-perturbative calculations were also performed for the effective Fermi liquid parameters (quasiparticle weight, scattering rate and density of states at the Fermi level)\cite{Sherman2015,Acheche2017,Markov2019}, which can be considered relevant for longitudinal conductivity.

In this paper we study the effect of magnetic field on both the longitudinal and transversal conductivity in the square lattice Hubbard model in several parameter regimes: from weak to strong coupling, low to high temperature, and in the full range of the magnetic field.

We first lay out the general formalism for the calculation of
conductivity in the presence of the magnetic field and then describe
the simplifications that arise at the level of the dynamical mean field theory (DMFT), as
previously implemented in Refs.~\onlinecite{Acheche2017} and \onlinecite{Markov2019}.
Most importantly, we show that the Kubo bubble is gauge invariant and that the vertex corrections to the current-current correlation function cancel, analogously to the zero field case.
The latter is done by rederiving the well-known zero-field proof
from Ref.~\onlinecite{khurana1990} in real space, and then generalizing it to the case of external magnetic fields.
Cancellation of vertex corrections at the level of DMFT was previously shown only for the transversal conductivity\cite{Markov2019}, and here we give a different, fully general proof.

We perform extensive DMFT calculations to cover a large part of the phase diagram.
Our numerical results show that the oscillatory behavior of
conductivity is restricted to a finite range of temperature which is
mainly determined by the amount of dynamic correlations (which are
promoted by interactions, yet hindered by doping). The amplitude of
oscillations decays exponentially with temperature, as expected from
the Lifshitz-Kosewich theory~\cite{LK}. Above a certain characteristic
temperature, no non-monotonic behavior can be induced no matter how
strong the magnetic field is.
We also observe that in a big range of magnetic fields and interaction strengths, the conductivity follows a scaling law, with the temperature scale set by the coupling strength. Most importantly, the $T$-linear dependence of resistivity in the high-temperature regime is not qualitatively modified by the magnetic field.

We investigate transverse conductivity in the non-interacting limit, and observe exponential decay of $\sigma^{xy}$ with temperature, and a power law divergence as $B_z\rightarrow 0$. The analytic behavior at $B_z=0$ is restored by interactions, which lead to a smooth decay of $\sigma^{xy}$ as the magnetic field is gradually turned off.

Finally, we investigate the oscillatory behavior of conductivity in weak-to-moderate magnetic fields. 
In all the cases we studied, the longitudinal conductivity turns out to be dominated by the current-vertex factors, rather than the local density of states at the Fermi level or the effective scattering rate.
This leads to an important simplification: one can reliably calculate conductivity at an arbitrary field by using the zero-field DMFT calculation for the self-energy.
We further observe that at high temperature, moderate-to-high
interactions and moderate fields, the oscillation occurs at two
separate frequencies: one that corresponds to the area of the Fermi
sea, as in the Shubnikov-de Haas effect, and the other which
corresponds to the full area of the two-dimensional Brillouin zone (BZ), and is therefore of higher frequency and independent of the doping level.
This finding is in excellent qualitative agreement with the recent experimental observations in graphene superlattices\cite{Hunt2013,KrishnaKumar2017,KrishnaKumar2018,Barrier2020}.
The full discussion of the observed phenomenology of quantum
oscillations of conductivity in the Hubbard model is presented in a
separate publication, Ref.~\onlinecite{OURPRL}, while here we present the raw data and describe the basic mechanism behind the onset of the high-frequency oscillations.

The paper is organized as follows. We first describe the formalism:
the Hamiltonian, the gauge choice, reciprocal-space formulation,
gauge-invariant Green's function, current operators and the
correlation function, DMFT approach, calculation of the conductivity
tensor, and vertex factors. We then present and discuss the results.
For the benefit of the reader and for easy rederivation and validation
of the results presented in this work, we provide very detailed
proofs of all steps in the derivations in the Appendices. For reasons
of clarity and to facilitate dimensional analysis we maintain all
constants ($e$, $\hbar$, $k_\mathrm{B}$, and lattice constants $a$ in $c$) in
the equations.

\section{Formalism}
  \subsection{Model}
  We study the Hubbard model on the square lattice with lattice
  constant $a$, defined by the Hamiltonian
  \begin{equation}
   H = H_0 + H_\mathrm{int},
  \end{equation}
  where the non-interacting part $H_0$ is the tight-binding (TB) model that we
  discuss in the following sections, while the interacting part
  $H_\mathrm{int}$ is the local
  density-density coupling, i.e., the Hubbard interaction
  \begin{equation}
    H_\mathrm{int} = U\sum_i n_{i,\up} n_{i,\dn},
  \end{equation}
  where $i$ indexes the lattice sites, $n_{i,\sigma}$ are the density
  operators, and $U$ is the coupling constant. The electron spin is denoted $\sigma=\up,\dn$.
    \subsubsection{Orbital space}
The lattice sites are assumed to lie in the $z=0$ plane; where
convenient, we will treat the system as a three-dimensional stack of such
planes separated by the lattice constant $c$ in the perpendicular
direction.

The effect of the external magnetic field $\mathbf{B}$ in the TB model is
two-fold: it couples to the electrons' spin degree of freedom (Zeeman
term), as well as the momentum. The latter is approximated on the
lattice by means of the Peierls
substitution\cite{Peierls1933,Wannier1962}.
The resulting Hamiltonian is
\begin{eqnarray} \nonumber
 H_0 &=&-\mu\sum_{i,\sigma} n_{i,\sigma} \\
    &&+g \mu_\mathrm{B} \sum_i \mathbf{B}(\mathbf{r}_i)\cdot \mathbf{S}_i \\ \nonumber
    &&-\sum_{i,j,\sigma} t_{ij} e^{if_{ij}} c^\dagger_{i,\sigma}
    c_{j,\sigma},
\end{eqnarray}
where $\mu$ is the chemical potential, $g$ is the gyromagnetic factor,
$\mu_\mathrm{B}$ is the Bohr magneton, the position vector of site $i$ is $\mathbf{r}_i$, and the operator of the electron SU(2) spin is
\begin{equation}
 S^\eta_i = \frac{1}{2}( c^\dagger_{i,\up},\; c^\dagger_{i,\dn} ) \hat{\sigma}^\eta 
 \begin{pmatrix} c_{i,\up} \\c_{i,\dn} \end{pmatrix},
\end{equation}
where $\hat{\sigma}^\eta$ are the Pauli matrices, and $\eta$ enumerates the spatial directions $x,y,z$.
The $t_{ij}$ is the hopping amplitude between the sites $i$ and $j$.
The Peierls substitution introduces a phase shift $f_{ij}$ that is
picked up by an electron on the path from site $i$ to site $j$:
\begin{eqnarray}\label{eq:Peirels_phase}
f_{ij} &=& \frac{e}{\hbar}\int_{\mathbf{r}_i}^{\mathbf{r}_j} \mathbf{A}(\mathbf{r}) \cdot \mathrm{d}\mathbf{r}.
\end{eqnarray}
Here $\mathbf{A}$ is the vector potential, $e$ the elementary charge,
and $\hbar$ the reduced Planck's constant.
In matrix notation in site space, the effect of the gauge field
corresponds to element-wise multiplication of the bare Hamiltonian:
\begin{equation}
 \mathbf{H}_0[\mathbf{A}] = \mathbf{H}_0[\mathbf{A}=0] \circ
 e^{i\mathbf{f}},
\end{equation}
where $e^{i\mathbf{f}}$ is simply a matrix constructed out of $e^{if_{ij}}$ elements.

We are interested in the effects of a uniform magnetic field perpendicular to the 2D lattice:
$\mathbf{B} = (0, 0, B_z)$. The vector potential $\mathbf{A}$ is not uniquely determined by $\mathbf{B}$. 
There are two obvious choices: the Landau gauge
\begin{equation}\label{eq:Landau_gauge}
 \mathbf{A}(\mathbf{r}) = (0, xB_z, 0),
\end{equation}
and the symmetric gauge
\begin{equation}\label{eq:symmetric_gauge}
 \mathbf{A}(\mathbf{r}) = (-\frac{y}{2}B_z, \frac{x}{2}B_z, 0).
\end{equation}
Throughout this paper, we work in the Landau gauge.

In the rest of the paper, we define lattice site coordinates $x_i$ and $y_i$ as integers,
and define $\mathbf{r}_i=(x_i,y_i,0)$. The physical position vector of the lattice site $i$ is then $a\mathbf{r}_i$
and we give spatial indices in terms of $\mathbf{r}$ as, e.g. $\mathbf{A}_\mathbf{r}\equiv\mathbf{A}(a\mathbf{r})$.

Plugging the Landau gauge field $\mathbf{A}$
(Eq.~\ref{eq:Landau_gauge}) in the expression for the Peierls phase
(Eq.~\ref{eq:Peirels_phase}) one obtains\cite{OURPRL,Acheche2017} (see Appendix~\ref{app:proof_Landau_Peirels_phase} for proof)
\begin{equation} \label{eq:Landau_Peirels_phase}
\begin{split}
f_{ij}\equiv f_{\mathbf{r}_i,\mathbf{r}_j} &= \frac{e}{\hbar} (B_z a^2) \frac{(y_j-y_i)(x_i+x_j)}{2} \\
&=2\pi \frac{\Phi}{\Phi_0} \frac{(y_j-y_i)(x_i+x_j)}{2},
\end{split}
\end{equation}
where $\Phi_0=h/e$ is the unit flux, and $\Phi=B_z a^2$ is the flux per
lattice plaquette.

To be able to define a finite-sized (commensurate) magnetic unit cell, the values of $B_z$ must satisfy 
\begin{eqnarray}\label{eq:condition_infinite}
 \frac{e}{\hbar}(B_z a^2) = 2\pi\frac{p}{q},
\end{eqnarray}
where $p$ and $q$ are coprime integers. $q$ is then the size of the
unit cell in the $x$-direction (for proof see
Appendix~\ref{app:proof_unit_cell_size}). In the other direction the
size of the unit cell is 1, as the translational invariance is not
broken along the $y$ axis; this is obvious as $f_{ij}$ depends only on the difference $y_i-y_j$. 


The effect of $B_z$ on the kinetic energy term is periodic. 
As $B_z$ enters the kinetic energy through $e^{2\pi i
\frac{p}{q}(y_j-y_i)(x_i+x_j)/2}$, if $(y_j-y_i)(x_i+x_j)/2$ is
an integer for all $(i,j)$ connected by hopping (as is the case with nearest-neighbor hopping),
increasing $p/q$ by an integer makes no difference. 
Therefore, the effect of $\frac{p}{q}$ is the same as that of
$\frac{p+mq}{q}$, with $m$ integer. The inversion symmetry of the
lattice implies that the effect of $\frac{p}{q}$ is the same as that
of $\frac{q-p}{q}$. When it comes to the kinetic energy term, all
physically discernible magnetic fields (that satisfy
Eq.~\ref{eq:condition_infinite}) can be mapped onto the range $0 \leq
p/q\leq 1/2$. The field $p/q=1$ is then a characteristic value of the
field, the lowest one (other than zero) that does not couple with
electron motion.

In numerics, we will consider a finite $L \times L$ cyclic lattice,
which must fit an integer number of magnetic unit cells of size $q
\times 1$. 
We can rewrite the condition Eq.~\ref{eq:condition_infinite} as:
\begin{eqnarray} \label{eq:condition_finite}
 \frac{e}{\hbar}(B_z a^2) = 2\pi\frac{p}{q} = 2\pi\frac{n}{L}
\end{eqnarray}
with the size of magnetic unit cell being $L$ or smaller, as given by $L/\mathrm{gcd}(L,n)$, where ``gcd'' denotes the greatest common divisor, and $n$ is an arbitrary integer. 
In fact, the relation between the finite and infinite lattice is simply
$$ q = L/\mathrm{gcd}(L,n), \;\;\;\; p = n/\mathrm{gcd}(L,n).$$
The size of the lattice $L$ determines the resolution of $p/q$ that one can achieve in scanning the strength of the field in the model.


Under the assumption of only the nearest-neighbor hopping, we now
rewrite the full Hamiltonian as
\begin{eqnarray}\label{eq:ham_nnhop}
 H_0 &=&-\mu\sum_{i,\sigma} n_{i,\sigma} \\ \nonumber
   && + \frac{1}{2} g \mu_\mathrm{B} B_z \sum_{i,\sigma=\up,\dn} (-1)^{\delta_{\sigma,\dn}} n_{i,\sigma} \\ \nonumber
   && -t\sum_{i, \mathbf{u}\in\{\mathbf{e}_x,\mathbf{e}_y\},\sigma} e^{i\frac{e a^2 }{\hbar} x_i B_z \mathbf{u}\cdot\mathbf{e}_y} c^\dagger_{\mathbf{r}_i,\sigma} c_{\mathbf{r}_i+\mathbf{u},\sigma} +\mathrm{H.c}
\end{eqnarray}
\emph{Numeric scales.}
The importance of the Zeeman splitting depends on the ratio of the
Zeeman energy over the bandwidth. Both $g$ and the bandwidth are material specific. 
A quick estimate for cuprate compounds, under the assumption of $g=2$ and
half-bandwidth of around $10^5\mathrm{K}$, gives that $B_z$ of about
$50\mathrm{T}$ corresponds to a Zeeman energy of about
$ \approx 3 \times 10^{-3} D$, where $D=4t$ is the half-bandwidth.
While the effect of Zeeman splitting is interesting to study on its own, throughout this paper we restrict to only the gauge-field effects and set $g=0$.

The effect of the gauge field is determined by the lattice spacing. Assuming $a \sim 5 \times 10^{-10}\mathrm{m}$ which is relevant for cuprates, we see that the characteristic $p/q=1/2$ field corresponds to
$
 B_z = \frac{\pi\hbar}{e a^2 }\approx 8 \times 10^{3} \mathrm{T}
$.
At $B_z=50\mathrm{T}$, we therefore have $p/q\approx 1/300$, and we need at least the lattice size $L=300$ to describe this regime.
Clearly, the bigger the lattice spacing, the bigger the phase picked up upon traveling between the lattice sites, and the bigger the effect of the coupling to the gauge field.
The regime of large $p/q$ is therefore relevant for systems with a
larger lattice spacing (as in moir\' e heterostructures \cite{andrei2021}), or where high gauge fields can be introduced
artificially (as in optical lattices\cite{Cooper2019,Dalibard2011}).

    \subsubsection{Momentum space}
Rewriting the kinetic-energy in momentum space leads to the 
Harper equation~\cite{harper1955, HofstadterPRB1976}.
By applying to the kinetic energy term the Fourier transformation of the creation/annihilation operators,
\begin{equation}
 c^\dagger_i = \frac{1}{\sqrt{N}} \sum_\mathbf{k}
 e^{-i\mathbf{k}\cdot\mathbf{r}_i} c^\dagger_\mathbf{k},\;\;c_i = \frac{1}{\sqrt{N}} \sum_\mathbf{k} e^{i\mathbf{k}\cdot\mathbf{r}_i} c_\mathbf{k},
\end{equation}
where $N=L^2$ is the number of sites in the lattice, one obtains
\begin{eqnarray} \nonumber
 H_\mathrm{kin} &=& -t\sum_{i, \mathbf{u}\in\{\mathbf{e}_x,\mathbf{e}_y\},\sigma} e^{i 2\pi \frac{n}{L} x_i \mathbf{u}\cdot\mathbf{e}_y} c^\dagger_{\mathbf{r}_i,\sigma} c_{\mathbf{r}_i+\mathbf{u},\sigma} +\mathrm{H.c}  \\ \label{eq:H0_momentum_space}
&=& -2t\sum_{\mathbf{k},\sigma} \cos k_x n_{\mathbf{k},\sigma} \\ \nonumber
&& \;\;\;\;-t\sum_{\mathbf{k},\sigma}e^{ik_y}c^\dagger_{\mathbf{k},\sigma}c_{\mathbf{k}- 2\pi \frac{n}{L}\mathbf{e}_x,\sigma} +\mathrm{H.c.}
\end{eqnarray}
For a detailed proof see Appendix~\ref{app:proof_H0_momentum_space}. 

There is a coupling between the different $\mathbf{k}$-states which
results in a reduction of the Brillouin zone (BZ) by a factor of $q =
L/\mathrm{gcd}(L,n)$, where $q$ is the size of the magnetic unit cell.
We define $\tilde{\mathbf{k}}$ as $\mathbf{k}$ within the reduced BZ (RBZ). Now, $\tilde{k}_x \in [0,2\pi/q)$, while $\tilde{k}_y \in [0,2\pi)$.
$\tilde{\mathbf{k}}$ is a good quantum number, but there is now an
additional degree of freedom that we denote $l$ such that $l\in[0,q)$.
A single-particle state is fully determined by a triplet $(\tilde{\mathbf{k}},l,\sigma)$, with $c_{\tilde{\mathbf{k}},l,\sigma}\equiv c_{\mathbf{k}=\tilde{\mathbf{k}}+l\frac{2\pi}{q}\mathbf{e}_x,\sigma}$.

On a finite cyclic lattice, the momentum space is discrete, with a step of the size $2\pi/L$. 
If $\mathrm{gcd}(L,n)=1$ there is only one $\tilde{k}_x$ value in the
RBZ (equal to 0), and for each momentum there are $q=L$ different values of $l$.

The Hamiltonian has a block diagonal structure.
For a given $(\tilde{\mathbf{k}},\sigma)$, the Hamiltonian in the space of $l$ is given by the Harper equation
\begin{eqnarray}\nonumber
 [H_{0,\tilde{\mathbf{k}},\sigma}]_{l,l'} &=& 
 \Bigg(-\mu_\sigma - 2t\cos \left(\tilde{k}_x+l\frac{2\pi}{q}\right)\Bigg)\delta_{l,l'} \\ \label{eq:harper_ham}
 &&- t\big(e^{i\tilde{k}_y}\delta_{l,l'\oplus p} + e^{-i\tilde{k}_y}\delta_{l',l\oplus p}\big)
\end{eqnarray}
where $\oplus$ denotes the cyclic addition modulo $q$ defined as
\begin{equation}
 l\oplus l' \equiv l+l'-q\mathrm{div}(l+l',q),
\end{equation}
and $p=n/\mathrm{gcd}(L,n)$. We also introduced $\mu_\sigma = \mu -
(-1)^{\delta_{\sigma,\dn}} g \mu_\mathrm{B} B_z/2$.

Each block of the Hamiltonian can be diagonalized to yield the eigenenergies $\varepsilon_{\tilde{\mathbf{k}},\sigma,m}$, with $m\in[0,q)$.
The basis change matrix elements are defined by
\begin{equation}
  c^\dagger_{\tilde{\mathbf{k}},l,\sigma} = \sum_m [\alpha_{\tilde{\mathbf{k}},\sigma}]_{l,m} c^\dagger_{\tilde{\mathbf{k}},m,\sigma}.
\end{equation}
Note that throughout this work, we distinguish between different operators ($c_{i,\sigma}\equiv c_{\mathbf{r}_i,\sigma}$, $c_{\mathbf{k},\sigma}$, $c_{\tilde{\mathbf{k}},l,\sigma}$, $c_{\tilde{\mathbf{k}},m,\sigma}$ etc.) only by the choice of the symbols in the subscript (e.g. $c_{\tilde{\mathbf{k}},l,\sigma}$ is not equal $c_{\tilde{\mathbf{k}},m,\sigma}$ even if $l=m$), and similarly for other functions.

The blocks of the Hamiltonian have several symmetries. One can invert the $x$ axis
\begin{equation}\label{eq:kx_flip}
[H_{0,(\tilde{k}_x,\tilde{k}_y),\sigma}]_{l,l'}=[H_{0,(-\tilde{k}_x,\tilde{k}_y),\sigma}]^*_{q-l,q-l'},
\end{equation}
which means that
\begin{equation}\label{eq:kx_flip2}
[\alpha_{(\tilde{k}_x,\tilde{k}_y),\sigma}]_{l,m}=[\alpha_{(-\tilde{k}_x,\tilde{k}_y),\sigma}]^*_{q-l,m} 
\end{equation}
and that the eigenenergies remain the same upon inverting the $x$-axis.

One can also invert the $y$-axis
\begin{equation}\label{eq:ky_flip}
[H_{0,(\tilde{k}_x,\tilde{k}_y),\sigma}]_{l,l'} = [H_{0,(\tilde{k}_x,-\tilde{k}_y),\sigma}]^*_{l,l'}.
\end{equation}
Again, inverting $k_y$ does not affect the eigenenergies, but merely flips the chirality of the eigenstates
\begin{equation}\label{eq:ky_flip2}
[\alpha_{(\tilde{k}_x,\tilde{k}_y),\sigma}]_{l,m}=[\alpha_{(\tilde{k}_x,-\tilde{k}_y),\sigma}]^*_{l,m}. 
\end{equation}

Inverting both axes at the same time therefore means
\begin{equation}\label{eq:k_flip}
[H_{0,(\tilde{k}_x,\tilde{k}_y),\sigma}]_{l,l'}=[H_{0,(-\tilde{k}_x,-\tilde{k}_y),\sigma}]_{q-l,q-l'}
\end{equation}
and 
\begin{equation}\label{eq:k_flip2}
[\alpha_{\tilde{\mathbf{k}},\sigma}]_{l,m}=[\alpha_{-\tilde{\mathbf{k}},\sigma}]_{q-l,m}. 
\end{equation}

There is an additional periodicity along the $y$-axis 
\begin{equation}\label{eq:ky_translate}
\varepsilon_{\tilde{\mathbf{k}},\sigma,m}=\varepsilon_{\tilde{\mathbf{k}}+(2\pi C/q) \mathbf{e}_y,\sigma,m}
\end{equation}
and 
\begin{equation}\label{eq:ky_translate2}
[\alpha_{\tilde{\mathbf{k}},\sigma}]_{lp\;\mathrm{mod}\;q,\;m}=e^{iC\frac{2\pi}{q}l}[\alpha_{\tilde{\mathbf{k}}+(2\pi C/q) \mathbf{e}_y,\sigma}]_{lp\;\mathrm{mod}\;q,\;m} 
\end{equation}
with $C$ integer.
This symmetry is important on a finite lattice, where $k_y$ takes
values of the form $C 2\pi/L$. If $L=q$ (i.e., $\mathrm{gcd}(n,L)=1$),
this means that the density of states and other relevant
quantities can be obtained by considering only the block
$\tilde{\mathbf{k}}=(0,0)$. Otherwise, $k_y$ values up to
$2\pi/q$ need to be considered.
For a proof see Appendix~\ref{app:periodicity_ky}.
              
  \subsection{Gauge invariant Green's function}
     
A uniform magnetic field does not break physical translational
invariance. However, at the formal level, the inclusion of the appropriate vector potential 
means that all correlators connecting 2 or more points in space
depend not only on the relative positions, but also on the absolute
positions. The spatial dependence of correlators can depend on the
gauge choice. Nevertheless, physical observables preserve both
translational and gauge invariance.

The quantity of primary interest is the Green's function. It is defined as a function of imaginary time
\begin{equation}
 G_{ij,\sigma}(\tau) = -\langle T_\tau c_{i,\sigma}(\tau) c^\dagger_{j,\sigma}(0) \rangle.
\end{equation}
As a function of complex frequency $z$, and as a matrix in the site-space, one can always write
\begin{equation}
 \mathbf{G}_{\sigma}(z) = [\hbar
 z\mathbf{I}-\mathbf{H}_0-\mathbf{\Sigma}(z)]^{-1},
\end{equation}
where $\mathbf{\Sigma}(z)$ is the self-energy.
The diagonal elements of the Green's function with $z=\omega+i0^+$ determine the local spectral function which is a physical observable. As such, the local Green's function is uniform in space. Nevertheless, the off-diagonal elements do not exhibit-translational invariance $G_{ij} = G_{\mathbf{r}_i-\mathbf{r}_j}$, but rather this equality is satisfied only up to a phase.

It can be shown that the quantity
\begin{equation}
 \bar{G}_{ij,\sigma}(z) \equiv e^{-if_{ij}}G_{ij,\sigma}(z)
\end{equation}
is gauge invariant, and \emph{preserves the full symmetry of the lattice}. 
We reproduce here a proof from Ref.~\onlinecite{Chen2011} which is valid in the non-interacting case, but is completely analogous in the case of a fully local and spatially uniform self-energy.
In orbital space we have
\begin{eqnarray} \nonumber
\mathbf{G} &=& [\mathbf{I}\hbar z - \mathbf{H}_0[\mathbf{A}] -
\mathbf{I}\Sigma(z)]^{-1},  \\ \nonumber
\mathbf{G}^{-1} &=& [\mathbf{I}\hbar z - \mathbf{H}_0\circ e^{i\mathbf{f}} -
\mathbf{I}\Sigma(z)],  \\ \label{eq:Gbar_proof_before_main_step}
\mathbf{I} &=&  [\mathbf{I}\hbar z - \mathbf{H}_0\circ e^{i\mathbf{f}} - \mathbf{I}\Sigma(z)] \mathbf{G}.
\end{eqnarray}
It is also easy to verify that
$
 \mathbf{I}\circ e^{i\mathbf{f}} = \mathbf{I}
$
so we can further write
\begin{eqnarray} \label{eq:Gbar_proof_main_step}
\mathbf{I}\circ e^{i\mathbf{f}} &=&  ([\mathbf{I}\hbar z - \mathbf{H}_0 - \mathbf{I}\Sigma(z)]\circ e^{i\mathbf{f}}) \mathbf{G}  \\ \nonumber
\mathbf{I}\circ e^{i\mathbf{f}} &=&  ([\mathbf{I}\hbar z - \mathbf{H}_0 - \mathbf{I}\Sigma(z)]\circ e^{i\mathbf{f}}) (\bar{\mathbf{G}}\circ e^{i\mathbf{f}}).
\end{eqnarray}
We now write the scalar form
\begin{eqnarray} \nonumber
\delta_{ij} e^{if_{ij}} &=& \sum_k [\mathbf{I}\hbar z - \mathbf{H}_0 - \mathbf{I}\Sigma(z)]_{ik} e^{if_{ik}} \bar{G}_{kj} e^{if_{kj}} \\ \nonumber
\delta_{ij} &=& \sum_k  [\mathbf{I}\hbar z - \mathbf{H}_0 - \mathbf{I}\Sigma(z)]_{ik} \bar{G}_{kj} e^{if_{ik}}  e^{if_{kj}} e^{-if_{ij}} \\ \nonumber
\delta_{ij} &=& \sum_k [\mathbf{I}\hbar z - \mathbf{H}_0 - \mathbf{I}\Sigma(z)]_{ik} \bar{G}_{kj} e^{if_{ik}}  e^{if_{kj}} e^{if_{ji}}. \\ \label{eq:Gbar_proof}
\end{eqnarray}
The expression $e^{if_{ik}}  e^{if_{kj}} e^{if_{ji}}$ is simply the
magnetic flux passing through the triangle defined by the lattice sites
$i$, $j$, and $k$. This quantity is gauge invariant. As Eq.~\ref{eq:Gbar_proof} is a defining relation for $\bar{G}_{ij}$, it means
that $\bar{G}_{ij}$ is gauge invariant. Furthermore, the quantity
$[\mathbf{I}\hbar z - \mathbf{H} - \mathbf{I}\Sigma(z)]_{ik}$ has full lattice symmetry, thus $\bar{G}_{ij}$ does as well.

However, it is interesting to consider the case of a general (possibly
non-local and non-uniform) self-energy $\Sigma$. In that case, the step performed between Eq.~\ref{eq:Gbar_proof_before_main_step} and Eq.~\ref{eq:Gbar_proof_main_step} reads
\begin{equation}
 [\mathbf{I}\hbar z - \mathbf{H}_0\circ e^{i\mathbf{f}} -
 \mathbf{\Sigma}(z)] = [\mathbf{I}\hbar z - \mathbf{H}_0 - \mathbf{\Sigma}(z)\circ e^{-i\mathbf{f}}]\circ e^{i\mathbf{f}}.
\end{equation}
The proof can proceed from there completely analogously, but only if the quantity
$\bar{\mathbf{\Sigma}}(z) \equiv \mathbf{\Sigma}(z)\circ e^{-i\mathbf{f}}$ is gauge invariant and preserves the full lattice symmetry.

A proof for the gauge invariance and lattice symmetry of $\bar{\mathbf{\Sigma}}$ can be given in terms of Feynman diagrams for a special case of local density-density interactions, as follows.
Each diagram's contribution is a product of a certain number of
fermionic loops. In case of local density-density interactions, a
single Green's function loop is just the density, and this quantity is
gauge invariant. Then we have a loop of arbitrary size $N$:
\begin{eqnarray}
 && G_{0,i_1 i_2} G_{0,i_2 i_3} ... G_{0,i_N i_1} \\ \nonumber
 && =  \bar{G}_{0,i_1 i_2} \bar{G}_{0,i_2 i_3} ... \bar{G}_{0,i_N i_1} \\ \nonumber
 && \;\;\;\;\times e^{if_{i_1i_2}}e^{if_{i_2i_3}}...e^{if_{i_Ni_1}},
\end{eqnarray}
which is gauge invariant for the same reason as we had above (note
that for the bare propagator $\bar{G}_0$, gauge invariance and symmetries have already been proven by Eq.~\ref{eq:Gbar_proof}). The closed fermionic loops are multiplied with the fermionic line connecting the terminals of the self-energy. Say, in case of $\Sigma_{i_1,i_N}$:
\begin{eqnarray}
 && G_{0,i_1 i_2} G_{0,i_2 i_3} ... G_{0,i_{N-1} i_N} \\ \nonumber
 && =  \bar{G}_{0,i_1 i_2} \bar{G}_{0,i_2 i_3} ... \bar{G}_{0,i_{N-1} i_N}  \\ \nonumber
 && \;\;\;\; \times e^{if_{i_1i_2}}e^{if_{i_2i_3}}...e^{if_{i_{N-1}i_N}}.
\end{eqnarray}
Clearly if we multiply now both sides with $e^{-if_{i_1,i_N}}$, we get on one side $\bar{\Sigma}_{i_1,i_N}$, and on the other
\begin{eqnarray}
 &&\bar{G}_{0,i_1 i_2} \bar{G}_{0,i_2 i_3} ... \bar{G}_{0,i_{N-1} i_N} \\ \nonumber
 &&\times e^{if_{i_1i_2}}e^{if_{i_2i_3}}...e^{if_{i_{N-1}i_N}} e^{-if_{i_1,i_N}} \\ \nonumber
 && = \bar{G}_{0,i_1 i_2} \bar{G}_{0,i_2 i_3} ... \bar{G}_{0,i_{N-1} i_N} \\ \nonumber
 && \;\;\;\; \times e^{if_{i_1i_2}}e^{if_{i_2i_3}}...e^{if_{i_{N-1}i_N}} e^{if_{i_N,i_1}}
\end{eqnarray}
and again the right-hand side is gauge invariant. This proves that the contribution to $\bar{\Sigma}_{ij}$ of each Feynman diagram individually is gauge invariant. Moreover, $\bar{\Sigma}$ is expressed solely in terms of objects with full lattice symmetry, thus it must itself exhibit full lattice symmetry. 

\subsubsection{Efficient calculation of $\bar{G}$}

A straightforward calculation of $\bar{G}$ performed in site-space
would involve an inverse of the $N\times N$ matrix
\begin{equation}
 \bar{G}(z) = e^{-i\mathbf{f}}\circ [\hbar z\mathbf{I} - \mathbf{H}_0[\mathbf{A}] - \mathbf{\Sigma}(z)]^{-1}.
\end{equation}
(Note that here the Peierls phase needs to be taken as Eq.\ref{eq:fij_finite_lattice}, see Appendix \ref{app:proof_Landau_Peirels_phase}.)
This operation scales as $O(N^3)$ and the size of the lattice one can treat this way is limited to $N\sim 1000$. 
A more efficient approach can be formulated, and especially so in the non-interacting
case, and the case when the self-energy is fully local, i.e., whenever
the Green's function is fully diagonal in the eigenbasis of $\mathbf{H}_0$, i.e. $G_{(\tilde{\mathbf{k}},m),(\tilde{\mathbf{k}}',m')}=\delta_{\tilde{\mathbf{k}},\tilde{\mathbf{k}}'}\delta_{mm'}G_{\tilde{\mathbf{k}},m}$. This is precisely the case relevant for our DMFT calculations.
We will make use of the basis change matrix elements to go from eigenbasis $|\tilde{\mathbf{k}},m,\sigma \rangle$ to orbital-basis $| i,\sigma \rangle$
\begin{eqnarray} \label{eq:eigenbasis_to_local}
 | i,\sigma \rangle &=& \frac{1}{\sqrt{N}} \sum_{\mathbf{k}}e^{-i\mathbf{k}\cdot \mathbf{r}_i}| \mathbf{k},\sigma \rangle \\ \nonumber
                    &=& \frac{1}{\sqrt{N}} \sum_{\tilde{\mathbf{k}},l }e^{-i\big(\tilde{\mathbf{k}}+l\frac{2\pi}{q}\mathbf{e}_x\big)\cdot \mathbf{r}_i}
                    | \tilde{\mathbf{k}},l,\sigma \rangle\\ \nonumber
                    &=& \frac{1}{\sqrt{N}} \sum_{\tilde{\mathbf{k}},l }e^{-i\big(\tilde{\mathbf{k}}+l\frac{2\pi}{q}\mathbf{e}_x\big)\cdot \mathbf{r}_i}
                    \sum_{m}[\alpha_{\tilde{\mathbf{k}},\sigma}]_{l,m}|\tilde{\mathbf{k}},m,\sigma \rangle.
\end{eqnarray}
Therefore,
\begin{equation}
 G_{\mathbf{r},\mathbf{r}',\sigma}(z) = \frac{1}{N}\sum_{\tilde{\mathbf{k}},m} W^*_{\tilde{\mathbf{k}},m,\mathbf{r},\sigma} W_{\tilde{\mathbf{k}},m,\mathbf{r}',\sigma} G_{\tilde{\mathbf{k}},m,\sigma}(z)
\end{equation}
with 
\begin{equation}
 W_{\tilde{\mathbf{k}},m,\mathbf{r},\sigma} = \sum_l e^{-i(\tilde{\mathbf{k}}+l\frac{2\pi}{q}\mathbf{e}_x)\cdot \mathbf{r}} [\alpha_{\tilde{\mathbf{k}},\sigma}]_{l,m}.
\end{equation}

Calculation of $[\alpha_{\tilde{\mathbf{k}},\sigma}]$ scales as
$O(q^3)$. As there is $N/q$ different $\tilde{\mathbf{k}}$ to
consider, the first step scales as $O(Nq^2)$, with $q\leq L$, i.e., at
most $O(N^2)$. Then the calculation of
$W_{\tilde{\mathbf{k}},m,\mathbf{r},\sigma}$ scales as $O(q)$ but
there is $N$ different $\tilde{\mathbf{k}},m$ to consider, and we need
$N$ different $\mathbf{r}$, which is in total $O(N^2 q)$, i.e., at most $O(N^2 L)$, which is the bottleneck in the calculation. 
The calculation of each $G_{\mathbf{r},\mathbf{r}',\sigma}(z)$ then
scales as $O(N)$, but only if $G$ is diagonal in
$\tilde{\mathbf{k}},m$; if it is only diagonal in $\tilde{\mathbf{k}}$ but not in $m$, this scales as $O(Nq)$.
As
$\bar{G}_{\mathbf{r},\mathbf{r}'}=\bar{G}_{\mathbf{r}-\mathbf{r}'}$,
we only need to calculate $N$ different elements of the $\mathbf{G}$
matrix rather than all $N^2$ of them:
\begin{equation}
 \bar{G}_\mathbf{r} = e^{-if_{\mathbf{r},\mathbf{r}'=0}} G_{\mathbf{r},\mathbf{r}'=0}.
\end{equation}
In total, this scales as $O(N^2)$. Again, if $G$ is diagonal in $\tilde{\mathbf{k}}$
(as we expect it to be in the absence of translational symmetry breaking),
but not in $m$, then the scaling is $O(N^2q)$, which is still better than the direct matrix inverse.
When there is no translational symmetry (e.g. there is disorder), then the scaling is $O(N^4)$, which is worse than the direct matrix inverse.
In that case $\bar{G}$ is still gauge invariant, but is not translationally invariant, and all $N^2$ $\mathbf{r},\mathbf{r}'$-components need to be calculated.

Finally, we are interested in the spatial Fourier transform
\begin{equation}
 \bar{G}_\mathbf{k} = \sum_\mathbf{r} e^{i\mathbf{k}\cdot\mathbf{r}}
 \bar{G}_\mathbf{r},
\end{equation}
which will be discussed in Section~\ref{sec:Gbar_results}.

We note that other approaches might be possible for the efficient calculation of $\bar{G}$, e.g. the recursive scheme from Ref.~\onlinecite{berciu2010}.
     
  \subsection{Current density operator and the current-current correlation function}
     \subsubsection{Orbital space}

We will be interested in the direct current conductivity with respect to an infinitesimal uniform electric field. Such electric field $\mathbf{E}=\partial_t \mathbf{A}^\mathrm{ext}$, can be introduced with an additional vector potential $\mathbf{A}^\mathrm{ext}$ pointing uniformly in a given direction, and growing linearly with time. 
For the purposes of a linear-response calculation, the current couples to such vector potential instantaneously through
$-\int \mathbf{j}(\mathbf{r}) \cdot \mathbf{A}^\mathrm{ext}(\mathbf{r}) \mathrm{d}^3\mathbf{r}
= - v_\mathrm{cell} \sum_i \mathbf{j}_{\mathbf{r}_i}\cdot
\mathbf{A}^\mathrm{ext}_{\mathbf{r}_i}$, where $v_\mathrm{cell}=a^2c$
is the volume of the unit cell.
The additional Peierls phase coming from a $\mathbf{A}^\mathrm{ext}$ can therefore be safely rewritten within the slowly-varying field approximation
\begin{equation} \label{eq:Peirels_phase_nn_hopping}
  \frac{e}{\hbar}\int_{a\mathbf{r}_i}^{a\mathbf{r}_j} \mathbf{A}^\mathrm{ext}(\mathbf{r}) \cdot \mathrm{d}\mathbf{r} \approx \frac{ea}{\hbar} \mathbf{A}^\mathrm{ext}\cdot(\mathbf{r}_j-\mathbf{r}_i).
\end{equation}
In the case when we have just the nearest-neighbor hoppings (as in Eq.~\ref{eq:ham_nnhop}), the kinetic term in the Hamiltonian can be rewritten as
\begin{eqnarray} \label{eq:hkin_nngop_with_Aext}
 &&H_\mathrm{kin} \\ \nonumber
 &&= -t\sum_{i, \mathbf{u}\in\{\mathbf{e}_x,\mathbf{e}_y\},\sigma} e^{i\left(f_{\mathbf{r}_i,\mathbf{r}_i+\mathbf{u}}+\frac{e a}{\hbar} \mathbf{A}^\mathrm{ext}_{\mathbf{r}_i}\cdot\mathbf{u}\right)} c^\dagger_{\mathbf{r}_i,\sigma} c_{\mathbf{r}_i+\mathbf{u},\sigma} +\mathrm{H.c.}
\end{eqnarray}
without any additional approximation.

We can now derive the expression for the current {\it density}
operator (with units of $\unit{A/m^2}$) in the absence of electric field, by employing
\begin{eqnarray} \label{eq:local_current_operator}
 \mathbf{j}_\mathbf{r} &=& -\left.\frac{1}{v_\mathrm{cell}} \frac{\partial H}{\partial \mathbf{A}^\mathrm{ext}_\mathbf{r}}\right|_{\mathbf{A}^\mathrm{ext}\rightarrow 0} \\ \nonumber
  &=& it\frac{1}{ac} \frac{e}{\hbar}\sum_{\mathbf{\mathbf{u}\in\{\mathbf{e}_x,\mathbf{e}_y\},\sigma}}\mathbf{u} e^{if_{\mathbf{r}_i,\mathbf{r}_i+\mathbf{u}}} c^\dagger_{\mathbf{r}_i,\sigma} c_{\mathbf{r}_i+\mathbf{u},\sigma} +\mathrm{H.c.} \\
\end{eqnarray}
The vector component $\eta$ can be written as
\begin{equation} \label{eq:jetagamma}
 j^\eta_\mathbf{r} = it\frac{1}{ac} \frac{e}{\hbar}\sum_\sigma \gamma^\eta(\mathbf{r})\, c^\dagger_{\mathbf{r}_i,\sigma}c_{\mathbf{r}_i+\mathbf{e}_\eta,\sigma} + \mathrm{H.c.} 
\end{equation}
with $\boldsymbol{\gamma}(\mathbf{r}) = (1,e^{i\frac{e a^2}{\hbar} B_z x})$.

The current is an observable and it should be zero even in the presence of a magnetic field.
Commonly, one separates the current into the paramagnetic and
diamagnetic parts. In magnetic field they may be non-zero even in thermal equilibrium, but they must cancel. 
See Appendix~\ref{app:paramagnetic_and_diamagnetic} for details.

For the sake of generality, we define the current-current correlation function without assuming zero persistent currents
\begin{widetext}
\begin{eqnarray}
 \Lambda^{\eta\eta'}_{\mathbf{r},\mathbf{r}'}(\tau) 
 &=& \langle j^\eta_\mathbf{r}(\tau) j^{\eta'}_{\mathbf{r}'}(0)
 \rangle - \langle j^\eta_{\mathbf{r}} \rangle\langle j^{\eta'}_{\mathbf{r}'} \rangle \\ \nonumber
 &=& -t^2 \frac{1}{a^2c^2} \frac{e^2}{\hbar^2} \sum_{\sigma,\sigma'} \sum_{b,b'\in\{0,1\}} (-1)^{b+b'} C^b[\gamma_\eta(\mathbf{r})]  C^{b'}[\gamma_{\eta'}(\mathbf{r}')]\times \\ \nonumber
 && \times \langle c^\dagger_{\mathbf{r}+b\mathbf{e}_\eta,\sigma}(\tau^{+})
         c_{\mathbf{r}+(1-b)\mathbf{e}_\eta,\sigma}(\tau)
         c^\dagger_{\mathbf{r}'+b'\mathbf{e}_{\eta'},\sigma'}(0^{+})
         c_{\mathbf{r}'+(1-b')\mathbf{e}_{\eta'},\sigma'}(0)
         \rangle
 - \langle j^\eta_\mathbf{r} \rangle\langle j^{\eta'}_{\mathbf{r}'}
 \rangle,
\end{eqnarray}
\end{widetext}
where $C[...]$ is the operator of complex conjugation, and $C^0=1$.

We are interested in calculating the Kubo bubble, i.e. the
disconnected part. The disconnected part will have a static and a
dynamic term. The static one cancels the persistent current part, and
the dynamic term can be expressed in terms of the Green's function as
\begin{eqnarray} \label{eq:Lambda_disc_realspace}
 && \Lambda^{\eta\eta',\mathrm{disc}}_{\mathbf{r},\mathbf{r}'}(\tau) \\ \nonumber
  &&= t^2 \frac{1}{a^2 c^2} \frac{e^2}{\hbar^2} \sum_{\sigma} \sum_{b,b'\in\{0,1\}} (-1)^{b+b'} C^b[\gamma_\eta(\mathbf{r})]  C^{b'}[\gamma_{\eta'}(\mathbf{r}')]\\ \nonumber
 &&\;\;\; \times  G_{\mathbf{r}'+(1-b')\mathbf{e}_{\eta'},\mathbf{r}+b\mathbf{e}_\eta, \sigma}(-\tau)
         G_{\mathbf{r}+(1-b)\mathbf{e}_\eta,\mathbf{r}'+b'\mathbf{e}_{\eta'},\sigma}(\tau)
  .
\end{eqnarray}
We can now rewrite this expression in terms of $\bar{G}$. In the case of the longitudinal component 
\begin{eqnarray}
  && \Lambda^{xx,\mathrm{disc}}_{\mathbf{r},\mathbf{r}'}(\tau) \\ \nonumber
    && \;\; = t^2 \frac{1}{a^2c^2} \frac{e^2}{\hbar^2} \sum_{\sigma}
         \Bigg[
         \bar{G}_{\mathbf{r}'-\mathbf{r}+\mathbf{e}_{x}, \sigma}(-\tau)
         \bar{G}_{\mathbf{r}-\mathbf{r}'+\mathbf{e}_{x},\sigma}(\tau) \\ \nonumber
    &&\;\;\;\;\;                    
       +
         \bar{G}_{\mathbf{r}'-\mathbf{r}-\mathbf{e}_{x}, \sigma}(-\tau)
         \bar{G}_{\mathbf{r}-\mathbf{r}'-\mathbf{e}_{x},\sigma}(\tau)
          \\ \nonumber  
    && \;\;\;\;\;   - 2\cos\Big(\frac{ea^2B_z}{\hbar}(y-y')\Big)
         \bar{G}_{\mathbf{r}'-\mathbf{r}, \sigma}(-\tau)
         \bar{G}_{\mathbf{r}-\mathbf{r}',\sigma}(\tau)
         \Bigg].
         \\ \nonumber  
\end{eqnarray}
We see that the expression only depends on the distance which means
that it preserves translational symmetry, and is only expressed in
terms of gauge-invariant quantities. We have checked explicitly that
exactly the same expression is also obtained in the symmetric gauge.
Furthermore, this expression has all the expected spatial symmetries.
A completely analogous calculation for
$\Lambda^{yy,\mathrm{disc}}_{\mathbf{r},\mathbf{r}'}(\tau)$ yields the expression with $x,y \rightarrow y,x$. A general proof of the gauge invariance of $\Lambda^{\eta,\eta',\mathrm{disc}}_{\mathbf{r},\mathbf{r}'}(\tau)$ is given in Appendix \ref{app:Kubo_gauge_invariance}.

     \subsubsection{Momentum space}
As we have shown that the current-current correlation function satisfies all the desired spatial symmetries, we can proceed to discuss
the uniform current-current correlation function in a straight-forward manner by performing the spatial Fourier transform.
We have
\begin{equation}
 \Lambda^{\eta\eta'}_{\mathbf{q}=0}(\tau)
 = v_{\mathrm{cell}}\sum_{\mathbf{r}} \Lambda_{\mathbf{r},\mathbf{r}'=0}(\tau).
\end{equation}
This is followed by the Fourier transform in imaginary time to finally
obtain
\begin{equation}
 \Lambda^{\eta\eta'}_{\mathbf{q}=0}(i\nu)
 = v_{\mathrm{cell}}\sum_{\mathbf{r}} 
 \frac{1}{2\hbar} \int_{-\beta\hbar}^{\beta\hbar}\mathrm{d}\tau e^{i\nu\tau} \Lambda_{\mathbf{r},\mathbf{r}'=0}(\tau).
\end{equation}
We can rewrite this expression more conveniently using the uniform current operator as
\begin{equation}
 \Lambda^{\eta\eta'}_{\mathbf{q}=0}(i\nu)= \frac{V}{2\hbar}
 \int_{-\beta\hbar}^{\beta\hbar}\mathrm{d}\tau \langle j^\eta_{\mathbf{q}=0}(\tau)j^{\eta'}_{\mathbf{q}=0}(0) \rangle,
\end{equation}
where $V$ is the total volume $V=Nv_\mathrm{cell}$ and
\begin{equation} \label{eq:uniform_current_operator}
 j^\eta_{\mathbf{q}=0} = \frac{1}{N}\sum_\mathbf{r}j^\eta_\mathbf{r}.
\end{equation}
Note that we have here defined the uniform current operator as the average current (density) operator, rather than the spatial Fourier transform of the current operator.
Using the creation/annihilation operators in the eigenbasis of the non-interacting Hamiltonian, we can write
\begin{eqnarray} \nonumber
 j^\eta_{\mathbf{q}=0} &=& 
 \frac{it}{N}\frac{1}{ac}\frac{e}{\hbar}\sum_\sigma \sum_{\tilde{\mathbf{k}},m,m'}
 v^\eta_{\tilde{\mathbf{k}},m,m',\sigma}
 c^\dagger_{\tilde{\mathbf{k}},m,\sigma} c_{\tilde{\mathbf{k}},m',\sigma} \\
\end{eqnarray}
with
\begin{eqnarray}\label{eq:vx}
v^x_{\tilde{\mathbf{k}},m,m',\sigma} &=&
  \sum_{l}
   [\alpha_{\tilde{\mathbf{k}},\sigma}]_{l,m}
   [\alpha_{\tilde{\mathbf{k}},\sigma}]^*_{l,m'} \\ \nonumber
 && \;\;\;\times  \Big[
 e^{i\tilde{k}_x}
 e^{il\frac{2\pi}{q} }
-
 e^{-i\tilde{k}_x}
 e^{-il\frac{2\pi}{q} } \Big]
\end{eqnarray}
and
\begin{eqnarray}\label{eq:vy}
v^y_{\tilde{\mathbf{k}},m,m',\sigma} &=&
  \sum_{l} [\alpha_{\tilde{\mathbf{k}},\sigma}]_{l,m} 
  \\ \nonumber 
  &&\;\;\;\times \Big[
 e^{i\tilde{k}_y}    
   [\alpha_{\tilde{\mathbf{k}},\sigma}]^*_{l\ominus p,m'}    
   -e^{-i\tilde{k}_y}   
   [\alpha_{\tilde{\mathbf{k}},\sigma}]^*_{l\oplus p,m'}
 \Big].
\end{eqnarray}
The proof for the above expressions is given in Appendix~\ref{app:vx_proofs}.
Assuming no persistent currents, the uniform current-current correlation function is therefore
\begin{eqnarray} \label{eq:lambda_full_orbital_time}
 &&\Lambda^{\eta\eta'}_\mathbf{q=0}(i\nu) \\ \nonumber
 &&= -\frac{t^2e^2}{c\hbar^2} \frac{1}{N}
  \sum_{\sigma_1,\sigma_2}  \frac{1}{2\hbar} \int_{-\beta\hbar}^{\beta\hbar} \mathrm{d}\tau e^{i\nu\tau}
  \\ \nonumber
  &&\;\;\;\times\sum_{\tilde{\mathbf{k}}_1,m_1,m'_1} \sum_{\tilde{\mathbf{k}}_2,m_2,m'_2}
 v^\eta_{\tilde{\mathbf{k}}_1,m_1,m'_1,\sigma_1}
 v^{\eta'}_{\tilde{\mathbf{k}}_2,m_2,m'_2,\sigma_2} \\ \nonumber
 &&\;\;\;\times \langle 
 c^\dagger_{\tilde{\mathbf{k}}_1,m_1,\sigma_1}(\tau^+) c_{\tilde{\mathbf{k}}_1,m'_1,\sigma_1} (\tau)
 c^\dagger_{\tilde{\mathbf{k}}_2,m_2,\sigma_2}(0^+)c_{\tilde{\mathbf{k}}_2,m'_2,\sigma_2} (0)
 \rangle.
\end{eqnarray}

The disconnected part written as a function of bosonic Matsubara frequency reads (see Appendix~\ref{app:uniform_lambda_disc_freq} for proof):
\begin{eqnarray}\label{eq:bubble_freq2}
 && \Lambda^{\eta\eta', \mathrm{disc}}_\mathbf{q=0}(i\nu) = 
 \frac{t^2e^2}{c\hbar^2} \frac{1}{N} \sum_\sigma \sum_{\tilde{\mathbf{k}},m_1,m_1',m_2,m_2'} \\ \nonumber
 && \times 
 v^\eta_{\tilde{\mathbf{k}},m_1,m_1',\sigma}
 v^{\eta'}_{\tilde{\mathbf{k}},m_2,m_2',\sigma} \\ \nonumber
 && \times \frac{1}{\beta}\sum_{i\omega}G_{\tilde{\mathbf{k}},m_2',m_1\sigma}(i\omega)
 G_{\tilde{\mathbf{k}},m_1',m_2,\sigma}(i\omega+i\nu).
\end{eqnarray}

  \subsection{Method}
    \subsubsection{DMFT}

In dynamical mean field theory (DMFT), the lattice problem is mapped onto a set of self-consistent local impurity problems on each lattice site $i$, defined by the action
\cite{RevModPhys.68.13,potthoff1999,potthoff1999prb1,potthoff1999prb2}
\begin{eqnarray}\nonumber
S^{\imp}_i &=& \sum_\sigma \int \mathrm{d}\tau \mathrm{d}\tau' c^+_{i,\sigma}(\tau) [-{\cal G}_{0,i}^{-1} ] (\tau-\tau') c_{i,\sigma}(\tau')\\ 
 && + U \int \mathrm{d}\tau c^+_{i,\up}(\tau) c_{i,\up}(\tau) c^+_{i,\dn}(\tau) c_{i,\dn}(\tau).
\end{eqnarray}
The bare propagator ${\cal G}_{0,i}$ in the impurity problem $i$ is determined self-consistently, so that the Green's function in each impurity problem is equal to the local Green's function on the site of the impurity problem, assuming that the self-energy on the lattice is local and on each site equal to the self-energy of the corresponding impurity problem. This self-consistency condition can be written as
\begin{equation}
 {\cal G}_{0,i}(z) = 1/([G^{-1}]_{ii}(z) + \Sigma^{\imp}_i(z)),
\end{equation}
where $\Sigma^{\imp}_i(z)$ is the self-energy calculated in the
impurity problem at site $i$, and the lattice Green's function is calculated as a matrix in the site space as
\begin{equation}
 \mathbf{G}(z) = [\mathbf{I}\hbar z - \mathbf{H}_0[\mathbf{A}] -
 \mathrm{diag}(\Sigma^\imp(z))]^{-1},
\end{equation}
where $\mathrm{diag}(\Sigma^\imp)$ is a diagonal matrix, with
$\Sigma^{\imp}_i(z)$ entries on the diagonal. This construction is
general and can be used in the presence of translational symmetry breaking fields, disorder, and even used to probe spatially ordered phases.
The DMFT approximation notably becomes exact in the limit of infinite coordination number, where the self-energy can be shown to be fully local\cite{RevModPhys.68.13}, at least in the absence of magnetic fields.

We see that in the calculation of the bare propagator for the impurity problems, only the local Green's function plays a role, and this quantity is gauge invariant and spatially uniform. Therefore, even in the presence of the uniform magnetic field, all impurity problems are equivalent, and we may solve only one impurity problem and calculate the lattice Green's function as 
\begin{equation}
 \mathbf{G}(z) = [\mathbf{I}\hbar z - \mathbf{H}_0[\mathbf{A}]-\mathbf{I}\Sigma^\mathrm{imp}(z)]^{-1}.
\end{equation}
This leads to further simplifications.
First, a local and spatially uniform self-energy is diagonal in the non-interacting eigenbasis ($\Sigma_{\sigma,ij} = \delta_{ij} \Sigma_\sigma \Longrightarrow \langle \tilde{\mathbf{k}},\sigma,m|\mathbf{\Sigma}|\tilde{\mathbf{k}}',\sigma,m'\rangle = \delta_{\tilde{\mathbf{k}},\tilde{\mathbf{k}}'} \delta_{m,m'} \Sigma_\sigma$, see Appendix~\ref{app:dmft_Sigma_is_diagonal} for proof)
which means that the lattice Green's function is diagonal as well:
\begin{equation} \label{eq:dmft_Gk_diagonal_in_m_space}
 G_{\tilde{\mathbf{k}},m,m',\sigma}(z) = \delta_{mm'}
 G_{\tilde{\mathbf{k}},mm,\sigma}(z), \\
\end{equation}
thus we can drop the second eigenstate index and simply calculate the lattice Green's function as
\begin{equation}
G_{\tilde{\mathbf{k}},m,\sigma}(z) = \frac{1}{\hbar z-\varepsilon_{\tilde{\mathbf{k}},m,\sigma} - \Sigma_{\sigma}(z)}.
\end{equation}
The local Green's function can then be obtained at low numerical cost from the knowledge of the local density of states, $\rho_0(\varepsilon)$, as (see Appendix~\ref{app:dmft_gloc} for proof):
\begin{eqnarray} \label{eq:dmft_Gloc}
G_{ii,\sigma}(z)
&=& \int \mathrm{d}\varepsilon \frac{\rho_0(\varepsilon) }{\hbar
z-\varepsilon - \Sigma_{\sigma}(z)}.
\end{eqnarray}
Therefore, the DMFT calculation for the Hubbard model in the magnetic
field proceeds as the standard DMFT, and all the effects of the gauge
field are contained in the non-interacting density of states\cite{Acheche2017}. In all
our calculations we employ the NRG impurity solver
\cite{wilson1975,krishna1980a,bulla2008,resolution} which works
directly in real-frequency space, so no analytical continuation is needed to perform calculations of conductivity. 
The NRG solver has been previously thoroughly cross checked in
Refs.\onlinecite{deng,Vucicevic2019,Vranic2020}.

    \subsubsection{Calculation of conductivity in DMFT}  
      
The fact that the Green's function is diagonal in the non-interacting
eigenbasis leads to a simplification in the Kubo bubble,
Eq.~\ref{eq:bubble_freq2}. One is left with only two summations over eigenstates $m$:
\begin{eqnarray}\label{eq:dmft_bubble_imfreq0}
 && \Lambda^{\eta\eta', \mathrm{disc}}_\mathbf{q=0}(i\nu) =
 \frac{t^2e^2}{c\hbar^2} \frac{1}{N} \sum_\sigma \sum_{\tilde{\mathbf{k}},m,m'} \frac{1}{\beta}\sum_{i\omega} \\ \nonumber
 && \times v^\eta_{\tilde{\mathbf{k}},m,m',\sigma}
 v^{\eta'}_{\tilde{\mathbf{k}},m',m,\sigma} 
 G_{\tilde{\mathbf{k}},m\sigma}(i\omega+i\nu)
 G_{\tilde{\mathbf{k}},m',\sigma}(i\omega)
\end{eqnarray}
Furthermore, because the self-energy is local, the Green's function only depends on the energy of the eigenstate, so we can define $G(\varepsilon_{\tilde{\mathbf{k}},m,\sigma},i\omega) \equiv G_{\tilde{\mathbf{k}},m,\sigma}(i\omega)$ and rewrite
\begin{eqnarray}\label{eq:dmft_bubble_imfreq}
 \Lambda^{\eta\eta', \mathrm{disc}}_\mathbf{q=0}(i\nu) &=&
 \frac{t^2e^2}{c\hbar^2} \sum_\sigma \frac{1}{\beta}\sum_{i\omega} 
 \int \mathrm{d}\varepsilon\int \mathrm{d}\varepsilon' \\ \nonumber
 && \times v^{\eta,\eta'}_\sigma(\varepsilon,\varepsilon')
 G(\varepsilon,\; i\omega+i\nu)
   G(\varepsilon',\; i\omega)
\end{eqnarray}
with
\begin{eqnarray}
 &&v^{\eta,\eta'}_\sigma(\varepsilon,\varepsilon') \equiv \\ \nonumber 
 &&\;\;\frac{1}{N}\sum_{\tilde{\mathbf{k}},m,m'} 
 \delta(\varepsilon-\varepsilon_{\tilde{\mathbf{k}},m,\sigma})
 \delta(\varepsilon'-\varepsilon_{\tilde{\mathbf{k}},m',\sigma})
 v^\eta_{\tilde{\mathbf{k}},m,m',\sigma}
 v^{\eta'}_{\tilde{\mathbf{k}},m',m,\sigma}.
\end{eqnarray}
In Landau gauge, one has the symmetry $v^x_{\tilde{\mathbf{k}},m,m',\sigma} = -(v^x_{\tilde{\mathbf{k}},m',m,\sigma})^*$ (see Appendix \ref{app:vx_proofs}), and therefore
$v^x_{\tilde{\mathbf{k}},m,m',\sigma}
 v^{x}_{\tilde{\mathbf{k}},m',m,\sigma} = -|v^{x}_{\tilde{\mathbf{k}},m',m,\sigma}|^2
$, which means $v^{xx}_\sigma(\varepsilon,\varepsilon')$ is purely real.
On the contrary, as already noted in Ref.\onlinecite{Markov2019}, $v^{xy}_\sigma(\varepsilon,\varepsilon')$ is purely imaginary.

The {\it sheet conductance} is related to the current-current correlation function through:
\begin{equation}
\sigma^{\eta\eta'}(\nu) =
c\frac{\Lambda^{\eta\eta'}(\nu)-\Lambda^{\eta\eta'}(\nu=0)}{i\;\nu},
\end{equation}
The $z$-axis lattice constant $c$ cancels out $c$ from
$v_\mathrm{cell}=a^2c$ and its value is irrelevant. In the following
we will discard the difference between the sheet conductance and the
conductivity, and refer to $\sigma$ as conductivity, even though it is
actually sheet conductance and the units of the two quantities are
different ($(\Omega m)^{-1}$ vs. $\Omega^{-1}$, respectively); this is common practice in the field.

After several lines of algebra aimed at the analytical continuation to
the real-axis frequency (see Appendix~\ref{app:analytical_continuation_lambda}), we obtain
\begin{eqnarray} \label{eq:sigma_xx_final}
 &&\mathrm{Re}\sigma^{xx, \mathrm{disc}}_\mathbf{q=0}(\nu=0) \\ \nonumber
 &&= t^2 \frac{e^2}{\hbar} \frac{1}{\pi} \sum_\sigma 
 \int \mathrm{d}\varepsilon\int \mathrm{d}\varepsilon'
 v^{xx}_\sigma(\varepsilon,\varepsilon') \int \mathrm{d}\omega \\ \nonumber
 &&\;\;\;\times \mathrm{Im}G(\varepsilon,\; \omega)\mathrm{Im}G(\varepsilon',\; \omega)
 n'_\mathrm{F}(\omega),
\end{eqnarray}
where $n'_\mathrm{F}(\omega)=-\beta \hbar e^{\beta\hbar \omega}/(1+e^{\beta\hbar \omega})^2$ is the derivative of the Fermi function.

For Hall conductivity one obtains\cite{Markov2019}
\begin{eqnarray}\label{eq:sigma_xy_final}
 &&\mathrm{Re}\sigma^{xy, \mathrm{disc}}_\mathbf{q=0}(\nu=0) \\ \nonumber
 &&= -t^2 \frac{e^2}{\hbar} \frac{1}{\pi^2}\sum_\sigma 
 \int \mathrm{d}\varepsilon\int \mathrm{d}\varepsilon'
 \mathrm{Im} v^{xy}_\sigma(\varepsilon,\varepsilon')
 \int \mathrm{d}\omega \int \mathrm{d}\omega' \\ \nonumber
 &&\;\;\; \times\mathrm{Im}G(\varepsilon,\; \omega)\mathrm{Im}G(\varepsilon',\; \omega')
 \frac{n_\mathrm{F}(\omega)-n_\mathrm{F}(\omega')}{(\omega-\omega')^2}.
\end{eqnarray}
An additional simplification is possible in the case of Hall conductivity when $U=0$. In that case we have\cite{Thouless1982}
\begin{eqnarray}\label{eq:sigma_xy_U0}
 &&\mathrm{Re}\sigma^{xy, \mathrm{disc}}_\mathbf{q=0}(\nu=0; U=0) \\ \nonumber
 &&= -t^2 \frac{e^2}{\hbar} \sum_\sigma 
 \int \mathrm{d}\varepsilon\int \mathrm{d}\varepsilon'
 \mathrm{Im} v^{xy}_\sigma(\varepsilon,\varepsilon')
 \frac{n_\mathrm{F}(\varepsilon)-n_\mathrm{F}(\varepsilon')}{(\varepsilon-\varepsilon')^2}.
\end{eqnarray}
Finally, the resistivity is obtained as a matrix inverse
\begin{equation}
 \left(\begin{array}{cc}
  \rho^{xx} & \rho^{xy} \\
  \rho^{yx} & \rho^{yy} 
 \end{array}\right)
 =
 \left(\begin{array}{cc}
  \sigma^{xx} & \sigma^{xy} \\
  \sigma^{yx} & \sigma^{yy} 
 \end{array}\right)^{-1}.
\end{equation}
      
    \subsubsection{Vertex corrections in DMFT}
      
Finally, to calculate the full current-current correlation function,
one should in principle also compute the vertex corrections. At the
level of the DMFT, in the absence of an external magnetic field, the vertex corrections cancel due to the 
well-known argument due to Khurana \cite{khurana1990}. As we show in
the following, a generalized Khurana argument holds even in the
presence of the magnetic field. In the following we first derive the Khurana argument in real space, and then generalize it to the case of non-zero magnetic fields.     

\begin{figure}[ht!]
 \begin{center}
 \includegraphics[width=3.2in, trim=0cm 0cm 0cm 0cm, clip]{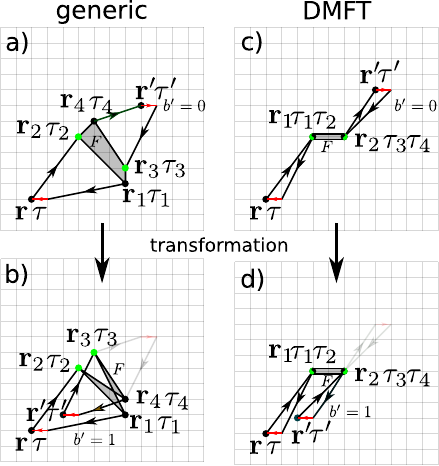}
  \end{center}
 \caption{Real-space diagrammatic representation of a contribution to the connected part of the current-current correlation function.
$F$ is the full vertex, red segments indicate a term in the current operator connecting two nearest neighbor sites in the $x$ direction; swapping the direction of the red arrow changes its sign.
Top panels: for generic case (left, Eq.~\ref{eq:lambda_conn_general})
and DMFT (right, Eq.~\ref{eq:lambda_conn_dmft}).
Bottom panels: after the transformation in
Eq.~\ref{eq:transformation_general} (left) and Eq.~\ref{eq:transformation_dmft} (right).
 }
 \label{fig:lambda_diags}
\end{figure}

We start first by writing the vertex corrections in orbital space, in the most general way.
\begin{eqnarray} \label{eq:lambda_conn_general}
 && \Lambda^{\eta\eta',\mathrm{conn}}_{\mathbf{q}=0}(\tau-\tau') \\ \nonumber 
  &&= t^2\frac{e^2}{a^2c^2\hbar^2}\sum_{\sigma,\sigma'} \sum_{b,b'\in\{0,1\}} (-1)^{b+b'} 
  \\ \nonumber
 &&\;\;\; \times \frac{1}{N^2}\sum_{\mathbf{r},\mathbf{r}'}
  C^b[\gamma_\eta(\mathbf{r})]  C^{b'}[\gamma_{\eta'}(\mathbf{r}')]  
   \sum_{\mathbf{r}_1,\mathbf{r}_2,\mathbf{r}_3,\mathbf{r}_4} \int \mathrm{d}\tau_1 \mathrm{d}\tau_2 \mathrm{d}\tau_3 \mathrm{d}\tau_4 \\ \nonumber
 &&\;\;\; \times  
         G_{\mathbf{r}_1,\mathbf{r}+b\mathbf{e}_\eta, \sigma}(\tau_1-\tau)
         G_{\mathbf{r}+(1-b)\mathbf{e}_\eta,\mathbf{r}_2, \sigma}(\tau-\tau_2)\\ \nonumber
 &&\;\;\; \times 
         F((\mathbf{r}_1,\tau_1),(\mathbf{r}_2,\tau_2),(\mathbf{r}_3,\tau_3),(\mathbf{r}_4,\tau_4))  \\ \nonumber
 &&\;\;\; \times
         G_{\mathbf{r}'+(1-b')\mathbf{e}_{\eta'},\mathbf{r}_3,\sigma'}(\tau'-\tau_3)
         G_{\mathbf{r}_4,\mathbf{r}'+b'\mathbf{e}_{\eta'},\sigma'}(\tau_4-\tau'),
\end{eqnarray}
as illustrated in Fig.~\ref{fig:lambda_diags}a. In the absence of external magnetic field or spontaneous symmetry breaking, $G_{\mathbf{r},\mathbf{r}'}$ has full lattice symmetry and depends only on the difference $G_{\mathbf{r}-\mathbf{r}'}$. 
However, there is no symmetry operation which guarantees cancellation of all terms.
There is only one symmetry operation that leaves the Green's functions intact and flips the overall sign, but it does change the full vertex.
It can be formulated either for internal variables 
$\mathbf{r}_1,\mathbf{r}_2,b$ or for $\mathbf{r}_3,\mathbf{r}_4,b'$,
and we illustrate the latter case in Fig.~\ref{fig:lambda_diags}b.
The transformation can be formulated as follows
\begin{eqnarray} \nonumber
\mathbf{r}'&\rightarrow& -\mathbf{r}'+\mathbf{r}_3+\mathbf{r}_4 -
\mathbf{e}_{\eta'}, \\ \label{eq:transformation_general}
b'&\rightarrow& 1- b', \\ \nonumber
(\mathbf{r}_3,\tau_3)&\leftrightarrow&(\mathbf{r}_4,\tau_4).
\end{eqnarray}
The flip of $b'$ changes the overall sign, but the exchange of terminals of the full vertex function changes its value in no obvious way, and there is no cancellation in the general case.

However, in DMFT there is an additional simplification that the full
vertex $F$ depends on only two spatial indices:
\begin{eqnarray} \label{eq:lambda_conn_dmft}
 && \Lambda^{\eta\eta',\mathrm{conn}}_{\mathbf{q}=0}(\tau-\tau') \\ \nonumber 
  &&= t^2\frac{e^2}{a^2c^2\hbar^2}\sum_{\sigma,\sigma'} \sum_{b,b'\in\{0,1\}} (-1)^{b+b'} 
  \\ \nonumber
 &&\;\;\; \times  \frac{1}{N^2}\sum_{\mathbf{r},\mathbf{r}'}
  C^b[\gamma_\eta(\mathbf{r})]  C^{b'}[\gamma_{\eta'}(\mathbf{r}')]   \sum_{\mathbf{r}_1,\mathbf{r}_2,} \int \mathrm{d}\tau_1 \mathrm{d}\tau_2 \mathrm{d}\tau_3 \mathrm{d}\tau_4 \\ \nonumber
 &&\;\;\; \times  
         G_{\mathbf{r}_1,\mathbf{r}+b\mathbf{e}_\eta, \sigma}(\tau_1-\tau)
         G_{\mathbf{r}+(1-b)\mathbf{e}_\eta,\mathbf{r}_1, \sigma}(\tau-\tau_2)\\ \nonumber
 &&\;\;\; \times 
         F((\mathbf{r}_1,\tau_1),(\mathbf{r}_1,\tau_2),(\mathbf{r}_2,\tau_3),(\mathbf{r}_2,\tau_4))  \\ \nonumber
 &&\;\;\; \times
         G_{\mathbf{r}'+(1-b')\mathbf{e}_{\eta'},\mathbf{r}_2,\sigma'}(\tau'-\tau_3)
         G_{\mathbf{r}_2,\mathbf{r}'+b'\mathbf{e}_{\eta'},\sigma'}(\tau_4-\tau').
\end{eqnarray}
In that case, a transformation 
\begin{eqnarray} \label{eq:transformation_dmft}
\mathbf{r}' &\rightarrow& -\mathbf{r}' - \mathbf{e}_{\eta'} +
2\mathbf{r}_2, \\ \nonumber
b'&\rightarrow& 1- b' 
\end{eqnarray}
keeps both the Green's functions and the full vertex intact, while changing the overall sign.
Then the two symmetry-connected terms together read (up to a prefactor)
\begin{eqnarray} \label{eq:vertex_correction_cancellation_criterion}
&&    G_{\mathbf{r}'+(1-b')\mathbf{e}_{\eta'},\mathbf{r}_2,\sigma'}(\tau'-\tau_3) \\ \nonumber
&&   \times      G_{\mathbf{r}_2,\mathbf{r}'+b'\mathbf{e}_{\eta'},\sigma'}(\tau_4-\tau') \\ \nonumber
&& -  G_{ -\mathbf{r}' + 2\mathbf{r}_2-(1-b')\mathbf{e}_{\eta'},\mathbf{r}_2,\sigma'}(\tau'-\tau_3) \\ \nonumber
&&   \times      G_{\mathbf{r}_2, -\mathbf{r}' + 2\mathbf{r}_2-b'\mathbf{e}_{\eta'},\sigma'}(\tau_4-\tau').
\end{eqnarray}
When the lattice preserves inversion symmetry, one has $G_{\mathbf{r},\mathbf{r}'}=G_{\mathbf{r}-\mathbf{r}'}=G_{\mathbf{r}'-\mathbf{r}}$, and the above two terms always cancel.
This is the real-space version of the Khurana argument.

However, when there is magnetic field, there are additional complications.
Nevertheless, the full vertex is a gauge invariant quantity, as the irreducible vertex in the particle-hole channel $\Gamma^\mathrm{ph}$ is fully local and therefore gauge invariant and spatially uniform $\Gamma^\mathrm{ph}_{ijkl} = \delta_{ij}\delta_{jk}\delta_{kl}\Gamma^\mathrm{ph}$. This comes as DMFT is the local approximation of the Luttinger-Ward functional\cite{RevModPhys.68.13,Vucicevic2018}, and $\Gamma^\mathrm{ph}_{ijkl}=\left.\frac{\partial^2 \Phi[\mathbf{G}]}{\partial G_{ij} \partial G_{kl}}\right|_{\mathbf{G}=\mathbf{G}_\mathrm{exact}}\approx\left.\frac{\partial^2 \Phi^{\mathrm{DMFT}}[\{G_{ii}\}_i]}{\partial G_{ij} \partial G_{kl}}\right|_{\mathbf{G}=\mathbf{G}_\mathrm{DMFT}}$. Therefore we have
\begin{eqnarray}
F_{ii,jj}&=&\delta_{ij}\Gamma^\mathrm{ph}+\Gamma^\mathrm{ph} G_{ij}G_{ji} \Gamma^\mathrm{ph} \\ \nonumber 
       && + \sum_l \Gamma^\mathrm{ph} G_{il}G_{li} \Gamma^\mathrm{ph} G_{lj}G_{jl} \Gamma^\mathrm{ph} + ... \\ \nonumber
       &=&\delta_{ij}\Gamma^\mathrm{ph}+\Gamma^\mathrm{ph} \bar{G}_{ij}\bar{G}_{ji} \Gamma^\mathrm{ph} \\ \nonumber 
       && + \sum_l \Gamma^\mathrm{ph} \bar{G}_{il}\bar{G}_{li} \Gamma^\mathrm{ph} \bar{G}_{lj}\bar{G}_{jl} \Gamma^\mathrm{ph} + ... 
\end{eqnarray}
and $F$ is clearly expressed entirely with gauge-invariant quantities.
Here we have omitted spin and temporal arguments and the corresponding sums/integrals for the sake of brevity, as they do not play a role in the proof.

In the presence of the magnetic field,
the Green's function does not satisfy $G_{\mathbf{r},\mathbf{r}'}=G_{\mathbf{r}-\mathbf{r}'}=G_{\mathbf{r}'-\mathbf{r}}$
and it is not a priori clear that the terms in
Eq.\ref{eq:vertex_correction_cancellation_criterion} cancel. We can,
however, rewrite them in terms of $\bar{G}$:
\begin{eqnarray}
&&         e^{if_{\mathbf{r}'+(1-b')\mathbf{e}_{\eta'},\mathbf{r}_2}}\bar{G}_{\mathbf{r}'+(1-b')\mathbf{e}_{\eta'},\mathbf{r}_2,\sigma'}(\tau'-\tau_3) \\ \nonumber
&& \times        e^{if_{\mathbf{r}_2,\mathbf{r}'+b'\mathbf{e}_{\eta'}}}\bar{G}_{\mathbf{r}_2,\mathbf{r}'+b'\mathbf{e}_{\eta'},\sigma'}(\tau_4-\tau') \\ \nonumber
&& -
         e^{if_{ -\mathbf{r}' + 2\mathbf{r}_2-(1-b')\mathbf{e}_{\eta'},\mathbf{r}_2}}\bar{G}_{ -\mathbf{r}' + 2\mathbf{r}_2-(1-b')\mathbf{e}_{\eta'},\mathbf{r}_2,\sigma'}(\tau'-\tau_3) \\ \nonumber
&&\;\; \times         e^{if_{\mathbf{r}_2, -\mathbf{r}' + 2\mathbf{r}_2-b'\mathbf{e}_{\eta'}}}\bar{G}_{\mathbf{r}_2, -\mathbf{r}' + 2\mathbf{r}_2-b'\mathbf{e}_{\eta'},\sigma'}(\tau_4-\tau').
\end{eqnarray}
As $\bar{G}$ satisfies $\bar{G}_{\mathbf{r},\mathbf{r}'}=\bar{G}_{\mathbf{r}-\mathbf{r}'}=\bar{G}_{\mathbf{r}'-\mathbf{r}}$, the products of $\bar{G}$ are the same in both terms, thus what determines whether there is cancellation or not is
\begin{eqnarray}
&&         e^{if_{\mathbf{r}'+(1-b')\mathbf{e}_{\eta'},\mathbf{r}_2}}
      e^{if_{\mathbf{r}_2,\mathbf{r}'+b'\mathbf{e}_{\eta'}}} \\ \nonumber
&& -
         e^{if_{ -\mathbf{r}' + 2\mathbf{r}_2-(1-b')\mathbf{e}_{\eta'},\mathbf{r}_2}}
         e^{if_{\mathbf{r}_2, -\mathbf{r}' + 2\mathbf{r}_2-b'\mathbf{e}_{\eta'}}}.
\end{eqnarray}
In Landau gauge and for, say, $\eta'=x$, we get
\begin{eqnarray}
&&         e^{i\frac{ea^2B_z}{2\hbar}(y_2-y')(x'+x_2+(1-b'))}
      e^{i\frac{ea^2B_z}{2\hbar}(y'-y_2)(x'+x_2+b')} \\ \nonumber
&& \;\;-  e^{i\frac{ea^2B_z}{2\hbar}(-y_2+y')(3x_2-x'-(1-b'))}
         e^{i\frac{ea^2B_z}{2\hbar}(-y'+y_2)(-x'-b' + 3x_2)}     \\ \nonumber
&&=   e^{i\frac{ea^2B_z}{2\hbar}(y_2-y')(1-2b')}      \\ \nonumber
&& \;\;-  e^{i\frac{ea^2B_z}{2\hbar}B_z(-y_2+y')(-1+2b')}    \\ \nonumber           
&&= 0.
\end{eqnarray}
This means that the vertex corrections cancel. This proof is immediately valid for both $\Lambda^{xx}$ and $\Lambda^{yx}$. Having in mind a completely analogous transformation of $\mathbf{r}$ and $b$, this proof holds also for $\Lambda^{xy}$ (a different proof was given for $\Lambda^{xy}$ in Ref.~\onlinecite{Markov2019}).
However, we also want to check what happens with $\eta'=y$, which is
relevant for $\Lambda^{yy}$. In that case, the transformation affects also $C^{b'}[\gamma_y(\mathbf{r}')]$ so we need to take that into account:
\begin{eqnarray}
&&   e^{i(1-2b')\frac{e a^2}{\hbar} B_z x'}
     e^{if_{\mathbf{r}'+(1-b')\mathbf{e}_{\eta'},\mathbf{r}_2}}
     e^{if_{\mathbf{r}_2,\mathbf{r}'+b'\mathbf{e}_{\eta'}}} \\ \nonumber
&& - e^{-i(1-2b')\frac{e a^2}{\hbar} B_z (-x'+2x_2)} \\ \nonumber
&&   \;\times e^{if_{ -\mathbf{r}' + 2\mathbf{r}_2-(1-b')\mathbf{e}_{\eta'},\mathbf{r}_2}}
     e^{if_{\mathbf{r}_2, -\mathbf{r}' + 2\mathbf{r}_2-b'\mathbf{e}_{\eta'}}} \\ \nonumber
&&=  e^{i(1-2b')\frac{e a^2}{\hbar} B_z x'} \\ \nonumber
&&   \;\;\;\;\times e^{i\frac{ea^2B_z}{2\hbar}(y_2-y'-(1-b'))(x_2+x')}
     e^{i\frac{ea^2B_z}{2\hbar}(y'+b'-y_2)(x_2+x')} \\ \nonumber
&& \;\;\;- e^{-i(1-2b')\frac{e a^2}{\hbar} B_z (-x'+2x_2)} \\ \nonumber
&&   \;\;\;\;\times e^{i\frac{ea^2B_z}{2\hbar}(y'-y_2+(1-b'))(3x_2-x')} \\ \nonumber
&&   \;\;\;\;\times e^{i\frac{ea^2B_z}{2\hbar}(-y'+y_2-b')(-x'+3x_2)} \\ \nonumber
&&=  e^{i(1-2b')\frac{e a^2}{\hbar} B_z x'} 
     e^{i\frac{ea^2B_z}{2\hbar}(-1+2b')(x_2+x')}
      \\ \nonumber
&& \;\;\;- e^{-i(1-2b')\frac{e a^2}{\hbar} B_z (-x'+2x_2)}
     e^{i\frac{ea^2B_z}{2\hbar}((1-2b')(3x_2-x')}     \\ \nonumber     
&&=  e^{i\frac{e a^2B_z}{2\hbar}  [2(1-2b')x' - (1-2b')(x_2+x')] }      
      \\ \nonumber
&& \;\;\;- e^{i\frac{e a^2B_z}{2\hbar}  [2(1-2b')x'-4(1-2b')x_2+(1-2b')(3x_2-x')]}
          \\ \nonumber          
&&=  e^{i\frac{e a^2B_z}{2\hbar}  (x' -2b'x'- x_2 + 2b'x_2) }      
      \\ \nonumber
&& \;\;\;- e^{i\frac{e a^2B_z}{2\hbar}  (x' -2b'x'- x_2 + 2b'x_2)}
          \\ \nonumber                  
&&=0.
\end{eqnarray}
Indeed, the vertex corrections for $\Lambda^{yy}$ cancel as well.
As we have shown that the Kubo bubble is gauge invariant (see Appendix \ref{app:Kubo_gauge_invariance}), and having that the full correlation function needs to be gauge invariant as it relates to observables, the proof given here is fully general, even though it is formulated in Landau gauge. The proof also does not depend on whether there is Zeeman term in the Hamiltonian or not.
      
\section{Results}
  \subsection{Density of states and spectral function}

 \subsubsection{Non-interacting density of states}
      
\begin{figure}[ht!]
 \begin{center}
 \includegraphics[page=1,width=3.2in, trim=0cm 0cm 0cm 0cm, clip]{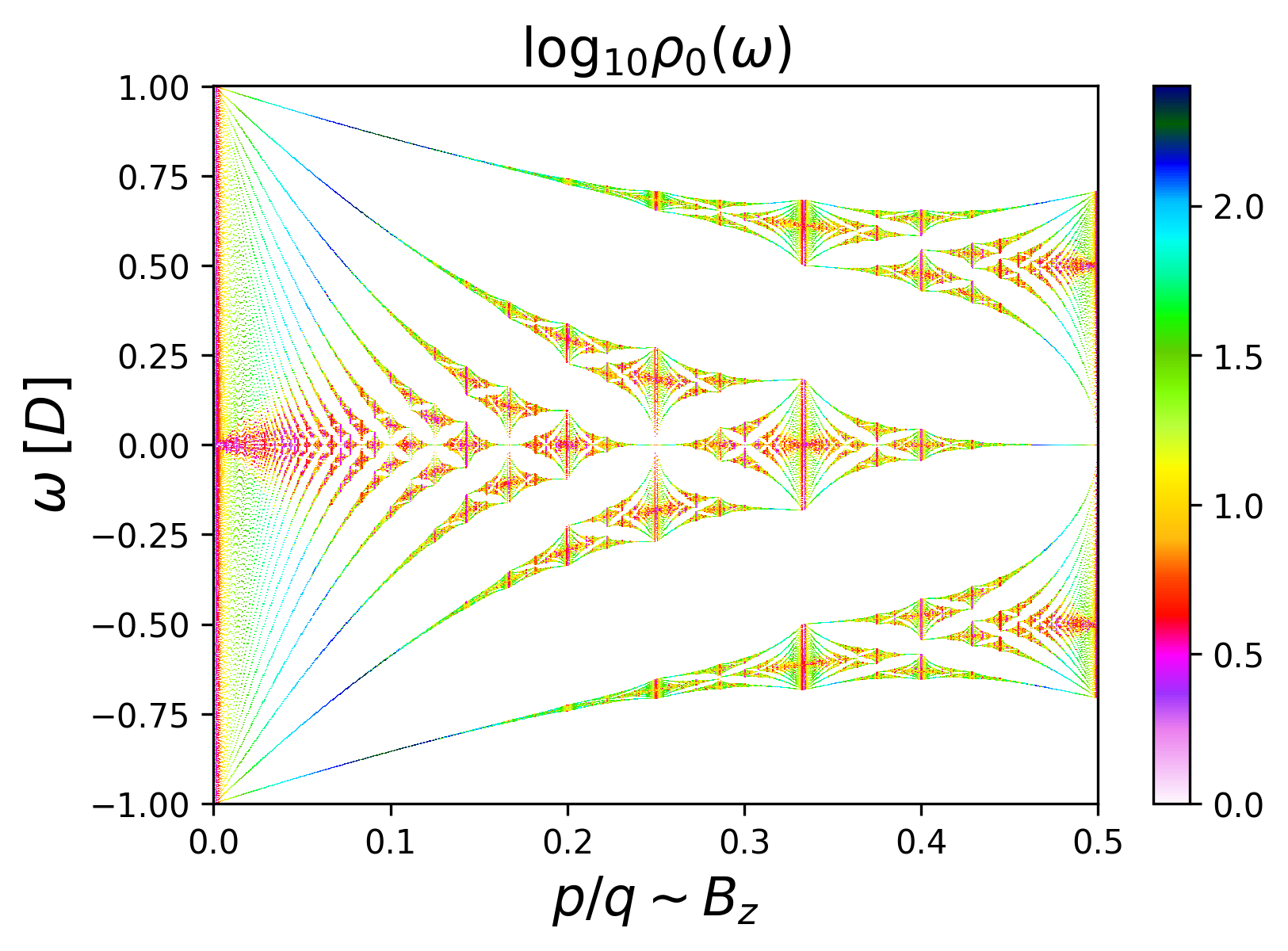}
  \end{center}
 \caption{Non-interacting density of states, as a function of frequency and magnetic field (the Hofstadter butterfly\cite{HofstadterPRB1976}).
 }
 \label{fig:dos}
\end{figure}

In the DMFT, the magnetic field enters through the non-interacting density of states, $\rho_0(\omega)$. The magnetic field dependence of $\rho_0(\omega)$ (the famous Hofstadter butterfly\cite{HofstadterPRB1976}) is shown in Figure~\ref{fig:dos}. This result was obtained with lattice size $L=1999$, and about $4000$ energy bins, which sets the resolution and the minimal size of an energy gap that one can observe.      
      
\begin{figure}[ht!]
 \begin{center}
 \includegraphics[page=1,width=3.2in, trim=0cm 0cm 0cm 0cm, clip]{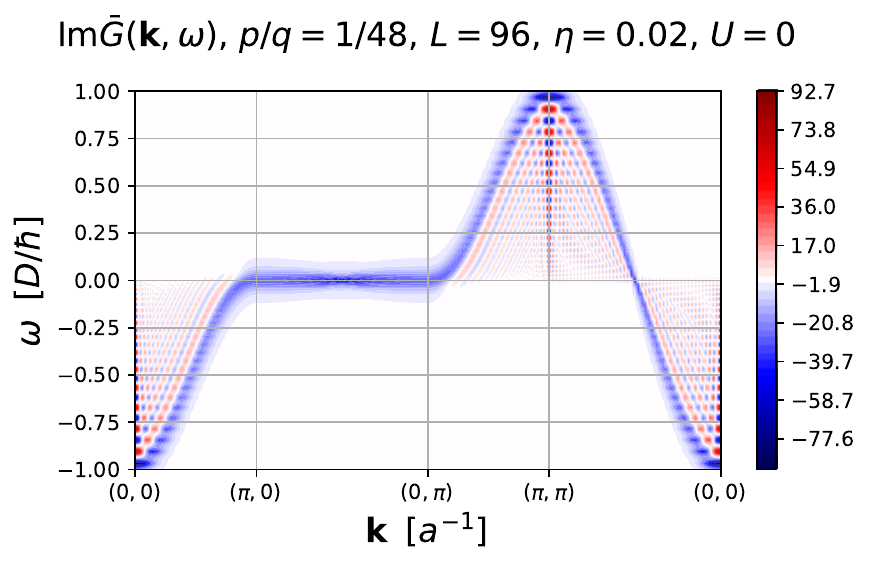}
 \includegraphics[page=1,width=3.2in, trim=0cm 0cm 0cm 0cm, clip]{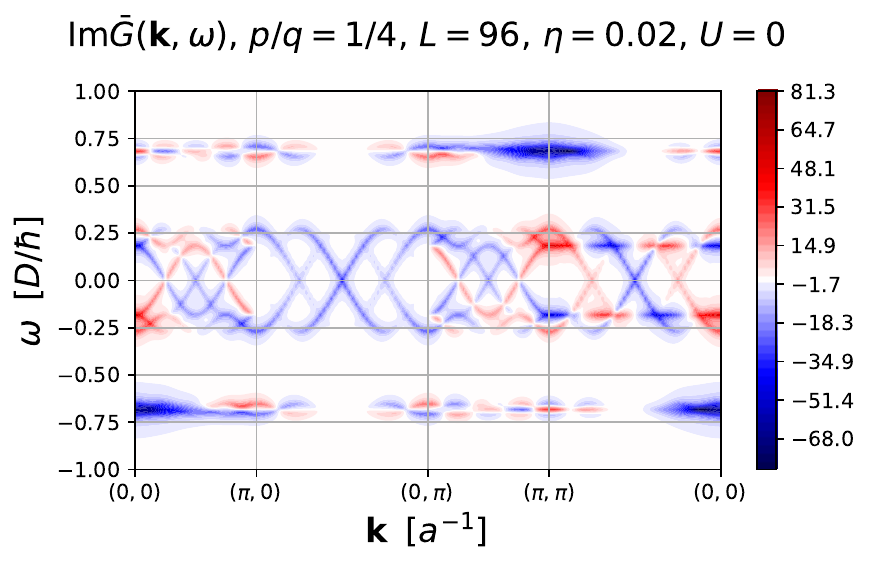}
 \includegraphics[page=1,width=3.2in, trim=0cm 0cm 0cm 0cm, clip]{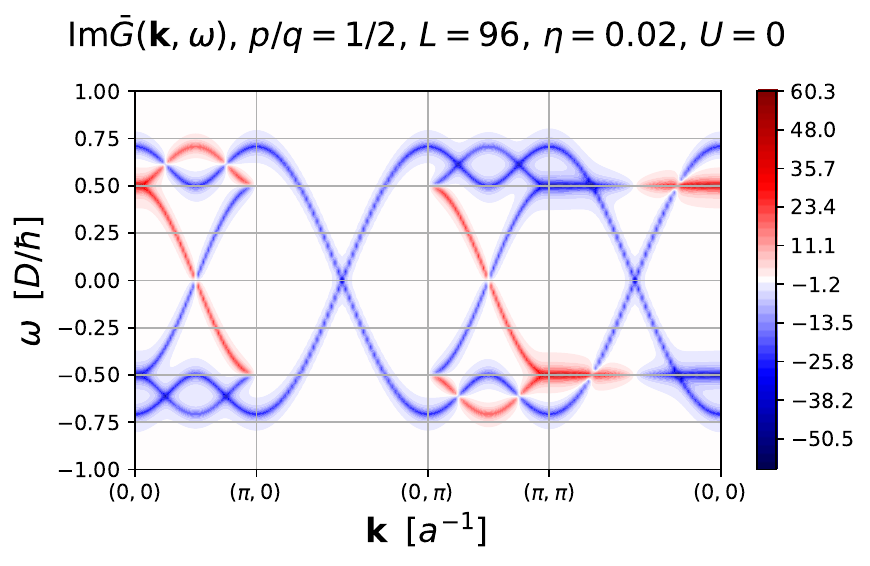}
  \end{center}
 \caption{Spectrum of the translation invariant Green's function, $\mathrm{Im}\bar{G}_\mathbf{k}(\omega)$. The examples are given for $U=0$ with a broadening $\eta=0.02D$, and at three different values of $p/q$, as indicated in panel titles. The spectrum is not negative-definite, and is therefore not indicative of the physical spectral function.
 }
 \label{fig:gbar}
\end{figure} 

\begin{figure*}[ht!]
 \begin{center}
 \includegraphics[page=1,width=3.2in, trim=0cm 0cm 0cm 0cm, clip]{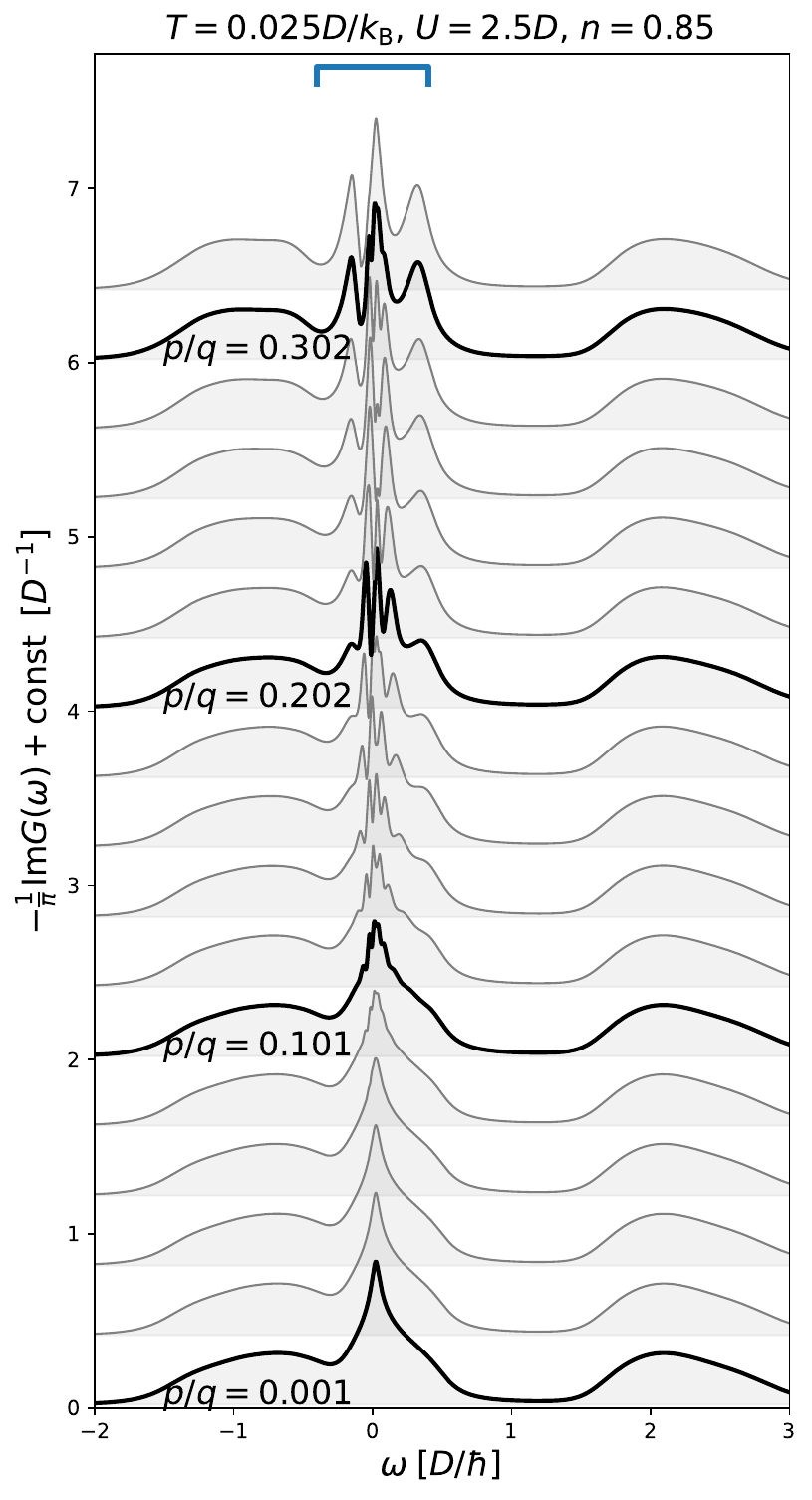}
 \includegraphics[page=1,width=3.2in, trim=0cm 0cm 0cm 0cm, clip]{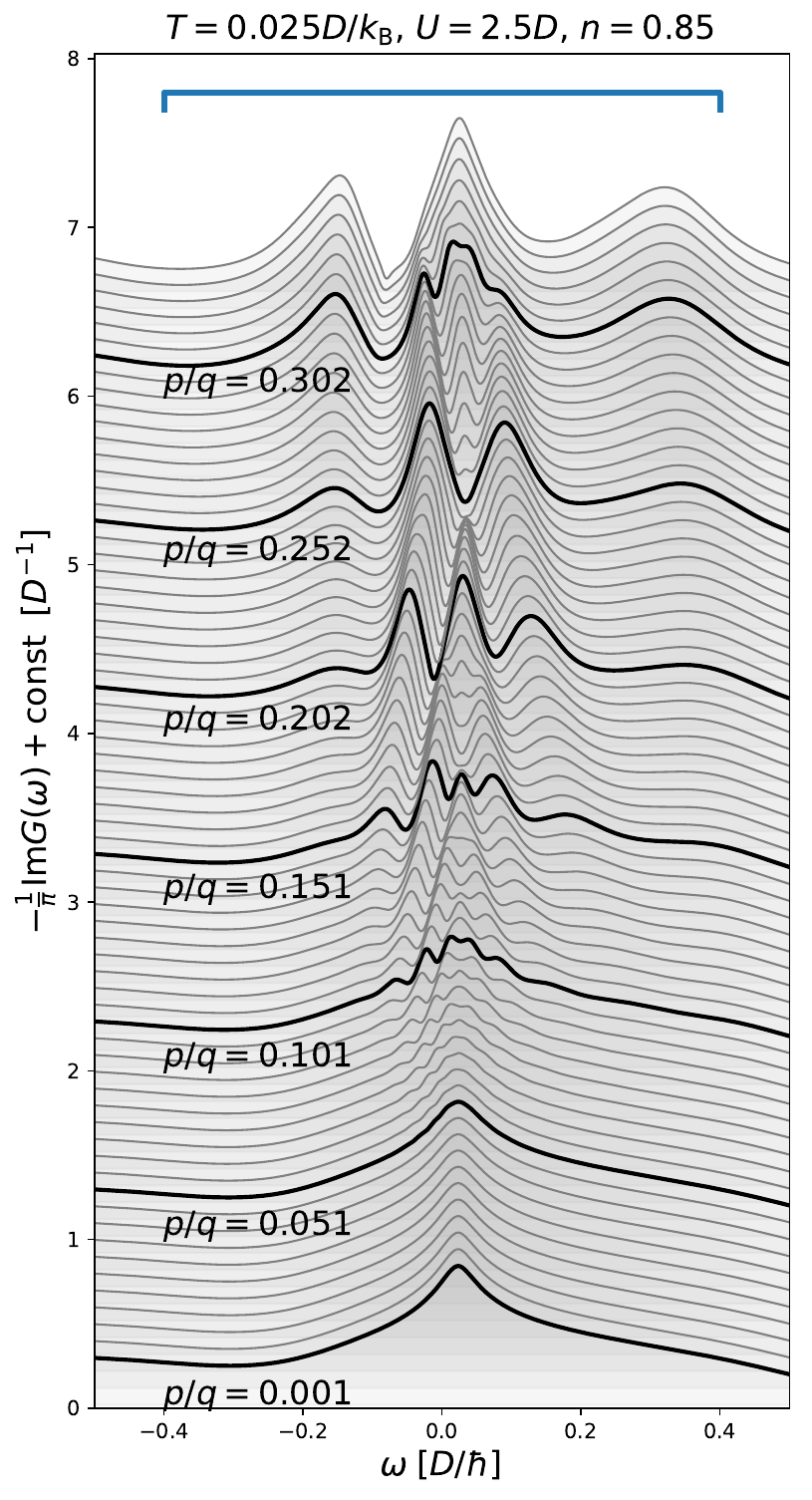}
  \end{center}
 \caption{Local density of states as a function of magnetic field.
 Left: full frequency range. Right: quasiparticle part of the
 spectrum. The result is obtained with the DMFT(NRG). The parameters are $U=2.5D$, $\langle n_\sigma \rangle = 0.425$, $T=0.025D/k_\mathrm{B}$. The calculation was performed with $L=q=997$.
 }
 \label{fig:spectrum}
\end{figure*} 

    \subsubsection{Translation invariant spectral function} \label{sec:Gbar_results}
As we have already proven, on a finite lattice $L\times L$ and $L=q$,
there is no dependence of the Hamiltonian on $k_y$. A straight-forward basis change of the Green's function to the original $\mathbf{k}$-states in the full BZ yields a meaningless result for the spectral function with no $k_y$-dependence whatsoever. Moreover, this result is gauge-dependent, as the choice of a slightly different gauge $\mathbf{A} \sim (-y,0)$ would yield a spectral function result with no $k_x$-dependence instead.
    
One is therefore interested in the translationally invariant
$\bar{G}$, as it has all the lattice symmetries, and can ultimately be
Fourier transformed into momentum space. We show this result in the
non-interacting case in Figure~\ref{fig:gbar}. We have used a small
broadening $\Sigma(z) = -i\mathrm{sgn}(\mathrm{Im}z)\eta$, $\eta=0.02$ to regularize the results. We
observe that the result for the imaginary part of $\bar{G}$ is not
necessarily negative, which is a signature of a breaking of causality,
and thus the result is not a proper physical spectral function. As
expected, the non-physical features subside as the magnetic field is
taken to zero. Note also that $\bar{G}_\mathrm{loc}=G_\mathrm{loc}$,
so the causality of the resulting local Green's function is restored
upon the summation over momenta.

\subsubsection{Local spectra from DMFT}

In Fig.~\ref{fig:spectrum} we present the DMFT(NRG) results for
$\mathrm{Im}G_\mathrm{loc}(\omega)$, at different values of magnetic
field, and fixed $U=2.5D$, $n=0.85$ and $T=0.025D$, which corresponds
to the regime of the doped Mott insulator. On the left panel we show
the full frequency range, while on the right panel we focus on the quasiparticle part of the spectrum. Increasing the magnetic field appears to affect an ever growing range of frequencies around $\omega=0$, but up to the highest fields the effect is restricted to the quasiparticle peak
and no significant change is observed in the Hubbard bands, apart from the lower Hubbard band getting flatter. No apparent change at all is observed below $p/q=0.1$ for these values of model parameters.     
     
  \subsection{Conductivity}
    \subsubsection{Longitudinal dc conductivity $\sigma^{xx}(\nu=0)$} \label{sec:sigmaxx}
    
\begin{figure}[ht!]
 \begin{center}
 \includegraphics[width=3.2in, trim=0cm 0cm 0cm 0cm, clip]{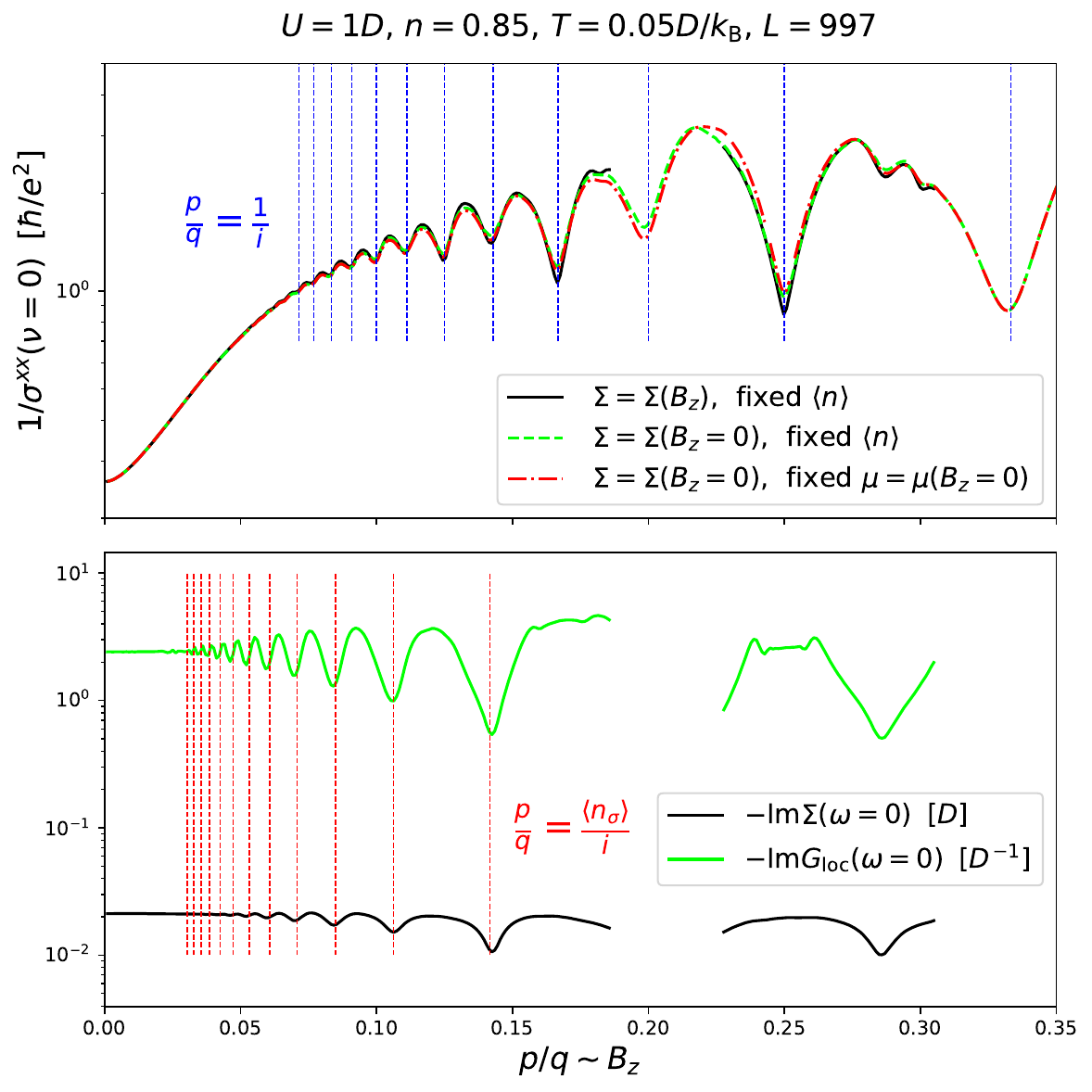}
  \end{center}
 \caption{Upper panel: conductivity obtained with the full DMFT(NRG) calculation (black line), and the simplified calculation where the self-energy is taken from the zero-field DMFT(NRG) calculation, and the chemical potential is either corrected to fix the overall density (lime line) or not (red line).
 Vertical lines indicate $p/q=1/i$ with $i$ integer, which coincides
 with the dips in the inverse conductivity.
 Lower panel: dependence on the magnetic field of the density of states at the Fermi level (lime line) and the scattering rate (black line).
 Vertical lines indicate $p/q=\langle n_\sigma \rangle /i$, which
 roughly coincides with the dips in both the scattering rate and the density of states at the Fermi level.
 }
 \label{fig:fixedSigma_check}
\end{figure}

We start by inspecting the effect of the magnetic-field dependence of the self-energy on the conductivity.
In the upper panel of Fig.~\ref{fig:fixedSigma_check} we show
$1/\sigma^{xx}(\nu=0)$ calculated with the self-energy
obtained from the DMFT(NRG) calculation for the given $B_z$ (black
curve), and compare it to the one obtained with the self-energy
obtained in the $B_z=0$ calculation; in that case the magnetic-field
enters the calculation only through the current vertex
$v^{xx}_{\tilde{\mathbf{k}},m,m',\sigma}$. Here, we can chose to fix the chemical potential to the one corresponding to the $B_z=0$ calculation, which will lead to some density variation as the magnetic field is increased (red curve); otherwise, we can correct the chemical potential for each given $B_z$ so that the overall occupancy is fixed (lime curve).

We see that there is excellent agreement between all three curves. The
parts of the black curve that are missing are due to our inability to
properly converge the DMFT(NRG) calculation at those values of $B_z$.
As finite-$B_z$ DMFT(NRG) calculations are difficult and require
significant computational time, it is a very important observation
that we can obtain solid finite $B_z$ results by using the self-energy
from the $B_z=0$ calculation. This way, the bottleneck of our
calculation becomes the calculation of the conductivity, rather than
the DMFT solver. In the regime of the main interest, this does not present a significant additional approximation. Therefore, in the remainder of the paper we fix $\Sigma=\Sigma(B_z=0)$ and correct $\mu$ at each $B_z$ so that the overall density is fixed, unless stated otherwise.

In the bottom panel of Fig.~\ref{fig:fixedSigma_check} we present the
density of states at the Fermi level ($\sim\mathrm{Im}G(\omega=0$), as
well as the effective scattering rate
($\sim\mathrm{Im}\Sigma(\omega=0)$). The oscillations in these two
quantities as functions of the magnetic field appear synchronous.
However, the oscillations in the conductivity follow a completely
different pattern. The oscillations in the density of states and the
scattering rate can be readily connected with the Shubnikov-de Haas
effect, where the period of oscillations in the space of inverse
magnetic field is inversely proportional to the surface area of the
Fermi see (roughly the density $\langle n_\sigma \rangle$), while the
oscillations of the conductivity appear to correspond to the full area
of the BZ. 
This mismatch in the oscillation frequencies of the spectral and transport properties is however restricted to only
certain parameter regimes.
The high-frequency oscillations have been previously identified in the experiment\cite{Hunt2013,KrishnaKumar2017,KrishnaKumar2018,Barrier2020}
and dubbed the Brown-Zak (BZ) oscillations.
We discuss this phenomenon in more detail in Section
\ref{sec:quantum_oscillations_sigmaxx}, and in Ref.~\onlinecite{OURPRL}
which is devoted to this very topic. In Appendix~\ref{app:thermodynamics}
we check that no $p/q=1$ oscillations are present in thermodynamic potentials.

\begin{figure}[ht!]
 \begin{center}
 \includegraphics[page=2, width=3.2in, trim=0cm 0cm 0cm 0cm, clip]{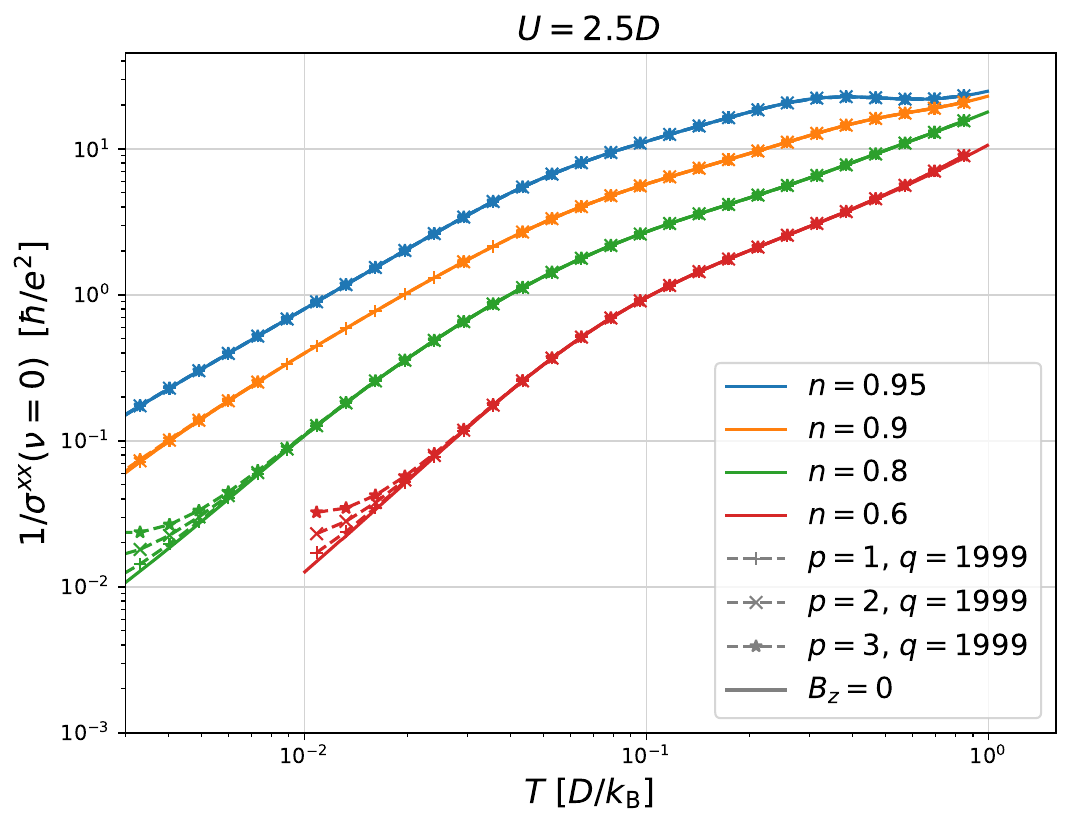}
  \end{center}
 \caption{Cross check between the zero-field formalism and the
 finite-field formalism at weak fields. Solid lines are the $B_z=0$
 DMFT(NRG) result, obtained within the zero-field formalism, at different values of density. We keep $U=2.5D$ fixed. Dashed lines with symbols are obtained within finite $B_z$ formalism, at three lowest values of the field, at $L=q=1999$.
 }
 \label{fig:p1_check}
\end{figure}

We also cross-check the results of our finite $B_z$ calculation
against the reference $B_z=0$ results, in the limit of low field. As
the results are obtained numerically in rather different ways, this is
a stringent test of our formalism and implementations. In
Figure~\ref{fig:p1_check} we show the DMFT results at $U=2.5D$, at 4
different levels of doping ($n=1$ is half-filling), at $B_z=0$ and at
three smallest possible fields in a finite $B_z$ calculation with
$L=q=1999$. At high temperatures, small field does not significantly
affect the result, and finite-field results are on top of the
zero-field result, thus validating our numerics. At low temperature,
the effect of the field becomes observable, but the results do tend towards the $B_z=0$ result as the field is decreased.

\begin{figure}[ht!]
 \begin{center}
 \includegraphics[width=3.2in, trim=0cm 0cm 0cm 0cm, clip]{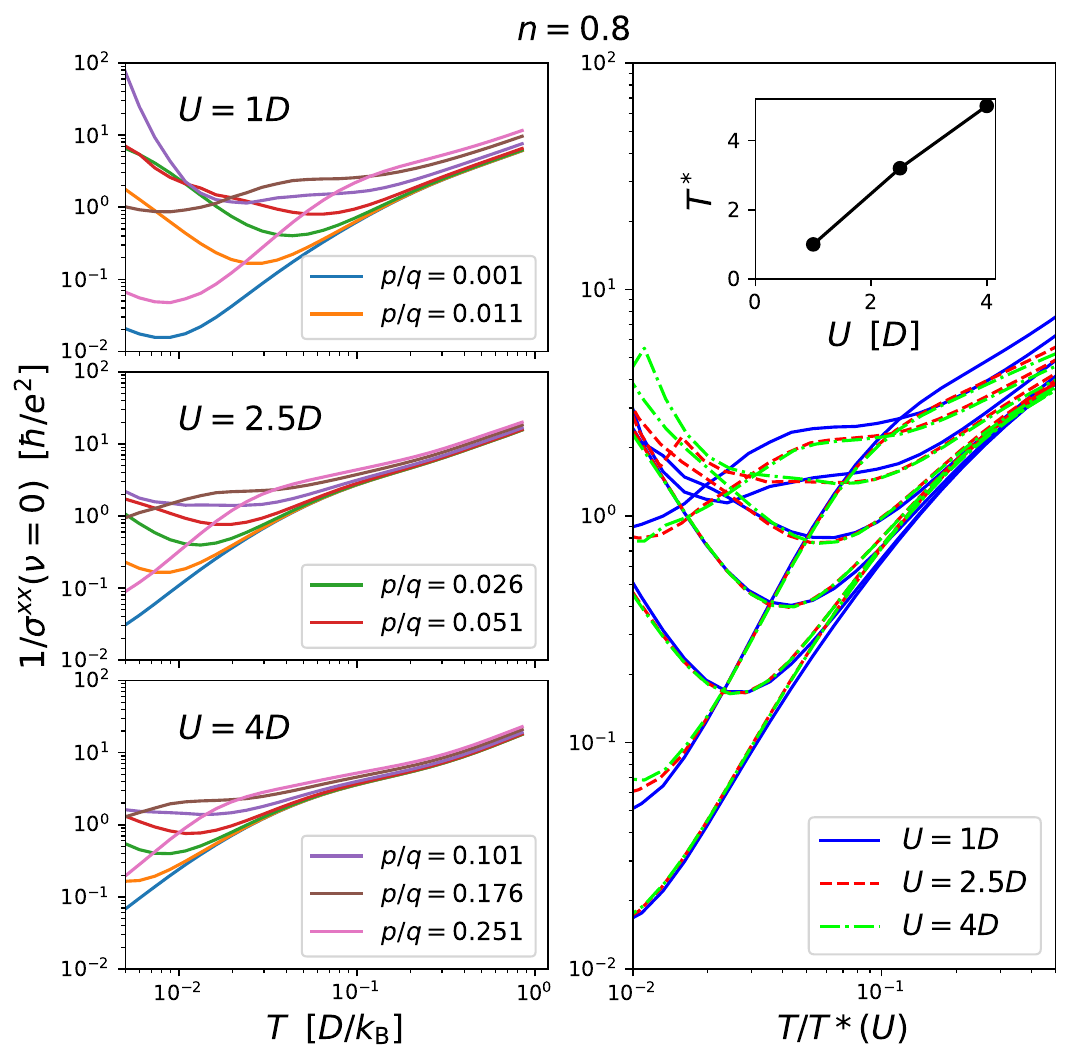}
  \end{center}
 \caption{Left panels: the $T$-dependence of the inverse conductivity at different values of the field and fixed coupling and density. Right panel: the same results with rescaled temperature $T\rightarrow T/T^*(U)$. $T^*(U)$ is given in the inset.
 }
 \label{fig:T_dependence}
\end{figure}

\begin{figure*}[ht!]
 \begin{center}
 \includegraphics[width=6.4in, trim=0cm 0cm 0cm 0cm, clip]{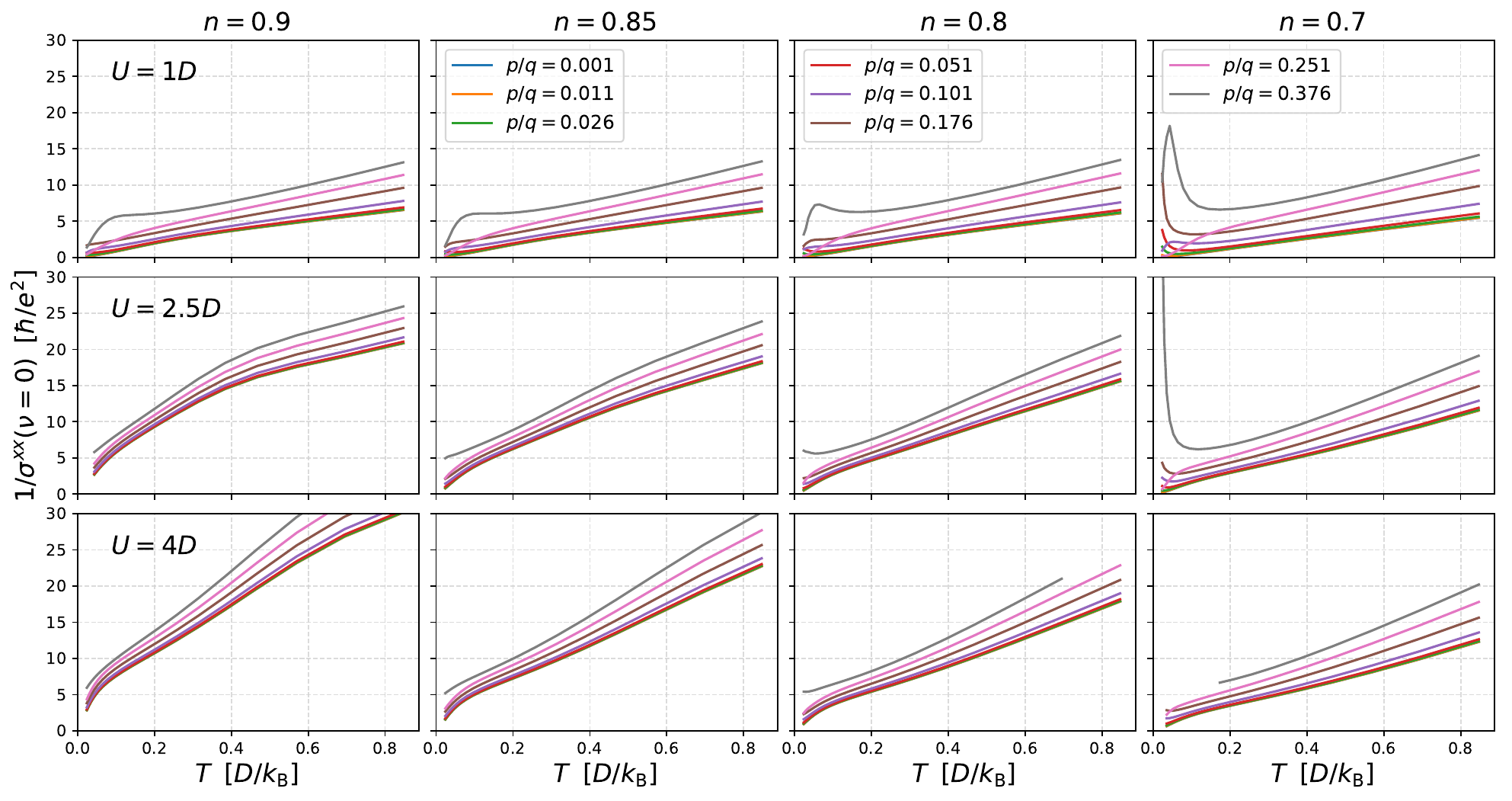}
  \end{center}
 \caption{Inverse conductivity on the linear scale. Columns are different dopings, rows are different values of interaction. Different curves are different values of the field.
 }
 \label{fig:linT_dependence}
\end{figure*}

Next, we inspect the effect of the magnetic field on the temperature dependence of $\sigma^{-1}_{xx}(\nu=0)$ on Figure~\ref{fig:T_dependence}. At low temperature, the behavior drastically depends on the precise choice of the magnetic field. At high temperature the behavior is weakly modified, and one still observes roughly linear dependence. At very high fields, the values appear increased by a constant prefactor, which means that the slope of the linear dependence is also increased. 
This can be more easily confirmed by looking at the linear scale plots in Fig.~\ref{fig:linT_dependence}.
The effect of the magnetic field appears somewhat insensitive to the strength of the interaction, and the overall trend appears similar in all three panels on the left side of Figure~\ref{fig:T_dependence}. We are able to roughly collapse the curves at three different values of $U$ by simply rescaling the temperature $T\rightarrow T/T^*(U)$, with $T^*(U)$ roughly a linear function of $U$. The scaling appears particularly valid between $U=2.5$ and $U=4$, while the $U=1$ curves somewhat deviate.

\begin{figure}[ht!]
 \begin{center}
 \includegraphics[width=3.2in, trim=0cm 0cm 0cm 0cm, clip]{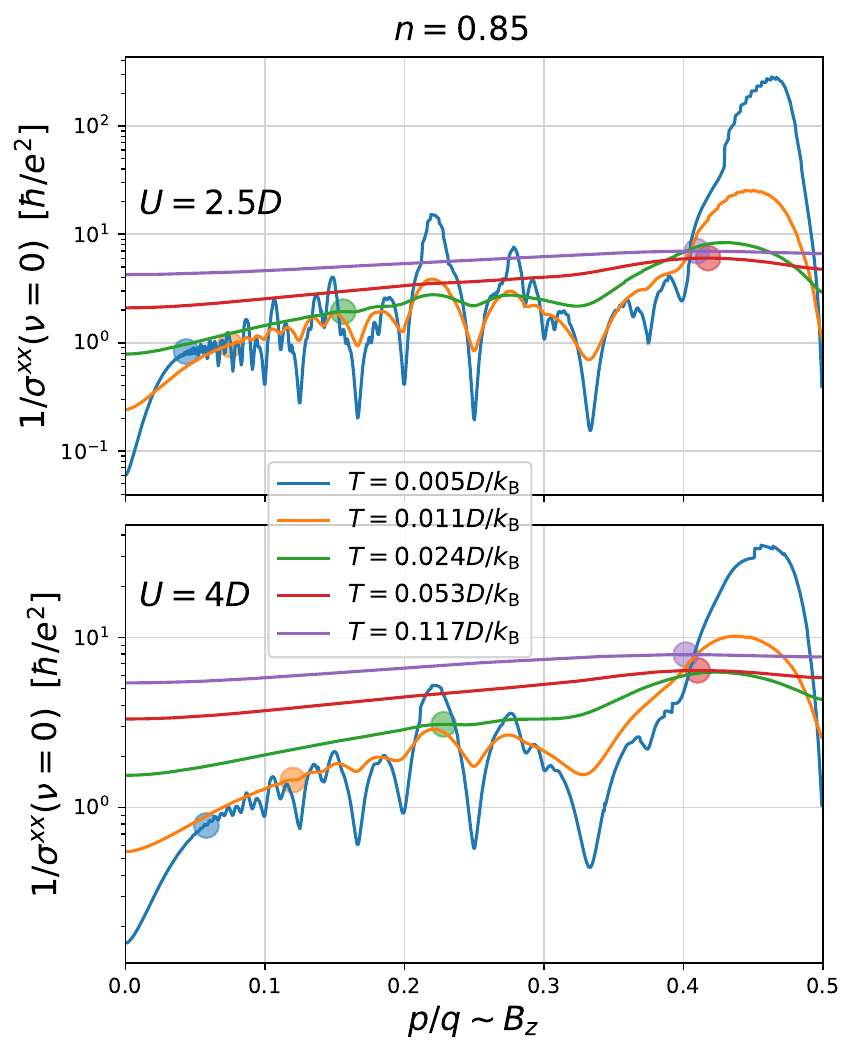}
  \end{center}
 \caption{Magnetic field dependence of the inverse conductivity at two values of coupling constant, and various temperatures. Shaded circles correspond to the first extremum $(p/q)^*$ (see text).
 }
 \label{fig:Bz_dependence}
\end{figure}  

Finally, in Figure~\ref{fig:Bz_dependence} we present the magnetic
field dependence of the conductivity, at a fixed temperature and
doping.  At low temperature one observes increasingly strong
oscillations as magnetic field is increased. At low fields, the
oscillations are relatively small and regular, which corresponds to
the SdH regime, while at strong fields the oscillations cover multiple
orders of magnitude, and exhibit no simple pattern as a function of
the magnetic field. (This is the quantum limit dominated by the lowest
Landau levels.) 

\begin{figure}[ht!]
 \begin{center}
 \includegraphics[page=1, width=3.2in, trim=0cm 0cm 0cm 0cm, clip]{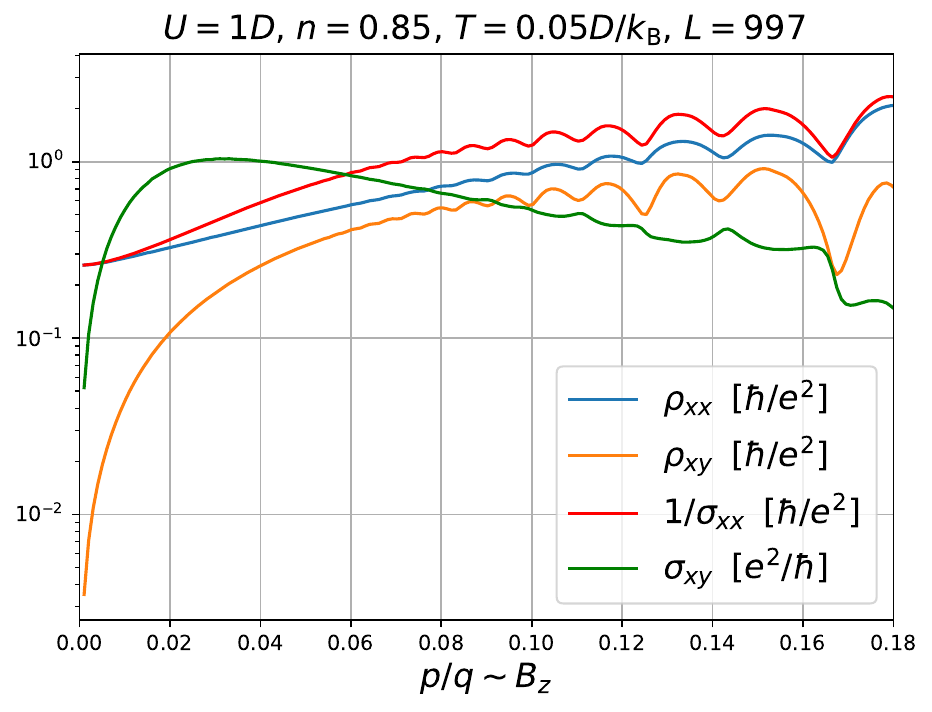}
  \end{center}
 \caption{Different components of the conductivity and resistivity
 tensors, showing the effect of the Hall component on the relation
 between the longitudinal resistivity $\rho^{xx}$ and the inverse
 longitudinal conductivity $1/\sigma^{xx}$.
 }
 \label{fig:sigma_xy_finiteU}
\end{figure}  

\subsubsection{Hall conductivity and resistivity}

In Figure~\ref{fig:sigma_xy_finiteU} we show results for
direct-current $\sigma^{xy}$, $1/\sigma^{xx}$ as well as $\rho^{xy}$
and $\rho^{xx}$. We see that $\sigma^{xy}$ exhibits a non-monotonic
dependence on the magnetic field, and also some oscillation, similar
to $\sigma^{xx}$.
The difference appears to be that when $\sigma^{xx}$ has a local maximum, $\sigma^{xy}$ has a local extremum in the value of its first derivative. 
The results presented in this plot correspond to the high-freqency oscillation regime, where the maxima in $\sigma^{xx}$ coincide with $p/q=1/q$. 
This behavior is in line with the experimental observations in Ref.~\onlinecite{KrishnaKumar2017}.
We also see that, as expected, $\sigma^{xy}$ tends to zero as magnetic
field is decreased. The effect of $\sigma^{xy}$ on $\rho^{xx}$ is not
negligible, and one clearly has $\rho^{xx}\neq 1/\sigma^{xx}$.
Nevertheless, the oscillatory behavior of $\rho^{xx}$ appears very
similar to that of $1/\sigma^{xx}$ and in phase with it.

\begin{figure}[ht!]
 \begin{center}
 \includegraphics[page=1, width=3.2in, trim=0cm 0cm 0cm 0cm, clip]{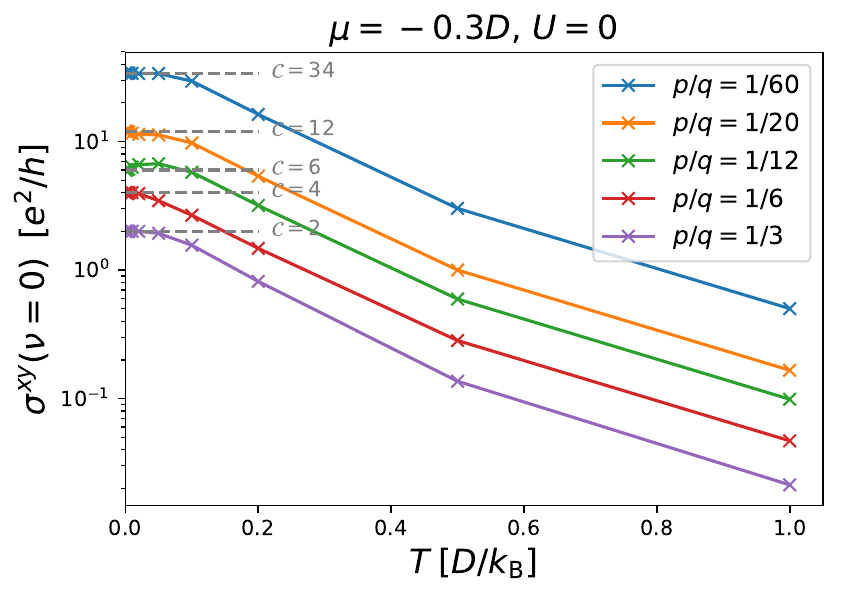}
 \includegraphics[page=1, width=3.2in, trim=0cm 0cm 0cm 0cm, clip]{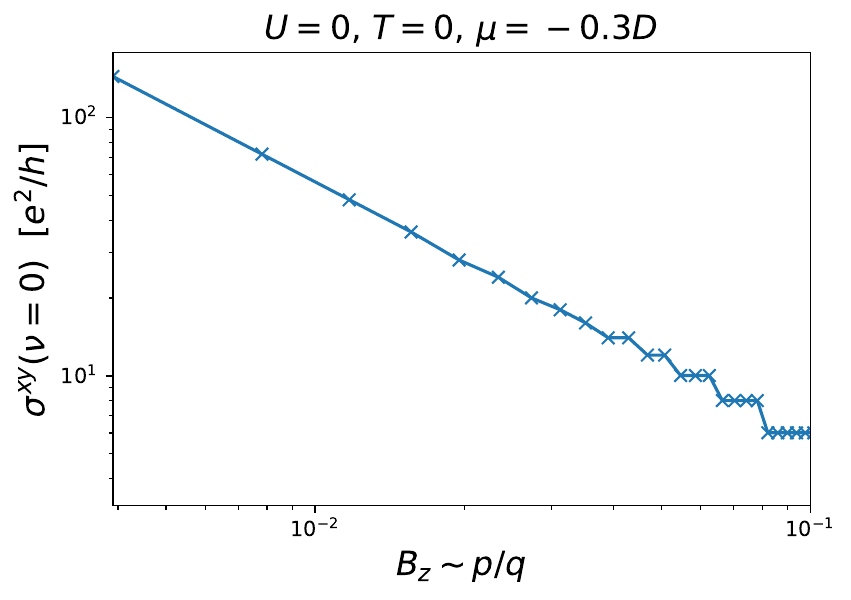}
 \includegraphics[page=1, width=3.2in, trim=0cm 0cm 0cm 0cm, clip]{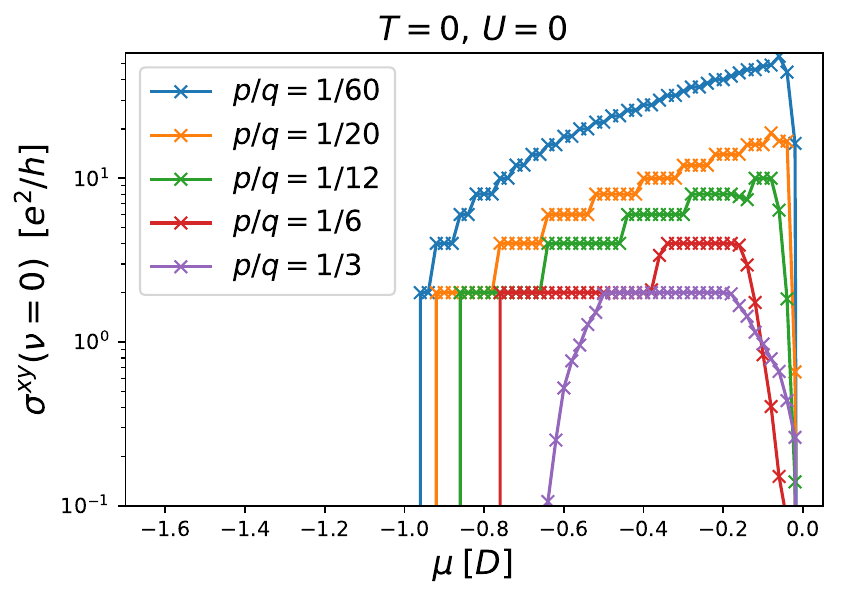}
 
  \end{center}
 \caption{Hall conductivity at $U=0$. Top: temperature dependence of
 the Hall conductivity for a selection of magnetic fields. Gray lines indicate the value at $T=0$, which is the Chern number for the corresponding topological insulator. Center: field dependence of the Hall
 conductivity at zero temperature. Bottom: chemical potential
 dependence of the Hall conductivity for a selection of magnetic fields.
 }
 \label{fig:sigma_xy_U0}
\end{figure}

We further study the behavior of $\sigma^{xy}$ in the limit of $U=0$
where the calculation can be performed at low numerical cost. 
The results are presented in Fig.~\ref{fig:sigma_xy_U0}. 
First we look at the temperature dependence (upper panel). 
We see that $\sigma^{xy}$ falls off exponentially with increasing temperature.
At $T=0$ the result corresponds to the Chern number of the topological
insulator, whenever the chemical potential falls in an energy gap. We
see that decreasing field produces gaps with ever larger Chern
numbers. On the middle panel we look at this dependence more closely,
and see that at a given chemical potential, the Chern number grows in
a power law fashion with $1/B_z$. It is clear that this law does not
have a well defined limiting behavior at $B_z=0$, as in the absence of
magnetic field there are no gaps and strictly $\sigma^{xy}=0$. This is an indication of the fractal structure of the density of states at low fields - the gaps become smaller and smaller, but more and more numerous, and fully disappear only strictly at $B_z=0$. However, this ill-defined behavior is corrected at finite values of the interaction (Fig.~\ref{fig:sigma_xy_finiteU}), and one observes a downturn of $\sigma^{xy}$ at a finite value of the field, and $\sigma^{xy}$ tends to zero smoothly as $B_z\rightarrow 0$.
    
Finally we study the dependence of $\sigma^{xy}$ on the chemical potential, at several different values of the field and $T=0$, in the bottom panel of Fig.~\ref{fig:sigma_xy_U0}. 
One has $\sigma^{xy}(\mu)=-\sigma^{xy}(-\mu)$. The plateaus in the value of Hall conductivity are indications of gaps in the density of states
and are always found at integer values.
We benchmark our implementation with the data in Ref.~\onlinecite{Markov2019} in Appendix~\ref{app:sigmaxy_benchmark}.

\begin{figure}[ht!]
 \begin{center}
 \includegraphics[width=3.2in, trim=0cm 0cm 0cm 0cm]{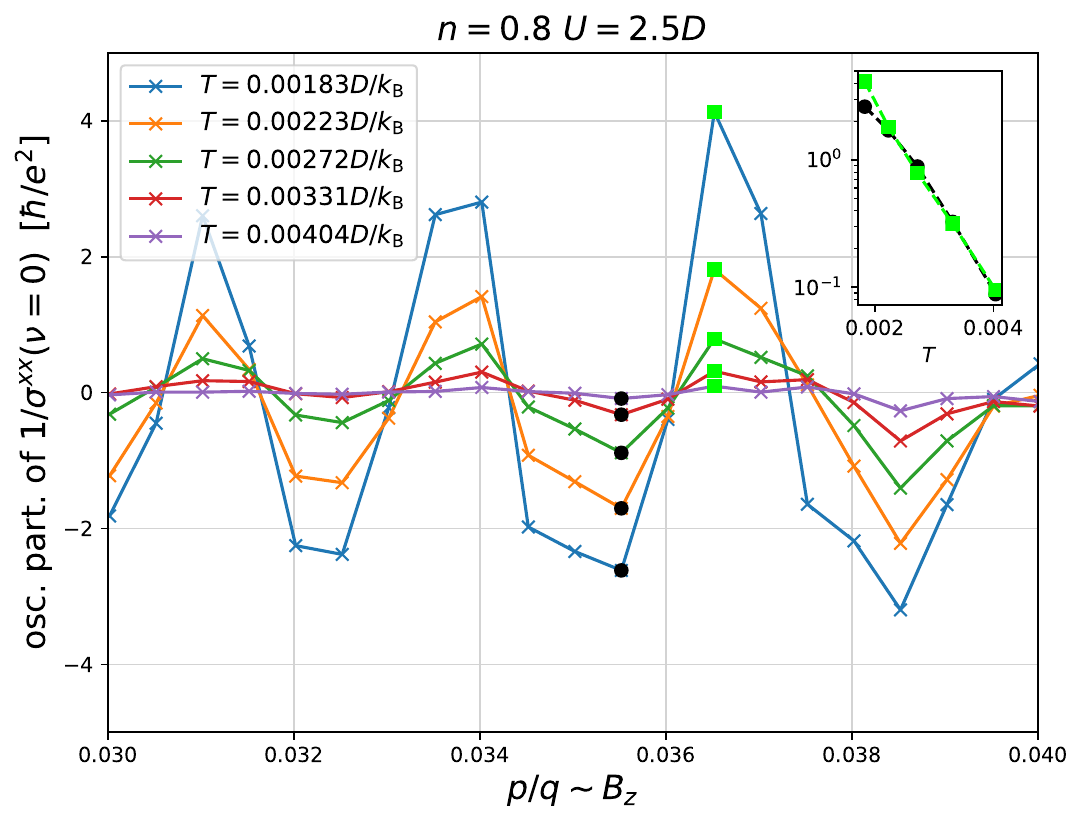}
  \end{center}
 \caption{Zoom in on the oscillatory part of the magnetic field dependence of inverse conductivity, in a range of weak fields. Different curves are different temperatures. Lime and black points denote the apparent anti-nodes of the wave, i.e. the amplitude of oscillation, up to a sign. Inset: dependence of the amplitude of oscillation vs. temperature, on the log scale, revealing exponential decay with $T$.
 }
 \label{fig:LK}
\end{figure}

\begin{figure*}[ht!]
 \begin{center}
 \includegraphics[width=5.5in, trim=0cm 0cm 0cm 0cm, clip]{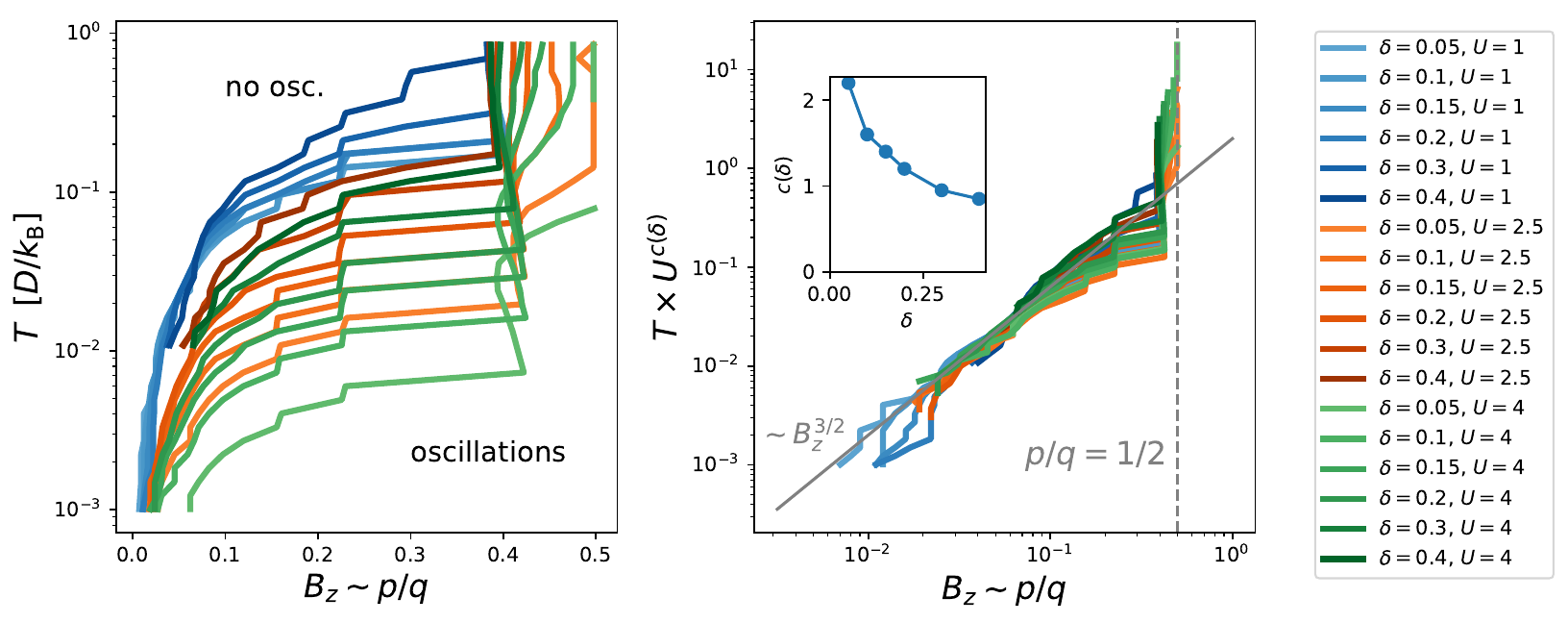}
  \end{center}
 \caption{The $(B_z,T)$ phase diagram for the quantum oscillations.
 The diagram indicates the minimal magnetic field for observing
 significant quantum oscillations (non-monotonic behavior) at a given temperature. Left: results for a set
 of model parameters. Right: rescaled results showing reasonable overlap.
 }
 \label{fig:pfirsts}
\end{figure*}

  \subsubsection{Quantum oscillations of $1/\sigma^{xx}$}  \label{sec:quantum_oscillations_sigmaxx}
 
As already noted in Section~\ref{sec:sigmaxx}, the conductivity displays oscillatory behavior as a function of magnetic field. 
We start by inspecting the region of the phase diagram where notable oscillatory behavior is present. We define a characteristic $(p/q)^*(T;n,U)$ (denoted by shaded circles in Fig.~\ref{fig:Bz_dependence}) as the value of magnetic field at which the first extremum in $1/\sigma^{xx}(B_z\sim p/q)$ occurs ($1/\sigma^{xx}$ initially grows, so the first extremum is always a maximum). 
Oscillations in the slope of $1/\sigma^{xx}(B_z\sim p/q)$ might
survive even below this characteristic value of the field (or
inversely at temperatures higher than $T((p/q)^*)$); the amplitude of
oscillations dies out with temperature exponentially, as in
Lifshitz-Kosewich law~\cite{LK} (see Fig.~\ref{fig:LK};
The oscillatory part of $1/\sigma^{xx}(p/q)$ is extracted by subtracting from the full result the average value in the range $[p/q-f(p/q),p/q+f(p/q)]$, where $f\sim p^{0.7}$).
The quantity $T((p/q)^*)$ is presented in Figure~\ref{fig:pfirsts}. We denote the doping with $\delta=1-n$.

We see that there is always roughly a plateau in $T((p/q)^*)$,
followed by a kink and a near saturation of $(p/q)^*(T)$ at high
temperature. This is because above a certain temperature, no oscillations are present at any value of magnetic field, and there remains at most a single maximum below $p/q=0.5$. The maximum can persist at a roughly fixed $p/q$ up to some temperature, and then ultimately moves to $p/q=1/2$. The shape of $(p/q)^*(T)$ appears nearly universal for all $\delta,U$. We are able to roughly collapse all the curves on the left panel in Fig.~\ref{fig:pfirsts} by rescaling the temperature as $T\rightarrow T\times U^{c(\delta)}$ (right panel), with $c(\delta)$ given in the inset. It is clear that the bigger the doping and the lower the interaction, the oscillations will persist up to a higher temperature, and start at a lower value of the field. It is interesting that roughly $(p/q)^*\sim T^{2/3}$ in the regime where multiple oscillation periods are observed.

\begin{figure*}[ht!]
 \begin{center}
 \includegraphics[width=3.2in, trim=0cm 0cm 0cm 0cm, clip]{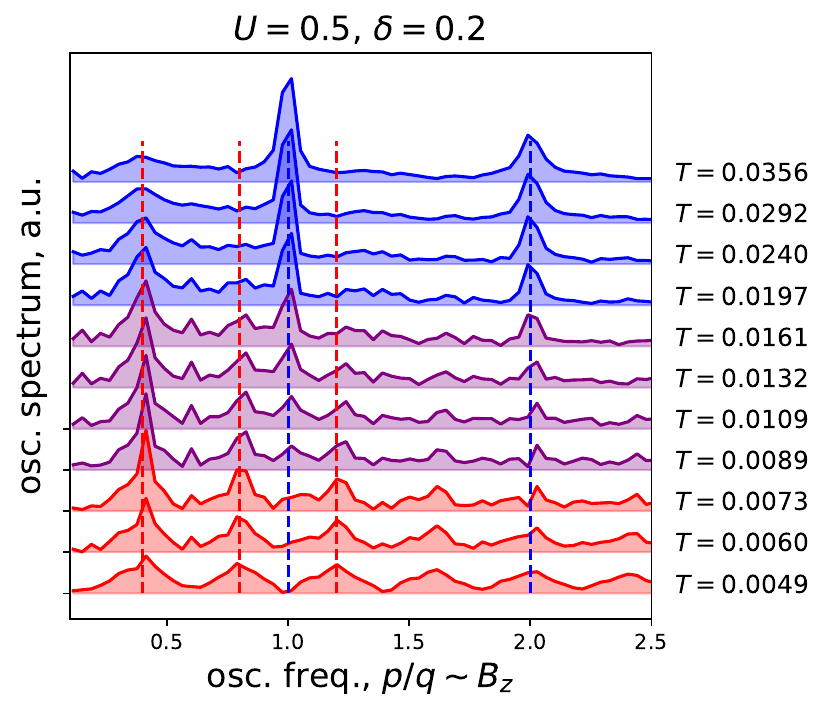}
 \includegraphics[width=2.5in, trim=0cm 0cm 0cm 0cm, clip]{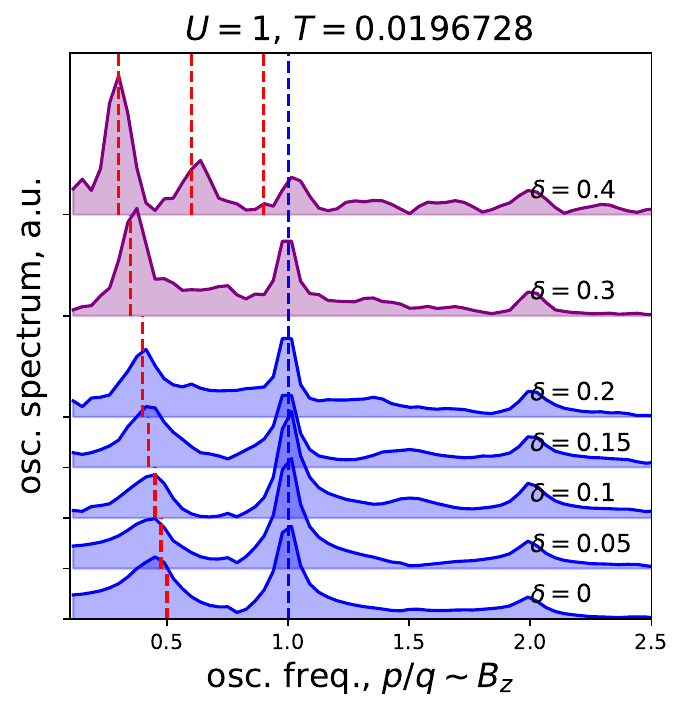}
  \end{center}
 \caption{Fourier spectra of the oscillatory component of the inverse
 dc conductivity. All spectra are normalized to 1. Left: temperature dependence at fixed electron
 density. 
 Right: density dependence at fixed temperature.
 Both panels: Vertical red dashed lines correspond to SdH frequency $p/q=\langle n_\sigma \rangle$ and its higher harmonics; 
 vertical blue dashed lines correspond to $p/q=1$ oscillation frequency and its higher harmonics.
 }
 \label{fig:quantum_oscillations}
\end{figure*}

The oscillations have a fixed period when $\sigma^{xx}$ is plotted as a function of inverse magnetic field. When this period is inversely proportional to the surface area of the Fermi sea, this corresponds to the well known Shubnikov-de Haas effect.
However, in our results, we observe in some regimes of parameters an
additional oscillation frequency. This is documented in
Fig.~\ref{fig:quantum_oscillations} where we present the Fourier
transform of $(\sigma^{xx})^{-1}(B_z^{-1}=q/p)$. The presented part of
each oscillation spectrum is normalized to 1. In the left panel we show the temperature dependence of the oscillation spectrum as a function of temperature, at fixed doping and coupling constant. At low temperature one observes peaks at roughly integer multiples of the density $\langle n_\sigma \rangle$ which corresponds to the SdH effect. However, at intermediate temperature, there is an additional frequency corresponding to the full area of the BZ, and its higher harmonics\cite{OURPRL}. At even higher temperature, before the oscillatory behavior is erased by thermal effects, the $p/q=1$ peak in the spectrum becomes dominant.
In the right panel we inspect the effect of doping on the oscillation
spectrum, at a fixed temperature. The SdH peak is present at all
dopings, and is always found at $p/q \approx \langle n_\sigma
\rangle$. The doping appears to reduce the $p/q=1$ peak, which is no
longer the dominant peak at $\delta>0.2$. At very low doping, it is
not possible to distinguish between the $p/q=1$ peak and the second
harmonic of the SdH peak, as $\langle n_\sigma \rangle$ approaches
0.5. However, it is unexpected in the SdH effect that the second
harmonic is stronger than the first harmonic (fundamental), which
indicates a presence of a separate mechanism which contributes to the
amplitude of the peak at $p/q=1$. In the next section, we are able to
trace the origin of these high-frequency oscillations; the full
description of the observed phenomenology and its relationship to experiments is presented in a separate
publication, Ref.~\onlinecite{OURPRL}.
    
\begin{figure*}[ht!]
 \begin{center}
 \includegraphics[width=6.4in, trim=0cm 0cm 0cm 0cm, clip]{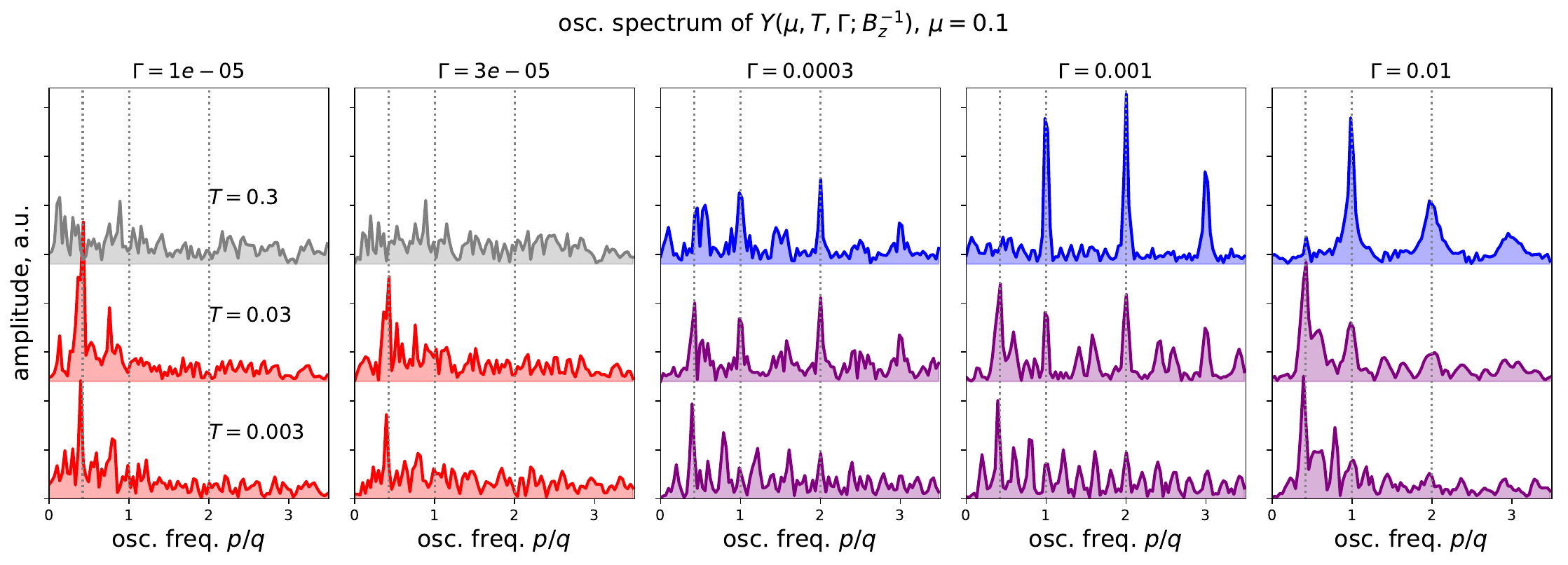}
  \end{center}
 \caption{Oscillation spectra of the current vertex (as quantified by $Y$) at various $\Gamma,T$. 
 Spectra are colored gray where no pronounced peaks are observed, red where the standard SdH $p/q=\langle n_\sigma \rangle$ peaks and its higher harmonics are observed, purple where we also observe the $p/q=1$ peaks, but the SdH peaks are dominant, and blue where $p/q=1$ peak is stronger than $p/q=\langle n_\sigma \rangle$ peak.
 }
 \label{fig:veep_oscillations_spectra}
\end{figure*} 

\begin{figure*}[ht!]
 \begin{center}
 \includegraphics[width=5.6in, trim=0cm 0cm 0cm 0cm, clip]{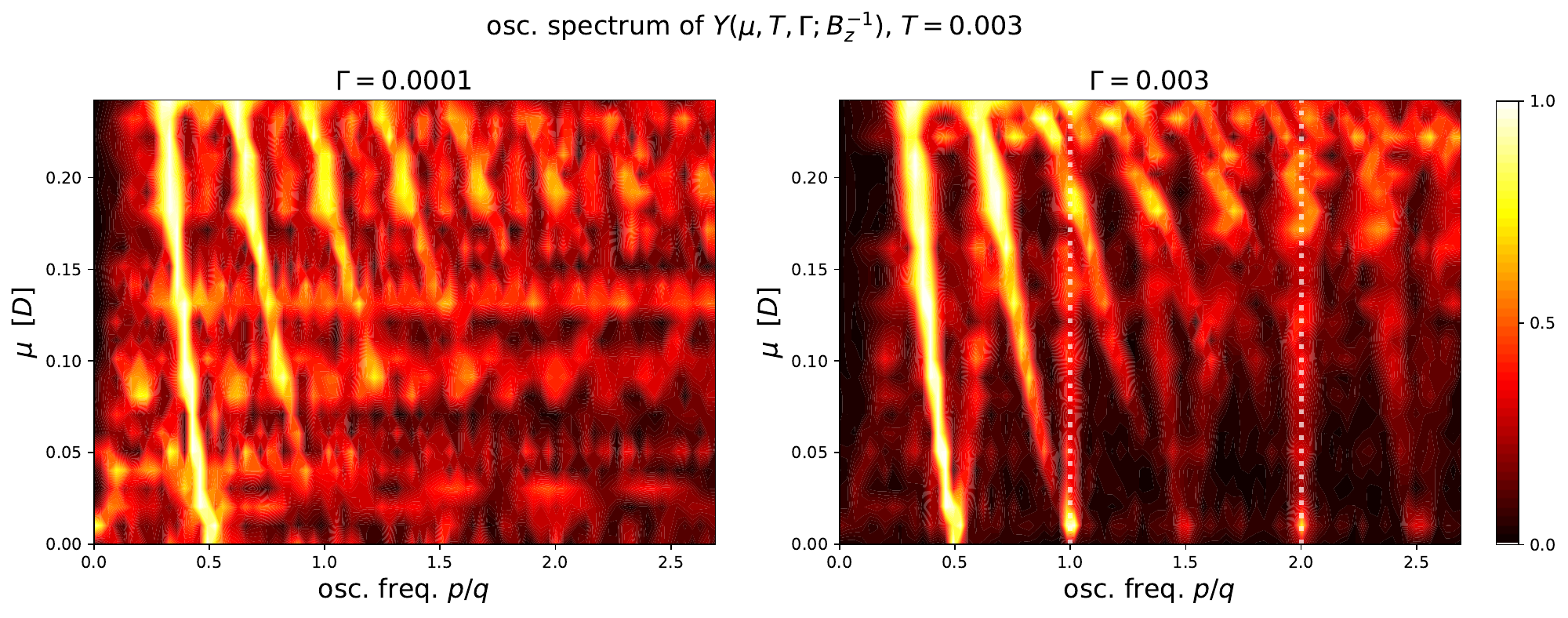}
  \end{center}
 \caption{Evolution of the oscillation spectrum of the current vertex (as quantified by $Y$) with chemical potential, for two values of $\Gamma$.
 }
 \label{fig:veep_oscillations_colorplot}
\end{figure*}    

  \subsubsection{Oscillations of the current vertex}  
    
At the level of the DMFT, the magnetic field enters the calculation of
conductivity through the self-energy and the current vertex $v$. As we
have concluded in the Section~\ref{sec:sigmaxx}, the dependence of
conductivity on the self-energy is of secondary importance, and cannot possibly account for the observed $p/q=1$-frequency oscillations, as the self-energy oscillates with the frequency $p/q=\langle n_\sigma \rangle$. The $p/q=1$ oscillations then must come from the current vertex. This was already suggested in previous experimental works\cite{KrishnaKumar2017,KrishnaKumar2018}, where the $p/q=1$ frequency oscillations have been linked to periodic changes in the velocity of magnetic minibands, featuring spikes at $p/q=1/q$.

We first discuss which part of $v^{xx}(\varepsilon,\varepsilon')$ plays a role at a given choice of parameters. First, by inspecting Eq.\ref{eq:sigma_xx_final} we see that the $\omega$-integrand will generally have two peaks, centered around $\omega^*/\omega^{*\prime}$ such that $\omega^{*(\prime)}+\mu-\varepsilon^{(\prime)}-\mathrm{Re}\Sigma(\omega^{*(\prime)})=0$. The width of those peaks is roughly proportional to $\mathrm{Im}\Sigma(\omega^{*(\prime)})$. When the two peaks are further apart than is their width, the contribution of the integral will be very small. Additionally, if they fall outside of the thermal window, they will not contribute. Assuming in a most simple way $\omega^{*(\prime)} = \mu-\varepsilon^{(\prime)}$, and that the contribution is negligible if $|\varepsilon-\varepsilon'|>\Gamma$, with $\Gamma$ playing the role of the width of the peaks (say roughly $\Gamma \approx  -\mathrm{Im}\Sigma(\omega=0)$), and taking that the thermal window is a hard cutoff $|\omega^{*(\prime)}|<T$, we can isolate the relevant values of $v^{xx}(\varepsilon,\varepsilon')$, which determine the value of the overall integral. We define a quantity
\begin{eqnarray}
 &&Y(\mu,T,\Gamma; B_z) \\ \nonumber
 &&= \int \mathrm{d}\varepsilon \int \mathrm{d}\varepsilon' v^{xx}(\varepsilon,\varepsilon'; B_z) \theta(\Gamma-|\varepsilon-\varepsilon'|)  \\  \nonumber
 &&\;\;\; \times \big(\theta(T-|\varepsilon-\mu|) + \theta(T-|\varepsilon'-\mu|)\big)
\end{eqnarray}
and inspect its oscillation spectrum as a function of $B_z^{-1}$. The results are presented in Fig.~\ref{fig:veep_oscillations_spectra}. We see that at low $T$ and low $\Gamma$, the oscillations resemble the SdH effect. However, increasing $\Gamma$ leads to an onset of $p/q=1$ oscillations, which, with an increase of $T$, eventually become dominant.
To better understand this behavior, in Figure~\ref{fig:veep_oscillations_colorplot} we show the spectrum as a function of $\mu$, at a fixed low temperature. We see how $p/q=\langle n_\sigma \rangle $ peak and its harmonics move with changing $\mu$, as expected. On the contrary, the $p/q=1,2,3...$ peaks (when present in the spectrum), do not move with changing $\mu$. The main insight is that roughly $ Y(\mu,T+\delta T) \approx Y(\mu,T,...)+Y(\mu+\delta\mu,T,...)+Y(\mu-\delta\mu,T,...)$. 
At high temperature, the contributions from different $\mu$ will
interfere destructively, and the $p/q=\langle n_\sigma \rangle$ peak
will wash out. On the other hand, the $p/q=1$ will accumulate, and
become the dominant peak, which is precisely what we find.

\subsubsection{Conductivity in the finite lifetime approximation (FLA)}

\begin{figure*}[ht!]
 \begin{center}
 \includegraphics[width=3.2in, trim=0cm 0cm 0cm 0cm, clip]{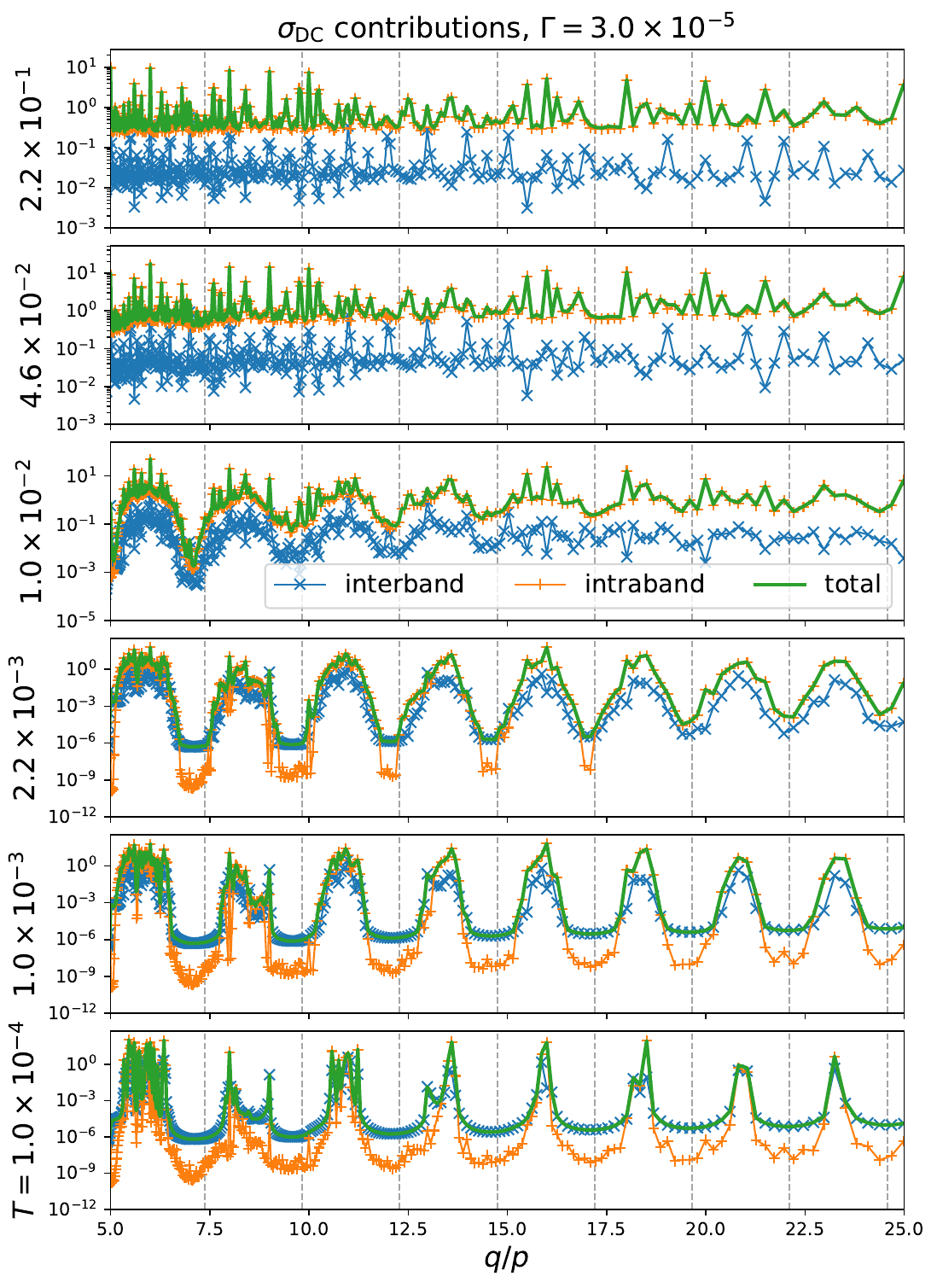}
 \includegraphics[width=3.2in, trim=0cm 0cm 0cm 0cm, clip]{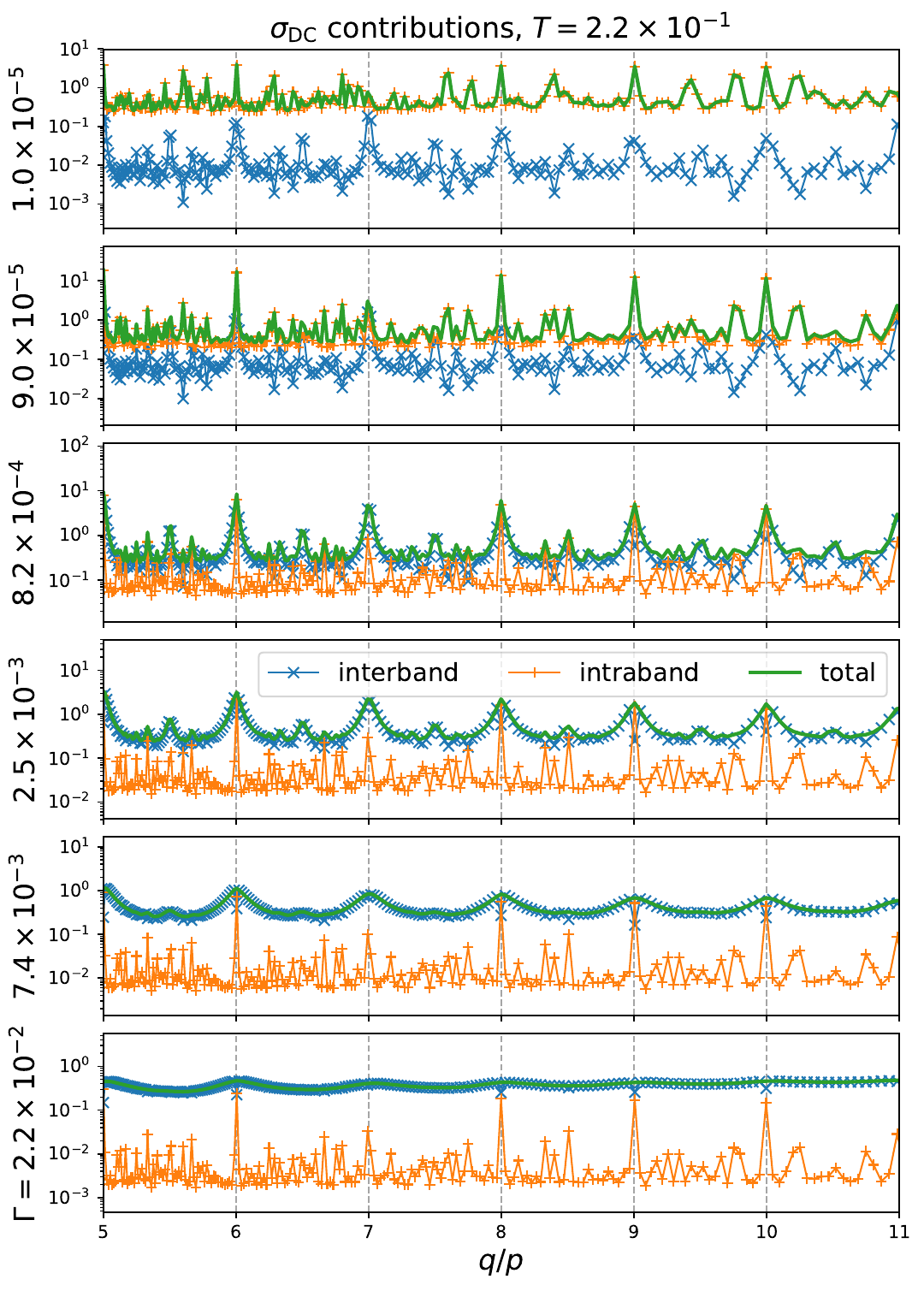}
  \end{center}
 \caption{Longitudinal dc conductivity within FLA - the total result and the contributions from interband and intraband processes. Left: different plots correspond to different temperatures at a fixed scattering rate. Right: different plots correspond to different $\Gamma$'s at a fixed temeprature.
 }
 \label{fig:FLA_trends}
\end{figure*} 

\begin{figure*}[ht!]
 \begin{center}
 \raisebox{-0.5\height}{\includegraphics[width=3.8in, trim=0cm 0cm 0cm 0cm, clip]{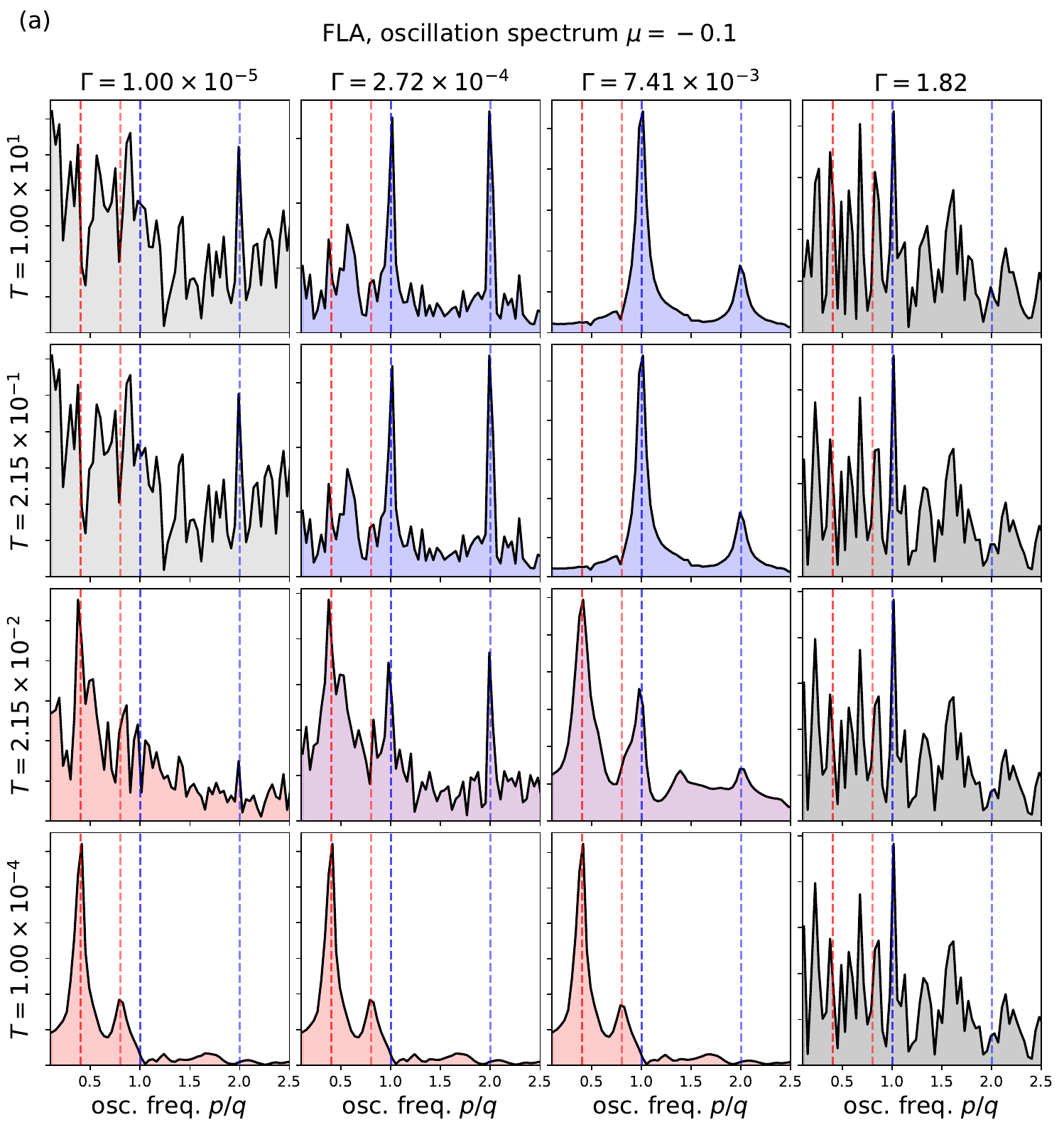}}
 \raisebox{-0.5\height}{\includegraphics[width=2.2in, trim=0cm 0cm 0cm 0cm, clip]{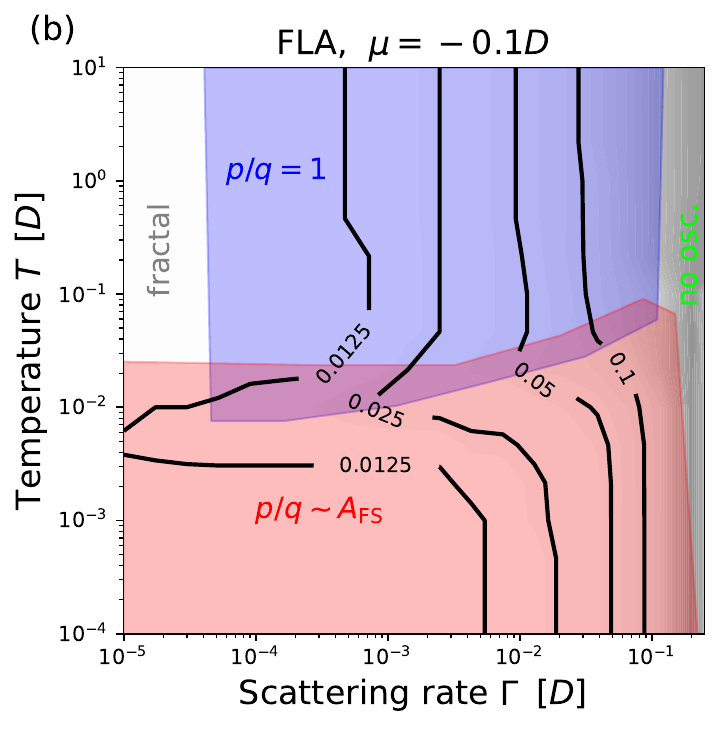}}
  \end{center}
 \caption{(a) Oscillation spectra of longitudinal dc conductivity obtained within the FLA, at different $\Gamma$ and $T$. Coloring is analogous to Fig.~\ref{fig:veep_oscillations_spectra}. (b) the phase diagram of the FLA toy model. Grayscale colorcoding in the background and the black contours correspond to the onset field for the non-monotonous behavior, $(p/q)^*$.
 }
 \label{fig:FLA_GammaT_phasediagram}
\end{figure*}

In the previous sections we have concluded that the high-frequency (Brown-Zak, BZ) oscillations of conductivity originate from the current vertex. Therefore, it is expected that BZ oscillations are observed even in the most simple models that feature no variability in the self-energy whatsoever. In this section we calculate conductivity in the finite-lifetime approximation (FLA), where the self-energy is assumed to be a local, frequency independent and purely imaginary quantity, i.e.
\begin{equation}
 \Sigma^\mathrm{FLA}_{ij}(\omega+i0^+) =  -i\Gamma \delta_{ij}
\end{equation}
where $\Gamma$ is the scattering rate.
At a fixed $\mu=-0.1D$, we construct the Green's function, and evaluate the Kubo bubble Eq.\ref{eq:sigma_xx_final}.
We are in particular interested in the trends with respect to temperature and scattering rate, and we wish to inspect the relative contributions of $\varepsilon\approx\varepsilon'$ and $\varepsilon\neq\varepsilon'$ terms in the double integral. We split the contributions by $d = |\varepsilon-\varepsilon'|$: the contributions with $d>0.0003$ we consider ``interband'' contributions, where the particle and the hole of the particle-hole pair reside in different bands; the contributions with $d<0.0003$ we consider ``intraband'' contributions, where the particle and the hole reside either in the same band, or two different bands which are very close in energy. The choice for the treshold value $0.0003$ is made based on the finite energy resolution that we can achieve and the systematic error made in energy levels due to finite lattice size.

The results are shown in Fig.~\ref{fig:FLA_trends} as a function of inverse magnetic field, i.e. $q/p$. On the plots on the left, we take a small value for the scattering rate $\Gamma\approx 3\times10^{-5}$, and show the two contributions to conductivity as well as the total result at several different temperatures. At low temperature, the dips in conductivity roughly coincide with $q/p=0.407/i$, with $i$ integer (denoted with vertical gray dashed lines). This clearly corresponds to SdH oscillations, and signals that the occupancy at $\mu=-0.1D$ is about $\langle n_\sigma \rangle = 0.407$. It is immediately clear that with increasing temperature, the SdH oscillations subside, and what is left is apparently a fractal-like behavior which cannot be fully resolved with our current resolution. At low temperature, both intraband and interband processes contribute, while at high temperature, the intraband processes are dominant.

On the plots on the right in Fig.~\ref{fig:FLA_trends}, we take a high temperature $T\approx 0.215$ and show results for different values of $\Gamma$. As $\Gamma$ is increased, the interband processes contribute increasingly, and ultimately become fully dominant; the fractal behavior is replaced by regular oscillations, with maxima coinciding with $q/p=q/1$. These are the high-frequency (or BZ) oscillations, which appear only when the scattering rate is sufficiently high.

We illustrate the trends with respect to temperature and the scattering rate on Fig.~\ref{fig:FLA_GammaT_phasediagram}(a) where we plot the oscillation spectra obtained by the Fourer transform of the data in the range of the field $p/q\in[0.03,0.15]$. The results show clearly that at low $\Gamma$, high-frequency oscillations are never observed, but that at sufficiently high $\Gamma$, they are observed above some treshold temperature, but up to indefinite temperature: note that the highest temperature that we show is 10 in units of half-bandwidth, with no sign of weakening of the high-frequency oscillations. In contrast, the SdH oscillations subside simply due to increasing temperature. All oscillations disappear at very high scattering rate, and there seems to be a well defined upper cutoff $\Gamma$ for the observation of any oscillations. These findings are summarized in the rough phase diagram of the FLA model in Fig.~\ref{fig:FLA_GammaT_phasediagram}(b).

\begin{figure}[ht!]
 \begin{center}
 \includegraphics[width=3.2in, trim=0cm 0cm 0cm 0cm, clip]{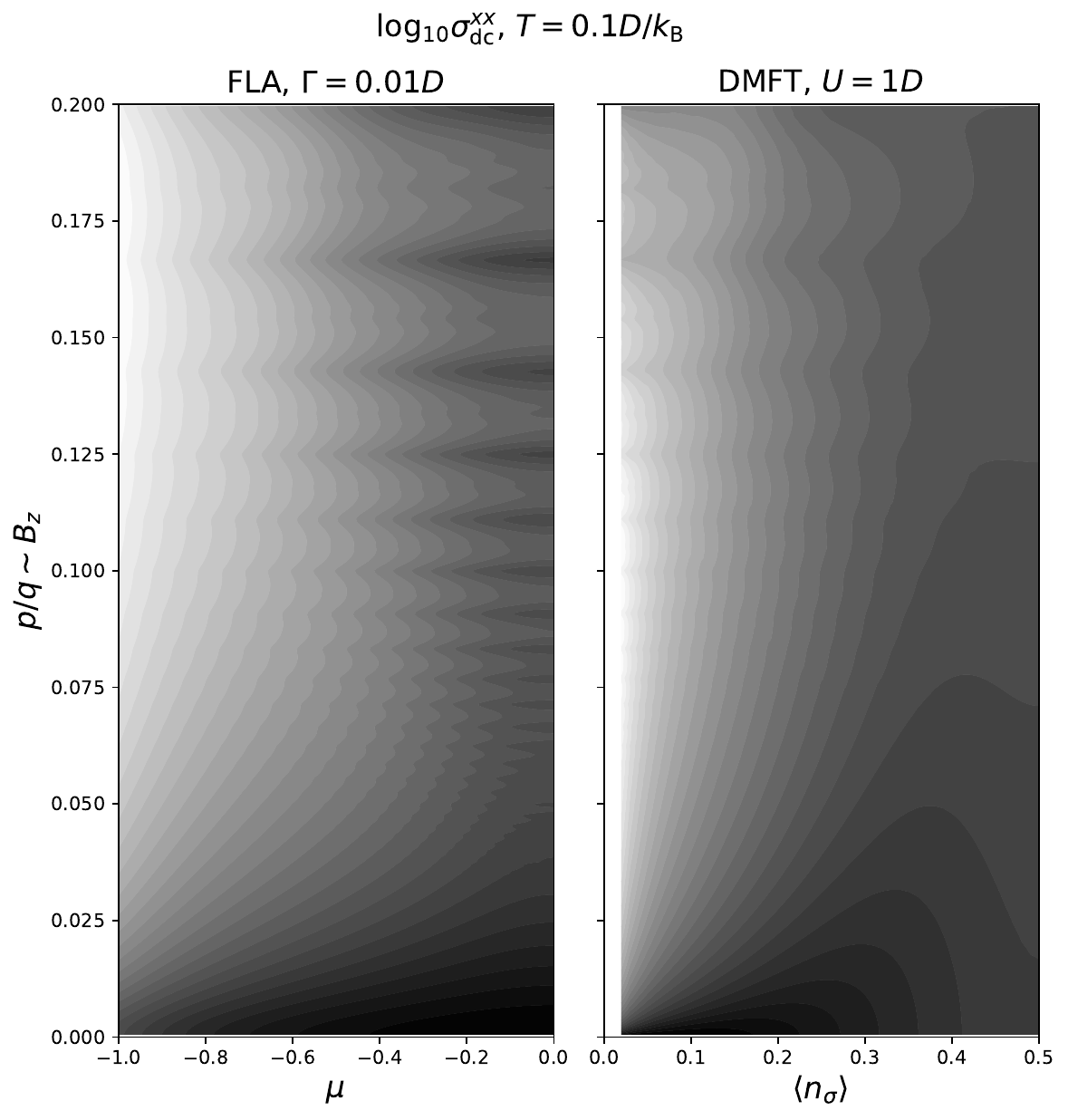} 
  \end{center}
 \caption{The doping-field dependence of longitudinal dc conductivity within FLA and DMFT. Colorcode is logarithmic: white is -2.40 and -3.12, respectively; black is 0.98 and 1.03, respectively.
 }
 \label{fig:FLA_vs_DMFT_doping_trend}
\end{figure} 

It is also interesting to compare the doping dependence of the quantum oscillation phenomenology in FLA and DMFT. On Fig.~\ref{fig:FLA_vs_DMFT_doping_trend} we show the color plot of $\log_{10}\sigma^{xx}(\nu=0)$ in the doping-field plane, at a high temperature where SdH oscillations are already thermally washed out. We see opposite trends in the two plots: in FLA, the oscillations are the strongest close to half-filling ($\mu=0$), while in DMFT, the oscillatory features become stronger in the empty-band limit ($\langle n_\sigma \rangle \rightarrow 0$). The difference must be due to the fact that in the Hubbard model, the scattering rate is maximal at half-filling and vanishes as the number of electrons goes to zero. In FLA, the scattering rate is simply held fixed at all dopings. The FLA result on Fig.~\ref{fig:FLA_vs_DMFT_doping_trend} is in solid agreement with the experimental results in Ref.~\onlinecite{KrishnaKumar2017}. This indicates that the scattering rate in experiment is not vanishing with doping, as one would have in the pure Hubbard model. The additional scattering in experiment probably comes from phonons, or even impurities.
    
\section{Conclusions and prospects}

In this paper we have identified several important features of the
DMFT results for conductivity in the square lattice Hubbard model in a perpendicular magnetic field.
First, the $T$-linear resistivity at high temperature is not strongly affected by magnetic field. 
At high temperature, varying the interaction also does not strongly affect the resistivity, but rather sets the temperature scale in a linear fashion.
Next, we observe that the effect of the magnetic field comes mainly
through the current-vertex factor (which only contains kinetic effects),
and not the self-energy (which involves dynamic effects and
defines the energy
and momentum
windows with significant contribution in the integration). 
We are able to reproduce the SdH effect and observe quantum
oscillations in $1/\sigma^{xx}(B^{-1})$ with the expected frequency
$B_z\sim p/q=\langle n_\sigma \rangle$. However, we also observe
oscillations on a different, higher frequency $B_z\sim p/q=1$,
independently of doping. For this behavior, the prerequisites appear
to be moderate scattering rate, moderate temperature and relatively high magnetic field flux per unit cell. 
Our observations are in line with the experimental results of recent experiments on moir\' e (graphene superlattice) systems\cite{Hunt2013,KrishnaKumar2017,KrishnaKumar2018}.

For the future work it will be necessary to investigate how much of the observed phenomenology is representative of the exact solution, and how much is an artifact of the DMFT approximation. At the level of DMFT the two important simplifications are: 1) the self-energy is fully local (which means that the Green's function is diagonal in the eigenbasis of the non-interacting Hamiltonian), and 2) the vertex corrections cancel. This question can in principle be addressed with cluster DMFT calculations\cite{MaierRMP2005}, but these may not be trivial to formulate or execute. Additionally, analytical continuation of self-energy data might be difficult (or an ED solver might be used, which introduces additional systematic error). Finally, calculation of vertex corrections in cluster DMFT schemes is notoriously difficult\cite{Lin2009,Lin2010,Lin2012,Sato2012,Sato2014}. Another possibility is to use recently developed DiagMC technique which requires no analytical continuation\cite{Taheridehkordi2019,Vucicevic2020,Taheridehkordi2020,Taheridehkordi2020b,Vucicevic2021}, and can access the thermodynamic limit directly. As the observed phenomena are not restricted to very high values of the coupling and can already be observed at $U=0.5-1D$, a DiagMC calculation with only several orders could be sufficient to work out the generic effect of self-energy non-locality and vertex corrections.

\appendix

\section{Peierls phase in Landau gauge (proof of Eq.\ref{eq:Landau_Peirels_phase})} \label{app:proof_Landau_Peirels_phase}

For the following derivation, we introduce 
$\mathbf{r}(\alpha)=a(x(\alpha),y(\alpha))=a[\mathbf{r}_i+\alpha (\mathbf{r}_j - \mathbf{r}_i)]$ 
and $\mathrm{d}\mathbf{r} = a \mathbf{d} \mathrm{d}\alpha$, 
with $\mathbf{d}=\mathbf{r}_j - \mathbf{r}_i$, while keeping $\mathbf{r}_i=(x_i,y_i)$ dimensionless and $x_i,y_i$ integers. 
The Peierls phase for the uniform magnetic field perpendicular to the lattice $\mathbf{B}=(0,0,B_z)$,
in the Landau gauge Eq.\ref{eq:Landau_gauge} can be expressed as
\begin{eqnarray} \nonumber
f_{ij} &=& \frac{e}{\hbar}\int_{\mathbf{r}_i}^{\mathbf{r}_j} \mathbf{A}(\mathbf{r}) \cdot \mathrm{d}\mathbf{r}\\  \nonumber
       &=& \frac{ea}{\hbar}\int_0^1 \mathrm{d}\alpha \;\mathbf{A}\big(\mathbf{r}(\alpha)\big)\cdot \mathbf{d}\\  \nonumber
       &=& \frac{ea^2}{\hbar}B_z\int_0^1 \mathrm{d}\alpha\; x(\alpha) \mathbf{e}_y\cdot \mathbf{d}\\  \nonumber
       &=& \frac{ea^2}{\hbar}B_z(y_j-y_i)\int_0^1 \mathrm{d}\alpha (x_i+\alpha d_x)  \\  \nonumber     
       &=& \frac{ea^2}{\hbar}B_z(y_j-y_i)\Bigg [x_i+ d_x\int_0^1 \mathrm{d}\alpha \alpha \Bigg]       \\  \nonumber
       &=& \frac{ea^2}{\hbar}B_z(y_j-y_i)\Bigg [ x_i+ d_x
       \frac{\alpha^2}{2}\Bigg|_0^1 \Bigg]. \\  \nonumber
\end{eqnarray}
We finally have
\begin{eqnarray} \nonumber
f_{ij} &=& \frac{ea^2}{\hbar}B_z(y_j-y_i)\Bigg [ x_i+ \frac{x_j-x_i}{2} \Bigg]\\
       &=& \frac{ea^2}{\hbar}\frac{B_z}{2}(y_j-y_i)(x_i+x_j).
\end{eqnarray}
When doing real-space calculations on a finite cyclic lattice, it is
necessary to always consider the shortest distance between the sites
and take instead the following periodicised expression:
\begin{eqnarray} \label{eq:fij_finite_lattice}
f_{ij} &\stackrel{\mathrm{fin.\;latt.}}{=}&
\frac{ea^2}{\hbar}B_z(y_j\ominus y_i)\Bigg [ x_i+ \frac{x_j\ominus
x_i}{2} \Bigg],
\end{eqnarray}
where $\ominus$ denotes the shortest distance on a finite cyclic lattice.

\section{Proof of unit cell size} \label{app:proof_unit_cell_size}

We prove now that in the Landau gauge, the size of the unit cell in the $x$ direction is $q$. Consider that we shift both $\mathbf{r}_i$ and $\mathbf{r}_j$ by $q\mathbf{e}_x=(q,0)$ (we remind the reader that we use dimensionless $\mathbf{r}_i=(x_i,y_i)$ as arguments of $f$). We must show that the additional phase shift must be an integer number times $2\pi$.
\begin{eqnarray} \nonumber
f_{\mathbf{r}_i+q\mathbf{e}_x,\mathbf{r}_j+q\mathbf{e}_x} &=& \frac{ea^2}{\hbar}\frac{B_z}{2}(y_j-y_i)(x_i+q+x_j+q) \\ 
 &=& f_{\mathbf{r}_i,\mathbf{r}_j}+\frac{ea^2}{\hbar}\frac{B_z}{2}(y_j-y_i)2q.
\end{eqnarray}
We now apply $B_z = 2\pi\frac{p}{q} \left(\frac{ea^2}{\hbar}\right)^{-1}$
\begin{eqnarray} \nonumber
f_{\mathbf{r}_i+q\mathbf{e}_x,\mathbf{r}_j+q\mathbf{e}_x} &=& f_{\mathbf{r}_i,\mathbf{r}_j}+\frac{2\pi\frac{p}{q}}{2}(y_j-y_i)2q \\
 &=& f_{\mathbf{r}_i,\mathbf{r}_j}+2\pi p(y_j-y_i).
\end{eqnarray}
As $p(y_j-y_i)$ is an integer, the condition Eq.\ref{eq:condition_infinite} indeed satisfies periodicity along the $x$-direction.

\section{Proof of Eq.\ref{eq:H0_momentum_space}} \label{app:proof_H0_momentum_space}
Here we rewrite in momentum space the kinetic term in the Hamiltonian Eq.~\ref{eq:ham_nnhop} to reach Eq.~\ref{eq:H0_momentum_space}.
\begin{widetext}
\begin{eqnarray} \nonumber
 H_{\mathrm{kin},\sigma} &=& 
 -t\frac{1}{N}\sum_{\mathbf{k},\mathbf{k}'}\sum_{i, \mathbf{u}\in\{\mathbf{e}_x,\mathbf{e}_y\}} e^{i 2\pi \frac{n}{L} x_i \mathbf{u}\cdot\mathbf{e}_y} e^{-i\mathbf{k}\cdot\mathbf{r}_i} c^\dagger_{\mathbf{k},\sigma} e^{i\mathbf{k}'\cdot(\mathbf{r}_i+\mathbf{u})} c_{\mathbf{k}',\sigma} +\mathrm{H.c} \\ \nonumber
 &=& -2t\sum_{\mathbf{k},\sigma} \cos k_x n_{\mathbf{k},\sigma} -t\frac{1}{N}\sum_{\mathbf{k},\mathbf{k}'}\sum_{i} e^{i 2\pi \frac{n}{L} x_i} e^{-i\mathbf{k}\cdot\mathbf{r}_i} c^\dagger_{\mathbf{k},\sigma} e^{i\mathbf{k}'\cdot(\mathbf{r}_i+\mathbf{e}_y)} c_{\mathbf{k}',\sigma} +\mathrm{H.c} \\ \nonumber
 &=& -2t\sum_{\mathbf{k},\sigma} \cos k_x n_{\mathbf{k},\sigma} -t\frac{1}{N}\sum_{\mathbf{k},\mathbf{k}'}\sum_{i} e^{i 2\pi \frac{n}{L} \mathbf{r}_i\cdot \mathbf{e}_x} e^{-i\mathbf{k}\cdot\mathbf{r}_i} c^\dagger_{\mathbf{k},\sigma} e^{i\mathbf{k}'\cdot(\mathbf{r}_i+\mathbf{e}_y)} c_{\mathbf{k}',\sigma} +\mathrm{H.c} \\ 
  &=& -2t\sum_{\mathbf{k},\sigma} \cos k_x n_{\mathbf{k},\sigma} -t\frac{1}{N}\sum_{\mathbf{k},\mathbf{k}'}e^{i\mathbf{k}'\cdot\mathbf{e}_y}c^\dagger_{\mathbf{k},\sigma}c_{\mathbf{k}',\sigma}\sum_{i,\sigma} e^{i  \mathbf{r}_i\cdot(\mathbf{k}' - \mathbf{k}+ 2\pi \frac{n}{L}\mathbf{e}_x)}  +\mathrm{H.c} \\ \nonumber
&=& -2t\sum_{\mathbf{k},\sigma} \cos k_x n_{\mathbf{k},\sigma} 
-t\sum_{\mathbf{k},\mathbf{k}'}e^{ik'_y}c^\dagger_{\mathbf{k},\sigma}c_{\mathbf{k}',\sigma}
\delta_{\mathbf{k}',\mathbf{k}- 2\pi \frac{n}{L}\mathbf{e}_x } +\mathrm{H.c} \\ \nonumber
&=& -2t\sum_{\mathbf{k},\sigma} \cos k_x n_{\mathbf{k},\sigma} 
-t\sum_{\mathbf{k},\sigma}e^{ik_y}c^\dagger_{\mathbf{k},\sigma}c_{\mathbf{k}- 2\pi \frac{n}{L}\mathbf{e}_x,\sigma} +\mathrm{H.c}
\\ \nonumber
\\ \nonumber
\\ \nonumber
\end{eqnarray}
\end{widetext}

\pagebreak
\pagebreak

\section{Periodicity with $k_y$} \label{app:periodicity_ky}

We prove here the periodicity of the non-interacting eigenproblem along the $k_y$ axis, as stated in Eqs.\ref{eq:ky_translate} and \ref{eq:ky_translate2}.

We start by reordering rows/columns of the Hamiltonian Eq.\ref{eq:harper_ham}: one can achieve that by redefining
$| \tilde{\mathbf{k}},l,\sigma \rangle \equiv | \mathbf{k}+2\pi l \frac{p}{q},\sigma \rangle$, or simply
$| \tilde{\mathbf{k}},lp\; \mathrm{mod}\; q,\sigma \rangle \rightarrow | \tilde{\mathbf{k}},l,\sigma \rangle$.
In that case, the Hamiltonian reads
\begin{eqnarray}\nonumber
 [\tilde{H}_{0,\tilde{\mathbf{k}},\sigma}]_{l,l'} &=& 
 \Bigg(-\mu_\sigma - 2t\cos \left(\tilde{k}_x+2\pi l\frac{p}{q}\right)\Bigg)\delta_{l,l'} \\
 &&- t\big(e^{ik_y} \delta_{l,l'\oplus 1} +e^{-ik_y} \delta_{l\oplus1,l'}\big).
\end{eqnarray}
This transformation does not affect the eigenvectors/eigenvalues.

Now we apply a unitary transformation defined by $[U_{\tilde{\mathbf{k}}}]_{ll'} = \delta_{ll'} e^{-ilk_y}$ and apply it to the Hamiltonian as 
\begin{equation}
\breve{\mathbf{H}}_{0,\tilde{\mathbf{k}},\sigma} \equiv 
  \mathbf{U}_{\tilde{\mathbf{k}}} 
   \tilde{\mathbf{H}}_{0,\tilde{\mathbf{k}},\sigma}
   \mathbf{U}^\dagger_{\tilde{\mathbf{k}}}.
\end{equation}
The transformed Hamiltonian $\breve{\mathbf{H}}_{0,\tilde{\mathbf{k}},\sigma}$ has the same eigenvalues as $\mathbf{H}_{0,\tilde{\mathbf{k}},\sigma}$, and the original eigenvectors can be obtained from the eigenvectors of $\breve{\mathbf{H}}_{0,\tilde{\mathbf{k}},\sigma}$ as $|m,\tilde{\mathbf{k}},\sigma\rangle=\mathbf{U}^\dagger_{\tilde{\mathbf{k}}}|\breve{m},\tilde{\mathbf{k}},\sigma\rangle$.
As $\mathbf{U}_{\tilde{\mathbf{k}}}$ is diagonal, the element-wise equation for $\breve{\mathbf{H}}_{\tilde{\mathbf{k}},\sigma}$ reads
\begin{equation}
[\breve{H}_{0,\tilde{\mathbf{k}},\sigma}]_{ll'} =
  [U_{\tilde{\mathbf{k}}}]_{ll} 
   [\tilde{H}_{0,\tilde{\mathbf{k}},\sigma}]_{ll'}
   [U_{\tilde{\mathbf{k}}}]^*_{l'l'}.
\end{equation}
The diagonal elements remain unchanged, and we must consider two
special cases for the off-diagonal elements: a) $|l-l'|=q-1$, and b) $|l-l'|=1$. The two cases correspond to whether the hopping between momenta winds around the BZ or not. In the latter case, we have for
\begin{eqnarray}
[\breve{H}_{0,\tilde{\mathbf{k}},\sigma}]_{l>1,l'=l-1} 
 &=&e^{-ilk_y} e^{ik_y} e^{i(l-1)k_y} \\ \nonumber
&=& e^{i(1-l)k_y} e^{i(l-1)k_y} \\ \nonumber
&=& 1 \\ \nonumber
&=& [\breve{H}_{0,\tilde{\mathbf{k}},\sigma}]_{l<q-1,l'=l+1}.
\end{eqnarray}
But in the case when there is winding around the BZ, we get
\begin{eqnarray}
[\breve{H}_{0,\tilde{\mathbf{k}},\sigma}]_{l=q-1,l'=0}  
&=& e^{-i(q-1)k_y} e^{-ik_y}  \\ \nonumber
&=& e^{-i q k_y}  \\ \nonumber
&=& [\breve{H}_{0,\tilde{\mathbf{k}},\sigma}]^*_{l=0,l'=q-1}.
\end{eqnarray}
These are the only elements of the matrix that depend on $\tilde{k}_y$, which means that the blocks of the Hamiltonian are invariant under a transformation
\begin{equation}
 \tilde{\mathbf{k}} \rightarrow (\tilde{k}_x,\tilde{k}_y+2\pi C/q)
\end{equation}
where $C \in \mathbb{Z}$. 

For the basis-change matrix elements, this periodicity means:
\begin{eqnarray}
 &&\mathbf{H}_{0,\tilde{\mathbf{k}},\sigma},\;[\alpha_{\mathbf{k},\sigma}]_{lp\;\mathrm{mod}\;q,m} \\ \nonumber
 &&\xrightarrow{\mathrm{permute}} \tilde{\mathbf{H}}_{0,\tilde{\mathbf{k}},\sigma},\;[\alpha_{\mathbf{k},\sigma}]_{l,m} \\ \nonumber
 &&\xrightarrow{\mathrm{unitary\;tr.}} \breve{\mathbf{H}}_{0,\tilde{\mathbf{k}},\sigma},\;e^{-ik_y l}[\alpha_{\mathbf{k},\sigma}]_{lm} \\ \nonumber
 &&\xrightarrow{\mathrm{translate}} \breve{\mathbf{H}}_{0,\tilde{\mathbf{k}}+(2\pi C/q)\mathbf{e}_y,\sigma},\;e^{-ik_y l}[\alpha_{\mathbf{k},\sigma}]_{lm} \\ \nonumber
 &&\xrightarrow{\mathrm{inv.\;unit.\;tr.}} \tilde{\mathbf{H}}_{0,\tilde{\mathbf{k}}+(2\pi C/q)\mathbf{e}_y,\sigma},\;
 e^{-ik_y l}[\alpha_{\mathbf{k},\sigma}]_{lm}e^{i(k_y+2\pi C/q) l} \\ \nonumber
 &&\;\;\;\;\;\;\;\;\;\;\;\;\;\;\;\;\;\;\;\;\;\;\;\;\;\;\;\;\;\;\;\;\;\;\;\;\;\;\;\;\;\;\;\;\;\;\;\;
 = e^{i(2\pi C/q)l}[\alpha_{\mathbf{k},\sigma}]_{lm} \\ \nonumber
 &&\xrightarrow{\mathrm{permute\;back}} \mathbf{H}_{0,\tilde{\mathbf{k}+(2\pi C/q)\mathbf{e}_y},\sigma},\; e^{i(2\pi C/q)l}[\alpha_{\mathbf{k},\sigma}]_{lp\;\mathrm{mod}\;q,\;m}.
\end{eqnarray}

\section{Decomposition into diamagnetic and paramagnetic currents} \label{app:paramagnetic_and_diamagnetic}

The current operator can be divided into the paramagnetic and diamagnetic parts
\begin{equation}
 \mathbf{j}_\mathbf{r} = \mathbf{j}_{\mathbf{r},\mathrm{P}}+\mathbf{j}_{\mathbf{r},\mathrm{D}},
\end{equation}
where we use Eq.~\eqref{eq:jetagamma} with
\begin{equation}
\boldsymbol{\gamma}_{\mathbf{r},\mathrm{P}} = (1,1),
\end{equation}
\begin{equation}
\boldsymbol{\gamma}_{\mathbf{r},\mathrm{D}} = \boldsymbol{\gamma}_{\mathbf{r}} - \boldsymbol{\gamma}_{\mathbf{r},\mathrm{P}}.
\end{equation}
Because the inversion symmetry is preserved along the $x$-axis, the diamagnetic part of $j^x$ is going to be zero.
This does not hold along the $y$-axis, so there will be non-zero
paramagnetic and diamagnetic parts in $j^y$, but $\langle
\mathbf{j}_{\mathbf{r},\mathrm{P}} \rangle = - \langle
\mathbf{j}_{\mathbf{r},\mathrm{D}} \rangle $ so that the total current
is zero. This must hold as the total current is a physical observable, and thus a gauge invariant quantity. Using the gauge-invariant Green's function, this can be easily proven: up to the constant prefactor, the thermal average of the paramagnetic part is simply 
\begin{eqnarray}
 \langle \mathbf{j}_{\mathbf{r},\mathrm{P}} \rangle 
 &=&
 iG_{\mathbf{r}+\mathbf{e}_y,\mathbf{r}}(\tau=0^-)-iG_{\mathbf{r},\mathbf{r}+\mathbf{e}_y}(\tau=0^-) \\ \nonumber
  &=&
 i\left(e^{if_{\mathbf{r}+\mathbf{e}_y,\mathbf{r}}}-e^{-if_{\mathbf{r}+\mathbf{e}_y,\mathbf{r}}}\right)\bar{G}_{\mathbf{r},\mathbf{r}+\mathbf{e}_y}(\tau=0^-) \\ \nonumber
   &=&
 i\left(e^{-i\frac{ea^2}{\hbar}B_z x}-e^{i\frac{ea^2}{\hbar}B_z
 x}\right)\bar{G}_{\mathbf{r},\mathbf{r}+\mathbf{e}_y}(\tau=0^-),
\end{eqnarray}
and similarly
\begin{eqnarray} \nonumber
 \langle \mathbf{j}_{\mathbf{r},\mathrm{D}} \rangle 
    &=&
 i\left(e^{i\frac{ea^2}{\hbar}B_z x}-e^{-i\frac{ea^2}{\hbar}B_z x}\right)\bar{G}_{\mathbf{r},\mathbf{r}+\mathbf{e}_y}(\tau=0^-)\\ 
 &=& -\langle \mathbf{j}_{\mathbf{r},\mathrm{P}} \rangle.
\end{eqnarray}

\section{Proof of Gauge invariance of the Kubo bubble} \label{app:Kubo_gauge_invariance}

We start with the expression for the site-space matrix (operator) for the current coupled to a vanishing external gauge field $\mathbf{A}^\mathrm{ext}$, and in the presence of a rotary gauge field $\mathbf{A}$ describing a perpendicular magnetic field. The total gauge field is $\mathbf{A}^\mathrm{tot}=\mathbf{A}^\mathrm{ext}+\mathbf{A}$. The corresponding contributions to the Peierls phase (which is additive as well) yield $\mathbf{f}^\mathrm{tot}=\mathbf{f}^\mathrm{ext}+\mathbf{f}$
\begin{eqnarray}
 j^\eta_\mathbf{r}&=& \left. -\frac{\partial \mathbf{H}_\mathrm{kin}}{\partial A^{\mathrm{ext},\eta}_\mathbf{r}}\right|_{\mathbf{A}^{\mathrm{ext}}\rightarrow 0} \\ \nonumber
                           &=& \left.-\frac{\partial \left( e^{i\mathbf{f}^\mathrm{tot}}\circ \mathbf{H}_\mathrm{kin}[\mathbf{A}^\mathrm{tot}=0]\right)}{\partial A^{\mathrm{ext},\eta}_\mathbf{r}}\right|_{\mathbf{A}^{\mathrm{ext}}_\mathbf{r}\rightarrow 0} \\ \nonumber
                           &=& \left.-\frac{\partial e^{i\mathbf{f}^\mathrm{tot}}}{\partial A^{\mathrm{ext},\eta}_\mathbf{r}}\right|_{\mathbf{A}^{\mathrm{ext}}\rightarrow 0} \circ \mathbf{H}_\mathrm{kin}[\mathbf{A}^\mathrm{tot}=0] \\ \nonumber
                           &=& \left.-i\frac{\partial \mathbf{f}^\mathrm{ext}}{\partial A^{\mathrm{ext},\eta}_\mathbf{r}}\right|_{\mathbf{A}^{\mathrm{ext}}\rightarrow 0} \circ e^{i\mathbf{f}} \circ \mathbf{H}_\mathrm{kin}[\mathbf{A}^\mathrm{tot}=0].
\end{eqnarray}
Expressed in terms of second-quantized operators, for the general TB Hamiltonian
\begin{eqnarray}
 j^\eta_\mathbf{r}&=& -\sum_\mathbf{u} t_{\mathbf{r},\mathbf{r}+\mathbf{u}} \left.\frac{\partial f^\mathrm{ext}_{\mathbf{r},\mathbf{r}+\mathbf{u}}}{\partial A^{\mathrm{ext},\eta}_\mathbf{r}}\right|_{\mathbf{A}^{\mathrm{ext}}\rightarrow 0} \\ \nonumber
 && \times
 \left(i e^{if_{\mathbf{r},\mathbf{r}+\mathbf{u}}}
 c^\dagger_\mathbf{r} c_{\mathbf{r}+\mathbf{u}} - i e^{-if_{\mathbf{r},\mathbf{r}+\mathbf{u}}} c^\dagger_{\mathbf{r}+\mathbf{u}} c_\mathbf{r}\right),
\end{eqnarray}
where the sum over $\mathbf{u}$ goes over all sites, and we have omitted the spin indices and the sum over spin for the sake of brevity. 

The current-current correlation function is then
\begin{eqnarray}
 && \Lambda^{\eta,\eta'}_{\mathbf{r},\mathbf{r}'}(\tau)  \\ \nonumber
 && = \sum_{\mathbf{u},\mathbf{u}'}t_{\mathbf{r},\mathbf{r}+\mathbf{u}}t_{\mathbf{r}',\mathbf{r}'+\mathbf{u}'} 
 \left.\frac{\partial f^\mathrm{ext}_{\mathbf{r},\mathbf{r}+\mathbf{u}}}{\partial A^{\mathrm{ext},\eta}_\mathbf{r}}\right|_{\mathbf{A}^{\mathrm{ext}}\rightarrow 0}
 \left.\frac{\partial f^\mathrm{ext}_{\mathbf{r}',\mathbf{r}'+\mathbf{u}'}}{\partial A^{\mathrm{ext},\eta'}_{\mathbf{r}'}}\right|_{\mathbf{A}^{\mathrm{ext}}\rightarrow 0}
 \\ \nonumber
 && \;\;\;\times \sum_{b,b'\in \{0,1\}} (-1)^{b+b'+1} 
 e^{i(-1)^b f_{\mathbf{r},\mathbf{r}+\mathbf{u}}}
 e^{i(-1)^{b'} f_{\mathbf{r},\mathbf{r}+\mathbf{u}}} 
 \sum_{\sigma,\sigma'}
  \\ \nonumber
 &&\;\;\; \times  
 \langle c^\dagger_{\mathbf{r}+b\mathbf{u},\sigma}(\tau) c_{\mathbf{r}+(1-b)\mathbf{u},\sigma}(\tau)
 c^\dagger_{\mathbf{r}'+b'\mathbf{u}',\sigma'}(0) c_{\mathbf{r}'+(1-b')\mathbf{u}',\sigma'}(0)\rangle.
\end{eqnarray}
The dynamic and disconnected part is
\begin{eqnarray}
 && \Lambda^{\eta,\eta'}_{\mathbf{r},\mathbf{r}'}(\tau)  \\ \nonumber
 && = \sum_{\mathbf{u},\mathbf{u}'}t_{\mathbf{r},\mathbf{r}+\mathbf{u}}t_{\mathbf{r}',\mathbf{r}'+\mathbf{u}'} 
 \left.\frac{\partial f^\mathrm{ext}_{\mathbf{r},\mathbf{r}+\mathbf{u}}}{\partial A^{\mathrm{ext},\eta}_\mathbf{r}}\right|_{\mathbf{A}^{\mathrm{ext}}\rightarrow 0}
 \left.\frac{\partial f^\mathrm{ext}_{\mathbf{r}',\mathbf{r}'+\mathbf{u}'}}{\partial A^{\mathrm{ext},\eta'}_{\mathbf{r}'}}\right|_{\mathbf{A}^{\mathrm{ext}}\rightarrow 0}
 \\ \nonumber
 && \;\;\;\times \sum_{b,b'\in \{0,1\}} (-1)^{b+b'+1} 
 e^{i(-1)^b f_{\mathbf{r},\mathbf{r}+\mathbf{u}}}
 e^{i(-1)^{b'} f_{\mathbf{r}',\mathbf{r}'+\mathbf{u}'}} 
  \\ \nonumber
 &&\;\;\; \times 
 \sum_{\sigma}
 (-G_{\mathbf{r}+(1-b)\mathbf{u},\mathbf{r}'+b'\mathbf{u}',\sigma}(\tau))
 G_{\mathbf{r}'+(1-b')\mathbf{u}',\mathbf{r}+b\mathbf{u},\sigma}(-\tau),
\end{eqnarray}
where the Green's functions are obtained with $\mathbf{A}^\mathrm{ext}=0$.
We now rewrite in terms of the gauge invariant Green's function
\begin{eqnarray}
 && \Lambda^{\eta,\eta'}_{\mathbf{r},\mathbf{r}'}(\tau)  \\ \nonumber
 && = \sum_{\mathbf{u},\mathbf{u}'}t_{\mathbf{r},\mathbf{r}+\mathbf{u}}t_{\mathbf{r}',\mathbf{r}'+\mathbf{u}'} 
\left.\frac{\partial f^\mathrm{ext}_{\mathbf{r},\mathbf{r}+\mathbf{u}}}{\partial A^{\mathrm{ext},\eta}_\mathbf{r}}\right|_{\mathbf{A}^{\mathrm{ext}}\rightarrow 0}
 \left.\frac{\partial f^\mathrm{ext}_{\mathbf{r}',\mathbf{r}'+\mathbf{u}'}}{\partial A^{\mathrm{ext},\eta'}_{\mathbf{r}'}}\right|_{\mathbf{A}^{\mathrm{ext}}\rightarrow 0}
  \\ \nonumber
 && \;\;\;\times \sum_{b,b'\in \{0,1\}} (-1)^{b+b'} 
 e^{i(-1)^b f_{\mathbf{r},\mathbf{r}+\mathbf{u}}}
 e^{i(-1)^{b'} f_{\mathbf{r}',\mathbf{r}'+\mathbf{u}'}} 
  \\ \nonumber
 &&\;\;\; \times  
  e^{i f_{\mathbf{r}+(1-b)\mathbf{u},\mathbf{r}'+b'\mathbf{u}'}}
  e^{i f_{\mathbf{r}+b\mathbf{u},\mathbf{r}'+(1-b')\mathbf{u}'}} 
  \\ \nonumber
 &&\;\;\; \times 
 \sum_{\sigma}
 \bar{G}_{\mathbf{r}+(1-b)\mathbf{u},\mathbf{r}'+b'\mathbf{u}',\sigma}(\tau)
 \bar{G}_{\mathbf{r}'+(1-b')\mathbf{u}',\mathbf{r}+b\mathbf{u},\sigma}(-\tau).
\end{eqnarray}
The sum over $b,b'$ yields four terms with exponential prefactors
\begin{eqnarray}
 b=0,b'=0:\;\;&& e^{i f_{\mathbf{r},\mathbf{r}+\mathbf{u}}}
           e^{i f_{\mathbf{r}',\mathbf{r}'+\mathbf{u}'}} 
           e^{i f_{\mathbf{r}+\mathbf{u},\mathbf{r}'}}
           e^{i f_{\mathbf{r}'+\mathbf{u}',\mathbf{r}}}  \\ \nonumber
 b=0,b'=1:\;\;&& e^{i f_{\mathbf{r},\mathbf{r}+\mathbf{u}}}
           e^{i f_{\mathbf{r}'+\mathbf{u}',\mathbf{r}'}} 
           e^{i f_{\mathbf{r}+\mathbf{u},\mathbf{r}'+\mathbf{u}'}}
           e^{i f_{\mathbf{r}',\mathbf{r}}}  \\ \nonumber
 b=1,b'=0:\;\;&& e^{if_{\mathbf{r}+\mathbf{u},\mathbf{r}}}
           e^{i f_{\mathbf{r}',\mathbf{r}'+\mathbf{u}'}} 
           e^{i f_{\mathbf{r},\mathbf{r}'}}
           e^{i f_{\mathbf{r}'+\mathbf{u}',\mathbf{r}+\mathbf{u}}}  \\ \nonumber
 b=1,b'=1:\;\;&& e^{i f_{\mathbf{r}+\mathbf{u},\mathbf{r}}}
           e^{i f_{\mathbf{r}'+\mathbf{u}',\mathbf{r}'}} 
           e^{i f_{\mathbf{r},\mathbf{r}'+\mathbf{u}'}}
           e^{i f_{\mathbf{r}',\mathbf{r}+\mathbf{u}}} 
\end{eqnarray}
and each factor above is gauge invariant.
Therefore what determines whether the Kubo bubble is gauge invariant are the factors of the type 
$\left.\frac{\partial f^\mathrm{ext}_{\mathbf{r},\mathbf{r}+\mathbf{u}}}{\partial A^{\mathrm{ext},\eta}_\mathbf{r}}\right|_{\mathbf{A}^{\mathrm{ext}}\rightarrow 0}$
which clearly do not depend on the choice of the gauge for $\mathbf{A}$.
A vanishing uniform electric field $\mathbf{E}=\partial_t \mathbf{A}^{\mathrm{ext}}$, can be achieved by letting $\mathbf{A}^\mathrm{ext}(\mathbf{r},t)\rightarrow \mathbf{A}^\mathrm{ext}$, i.e. by having a constant and uniform vector potential. The only gauge freedom for the external electric field then corresponds to chosing the inertial reference frame, which is a trivial transformation that our calculation is certainly invariant to; the slowly varying-field approximation holds, and we have
\begin{eqnarray} \nonumber
 \frac{\partial}{\partial A^{\mathrm{ext},\eta}_\mathbf{r}} \int_{\mathbf{r}}^{\mathbf{r}+\mathbf{u}} \mathbf{A}^\mathrm{ext}({\tilde{\mathbf{r}}})\cdot \mathrm{d}\tilde{\mathbf{r}}
 &=& \frac{\partial}{\partial A^{\mathrm{ext},\eta}_\mathbf{r}} (\mathbf{A}^\mathrm{ext} \cdot \mathbf{u}) = u^\eta \\ 
\end{eqnarray}
which clearly does not depend on the precise choice of the uniform $\mathbf{A}^{\mathrm{ext}}$.
We therefore conclude that for the calculation of the linear response to a spatially uniform $\mathbf{q}=0$ electric field, the Kubo bubble Eq.\ref{eq:Lambda_disc_realspace} is \emph{gauge invariant}.

We also emphasize that the bubble for the charge-charge correlation function is trivially gauge invariant because $G_{ij}G_{ji} = \bar{G}_{ij}\bar{G}_{ji}$.

\begin{widetext}

\section{Current operator in momentum space} \label{app:vx_proofs}

Here we derive the current operator in momentum space. The general form is
\begin{equation}
 j^\eta_{\mathbf{q}=0} =
 \frac{it}{N}\frac{e}{ac\hbar}\sum_\sigma \sum_{\tilde{\mathbf{k}},m,m'}
 v^\eta_{\tilde{\mathbf{k}},m,m',\sigma}
 c^\dagger_{\tilde{\mathbf{k}},m,\sigma} c_{\tilde{\mathbf{k}},m',\sigma}.
\end{equation}
The goal of this section is to get expressions for the vertex factors $v^\eta_{\tilde{\mathbf{k}},m,m',\sigma}$.

We start with the current along the $x$-direction. The local contribution is given in Eq.\ref{eq:jetagamma}. After plugging this in Eq.\ref{eq:uniform_current_operator} and applying the basis transformation from Eq.\ref{eq:eigenbasis_to_local}, we obtain
\begin{eqnarray} \nonumber
 j^x_{\mathbf{q}=0} &=& 
 \frac{it}{N}\frac{e}{ac\hbar} \sum_\sigma\sum_{\tilde{\mathbf{k}},m,\tilde{\mathbf{k}'},m'} 
 c^\dagger_{\tilde{\mathbf{k}},m,\sigma} c_{\tilde{\mathbf{k}}',m',\sigma} 
 \sum_{l,l'} 
   [\alpha_{\tilde{\mathbf{k}},\sigma}]_{l,m}
   [\alpha_{\tilde{\mathbf{k}}',\sigma}]^*_{l',m'} \frac{1}{N}\sum_\mathbf{r} 
 e^{-i\big(\tilde{\mathbf{k}}+l\frac{2\pi}{q}\mathbf{e}_x\big) \cdot \mathbf{r} }
 e^{i\big(\tilde{\mathbf{k}}'+l'\frac{2\pi}{q}\mathbf{e}_x\big)\cdot (\mathbf{r}+\mathbf{e}_x) }  + \mathrm{H.c.} \\ \nonumber
 &=& \frac{it}{N}\frac{e}{ac\hbar}\sum_\sigma \sum_{\tilde{\mathbf{k}},m,\tilde{\mathbf{k}'},m'} 
 c^\dagger_{\tilde{\mathbf{k}},m,\sigma} c_{\tilde{\mathbf{k}}',m',\sigma} 
 \sum_{l,l'} 
   [\alpha_{\tilde{\mathbf{k}},\sigma}]_{l,m}
   [\alpha_{\tilde{\mathbf{k}}',\sigma}]^*_{l',m'}
   e^{i\big(\tilde{\mathbf{k}}'+l'\frac{2\pi}{q}\mathbf{e}_x\big)\cdot \mathbf{e}_x } \delta_{\mathbf{k},\mathbf{k}'}\delta_{l,l'} 
   + \mathrm{H.c.} \\
  &=& \frac{it}{N}\frac{e}{ac\hbar}\sum_\sigma \sum_{\tilde{\mathbf{k}},m,m'} 
 c^\dagger_{\tilde{\mathbf{k}},m,\sigma} c_{\tilde{\mathbf{k}},m',\sigma} 
 \sum_{l} 
   [\alpha_{\tilde{\mathbf{k}},\sigma}]_{l,m}
   [\alpha_{\tilde{\mathbf{k}},\sigma}]^*_{l,m'}
      e^{i\big(\tilde{\mathbf{k}}+l\frac{2\pi}{q}\mathbf{e}_x\big)\cdot \mathbf{e}_x }
   + \mathrm{H.c.} \\ \nonumber
   &=& \frac{it}{N}\frac{e}{ac\hbar}\sum_\sigma \sum_{\tilde{\mathbf{k}},m,m'} 
 c^\dagger_{\tilde{\mathbf{k}},m,\sigma} c_{\tilde{\mathbf{k}},m',\sigma} 
 e^{i\tilde{k}_x}
 \sum_{l} 
   [\alpha_{\tilde{\mathbf{k}},\sigma}]_{l,m}
   [\alpha_{\tilde{\mathbf{k}},\sigma}]^*_{l,m'}
      e^{il\frac{2\pi}{q} }
   + \mathrm{H.c.}  \\ \nonumber
&=& \frac{it}{N}\frac{e}{ac\hbar}\sum_\sigma \sum_{\tilde{\mathbf{k}},m,m'} \Bigg[ 
 c^\dagger_{\tilde{\mathbf{k}},m,\sigma} c_{\tilde{\mathbf{k}},m',\sigma} 
 e^{i\tilde{k}_x}
 \sum_{l} 
   [\alpha_{\tilde{\mathbf{k}},\sigma}]_{l,m}
   [\alpha_{\tilde{\mathbf{k}},\sigma}]^*_{l,m'}
   e^{il\frac{2\pi}{q} }
-
 c^\dagger_{\tilde{\mathbf{k}},m',\sigma} c_{\tilde{\mathbf{k}},m,\sigma} 
 e^{-i\tilde{k}_x}
 \sum_{l} 
   [\alpha_{\tilde{\mathbf{k}},\sigma}]^*_{l,m}
   [\alpha_{\tilde{\mathbf{k}},\sigma}]_{l,m'}
   e^{-il\frac{2\pi}{q} }
\Bigg].
\end{eqnarray}
We are free to swap $m$ and $m'$ in the last term:
\begin{eqnarray} \nonumber
 j^x_{\mathbf{q}=0} &=& 
 \frac{it}{N}\frac{e}{ac\hbar}\sum_\sigma \sum_{\tilde{\mathbf{k}},m,m'} \Bigg[ 
 c^\dagger_{\tilde{\mathbf{k}},m,\sigma} c_{\tilde{\mathbf{k}},m',\sigma} 
 e^{i\tilde{k}_x}
 \sum_{l} 
   [\alpha_{\tilde{\mathbf{k}},\sigma}]_{l,m}
   [\alpha_{\tilde{\mathbf{k}},\sigma}]^*_{l,m'}
   e^{il\frac{2\pi}{q} }
-
 c^\dagger_{\tilde{\mathbf{k}},m,\sigma} c_{\tilde{\mathbf{k}},m',\sigma} 
 e^{-i\tilde{k}_x}
 \sum_{l}
   [\alpha_{\tilde{\mathbf{k}},\sigma}]_{l,m}
   [\alpha_{\tilde{\mathbf{k}},\sigma}]^*_{l,m'}
    e^{-il\frac{2\pi}{q} }
\Bigg] \\
&=& 
 \frac{it}{N}\frac{e}{ac\hbar}\sum_\sigma \sum_{\tilde{\mathbf{k}},m,m'}
 c^\dagger_{\tilde{\mathbf{k}},m,\sigma} c_{\tilde{\mathbf{k}},m',\sigma} 
  \sum_{l}
   [\alpha_{\tilde{\mathbf{k}},\sigma}]_{l,m}
   [\alpha_{\tilde{\mathbf{k}},\sigma}]^*_{l,m'}
   \Big[
 e^{i\tilde{k}_x}
 e^{il\frac{2\pi}{q} }
-
 e^{-i\tilde{k}_x}
 e^{-il\frac{2\pi}{q} }
\Big],
\end{eqnarray}
and we can simply read off Eq.\ref{eq:vx}.

Along the $y$-direction, similarly we have
\begin{eqnarray} \nonumber
 &&N \frac{ac\hbar}{e} j^y_{\mathbf{q}=0} \\ \nonumber
 &&=it \sum_\sigma\sum_{\tilde{\mathbf{k}},m,\tilde{\mathbf{k}'},m'} 
 c^\dagger_{\tilde{\mathbf{k}},m,\sigma} c_{\tilde{\mathbf{k}}',m',\sigma} 
 \sum_{l,l'} 
   [\alpha_{\tilde{\mathbf{k}},\sigma}]_{l,m}
   [\alpha_{\tilde{\mathbf{k}}',\sigma}]_{l',m}^{*}
   \frac{1}{N}\sum_\mathbf{r} 
 e^{2i\pi\frac{p}{q}\mathbf{r}\cdot\mathbf{e}_x}  
 e^{-i\big(\tilde{\mathbf{k}}+l\frac{2\pi}{q}\mathbf{e}_x\big) \cdot \mathbf{r} }
 e^{i\big(\tilde{\mathbf{k}}'+l'\frac{2\pi}{q}\mathbf{e}_x\big)\cdot (\mathbf{r}+\mathbf{e}_y) }  + \mathrm{H.c.} \\ \nonumber
 &&= 
 it \sum_\sigma\sum_{\tilde{\mathbf{k}},m,\tilde{\mathbf{k}'},m'} 
 c^\dagger_{\tilde{\mathbf{k}},m,\sigma} c_{\tilde{\mathbf{k}}',m',\sigma} 
 \sum_{l,l'} 
   [\alpha_{\tilde{\mathbf{k}},\sigma}]_{l,m}
   [\alpha_{\tilde{\mathbf{k}}',\sigma}]_{l',m}^{*}
   \frac{1}{N}\sum_\mathbf{r} 
 e^{i\big(\tilde{\mathbf{k}}'+l'\frac{2\pi}{q}\mathbf{e}_x - \tilde{\mathbf{k}}-l\frac{2\pi}{q}\mathbf{e}_x + p \frac{2\pi}{q}\mathbf{e}_x\big) \cdot \mathbf{r}}
 e^{i\big(\tilde{\mathbf{k}}'+l'\frac{2\pi}{q}\mathbf{e}_x\big)\cdot \mathbf{e}_y}  + \mathrm{H.c.} \\ \nonumber
  &&= 
 it \sum_\sigma\sum_{\tilde{\mathbf{k}},m,\tilde{\mathbf{k}'},m'} e^{i\tilde{k}'_y}
 c^\dagger_{\tilde{\mathbf{k}},m,\sigma} c_{\tilde{\mathbf{k}}',m',\sigma} 
 \sum_{l,l'} 
   [\alpha_{\tilde{\mathbf{k}},\sigma}]_{l,m}
   [\alpha_{\tilde{\mathbf{k}}',\sigma}]_{l',m}^{*}
   \delta_{\tilde{\mathbf{k}},\tilde{\mathbf{k}}'}\delta_{l', l\ominus p}   + \mathrm{H.c.} 
\\ \nonumber
  &&= 
 it \sum_\sigma\sum_{\tilde{\mathbf{k}},m,m'} e^{i\tilde{k}_y}
 c^\dagger_{\tilde{\mathbf{k}},m,\sigma} c_{\tilde{\mathbf{k}},m',\sigma} 
 \sum_{l} 
    [\alpha_{\tilde{\mathbf{k}},\sigma}]_{l,m}
    [\alpha_{\tilde{\mathbf{k}},\sigma}]_{l\ominus p,m'}^{*}
   + \mathrm{H.c.} 
\\ \nonumber
  &&= 
 it \sum_\sigma\sum_{\tilde{\mathbf{k}},m,m'} \Bigg[ 
 e^{i\tilde{k}_y}
 c^\dagger_{\tilde{\mathbf{k}},m,\sigma} c_{\tilde{\mathbf{k}},m',\sigma} 
 \sum_{l} 
    [\alpha_{\tilde{\mathbf{k}},\sigma}]_{l,m}
    [\alpha_{\tilde{\mathbf{k}},\sigma}]_{l\ominus p,m'}^{*}
 -  
 e^{-i\tilde{k}_y}
 c^\dagger_{\tilde{\mathbf{k}},m',\sigma} c_{\tilde{\mathbf{k}},m,\sigma} 
 \sum_{l} 
    [\alpha_{\tilde{\mathbf{k}},\sigma}]^{*}_{l,m}
    [\alpha_{\tilde{\mathbf{k}},\sigma}]_{l\ominus p,m'}
\Bigg]   
\\
  &&= 
 it \sum_\sigma\sum_{\tilde{\mathbf{k}},m,m'} c^\dagger_{\tilde{\mathbf{k}},m,\sigma} c_{\tilde{\mathbf{k}},m',\sigma}  \Bigg[ 
 e^{i\tilde{k}_y} 
 \sum_{l} 
    [\alpha_{\tilde{\mathbf{k}},\sigma}]_{l,m}
    [\alpha_{\tilde{\mathbf{k}},\sigma}]_{l\ominus p,m'}^{*}
 -  
 e^{-i\tilde{k}_y}
 \sum_{l} 
     [\alpha_{\tilde{\mathbf{k}},\sigma}]_{l,m}
     [\alpha_{\tilde{\mathbf{k}},\sigma}]_{l\oplus p,m'}^{*}   
\Bigg]   
\end{eqnarray}
and we can immediately recognize Eq.\ref{eq:vy}.

We can use the property of the basis change matrix elements Eq.\ref{eq:k_flip2} to work out a symmetry of $v^x$ with respect to momentum inversion.
\begin{eqnarray}
v^x_{-\tilde{\mathbf{k}},m,m',\sigma}
&=&
  \sum_{l}
   [\alpha_{-\tilde{\mathbf{k}},\sigma}]_{l,m}
   [\alpha_{-\tilde{\mathbf{k}},\sigma}]^*_{l,m'}
   \Big[
 e^{-i\tilde{k}_x}
 e^{il\frac{2\pi}{q} }
-
 e^{i\tilde{k}_x}
 e^{-il\frac{2\pi}{q} } \Big] \\
&=&
  \sum_{l}
  [\alpha_{\tilde{\mathbf{k}},\sigma}]_{q-l,m}
   [\alpha_{\tilde{\mathbf{k}},\sigma}]^*_{q-l,m'}
   \Big[
 e^{-i\tilde{k}_x}
 e^{il\frac{2\pi}{q} }
-
 e^{i\tilde{k}_x}
 e^{-il\frac{2\pi}{q} } \Big].
\end{eqnarray}
We now make a change of variables: $l'=q-l$, $l=q-l'$:
\begin{eqnarray}
v^x_{-\tilde{\mathbf{k}},m,m',\sigma} &=&
  \sum_{l'}
   [\alpha_{\tilde{\mathbf{k}},\sigma}]_{l',m}
   [\alpha_{\tilde{\mathbf{k}},\sigma}]^*_{l',m'}
   \Big[
 e^{-i\tilde{k}_x}
 e^{i(q-l')\frac{2\pi}{q} }
-
 e^{i\tilde{k}_x}
 e^{-i(q-l')\frac{2\pi}{q} } \Big] \\ \nonumber
 &=&
  \sum_{l'}
   [\alpha_{\tilde{\mathbf{k}},\sigma}]_{l',m}
   [\alpha_{\tilde{\mathbf{k}},\sigma}]^*_{l',m'}
   \Big[
 e^{-i\tilde{k}_x}
 e^{-il'\frac{2\pi}{q} }
-
 e^{i\tilde{k}_x}
 e^{il'\frac{2\pi}{q} } \Big].
\end{eqnarray}
\end{widetext}
Therefore
\begin{equation}
 v^x_{-\tilde{\mathbf{k}},m,m',\sigma} = -v^x_{\tilde{\mathbf{k}},m,m',\sigma}.
\end{equation}
Furthermore, by noting that $v$ can be more simply written as
\begin{equation}
v^{x}_{\tilde{\mathbf{k}},m,m',\sigma} =
  -2i\sum_{l}   
    [\alpha_{\tilde{\mathbf{k}},\sigma}]_{l',m}
   [\alpha_{\tilde{\mathbf{k}},\sigma}]^*_{l',m'}
 \sin\Big(\tilde{k}_x+l\frac{2\pi}{q} \Big),
\end{equation}
we can easily prove
\begin{equation}\label{eq:v_transpose}
 v^x_{\tilde{\mathbf{k}},m,m',\sigma} = -(v^x_{\tilde{\mathbf{k}},m',m,\sigma})^*.
\end{equation}
The matrix $[v^x_{\mathbf{k},\sigma}]_{mm'}$ is hence anti-hermitian, which
also implies $\mathrm{Re}v^x_{\tilde{\mathbf{k}},m,m}=0$.
Also, by using Eq.~\ref{eq:ky_flip2}, it is easily proven that
\begin{equation}\label{eq:v_kyinvert}
 v^x_{(\tilde{k}_x,-\tilde{k}_y),m,m',\sigma} = v^x_{(\tilde{k}_x,\tilde{k}_y),m',m,\sigma}.
\end{equation}
\pagebreak

\section{Derivation for $\Lambda^{\eta,\eta',\mathrm{disc}}_{\mathbf{q}=0}(i\nu)$} \label{app:uniform_lambda_disc_freq}

The disconnected part of Eq.\ref{eq:lambda_full_orbital_time} reads
\begin{eqnarray}\nonumber
 && \Lambda^{\eta\eta', \mathrm{disc}}_\mathbf{q=0}(i\nu) = 
 \frac{1}{2\hbar} \int_{-\beta\hbar}^{\beta\hbar} \mathrm{d}\tau e^{i\nu\tau}
 \frac{t^2e^2}{c\hbar^2} \frac{1}{N} \sum_\sigma \sum_{\tilde{\mathbf{k}},m_1,m_1',m_2,m_2'} \\ \nonumber
 && \times 
 v^\eta_{\tilde{\mathbf{k}},m_1,m_1',\sigma}
 v^{\eta'}_{\tilde{\mathbf{k}},m_2,m_2',\sigma} 
 G_{\tilde{\mathbf{k}},m_2',m_1\sigma}(-\tau)
 G_{\tilde{\mathbf{k}},m_1',m_2,\sigma}(\tau).\\ \label{eq:bubble_freq}
\end{eqnarray}
We now apply inverse Fourier transform to the Green's functions
\begin{equation}
 G(\tau) = \frac{1}{\beta}\sum_{i\omega} e^{-i\omega\tau} G(i\omega),
\end{equation}
to obtain
\begin{eqnarray}\label{eq:bubble_freq1}
 \Lambda^{\eta\eta', \mathrm{disc}}_\mathbf{q=0}(i\nu) &=& 
 \frac{t^2e^2}{c\hbar^2} \frac{1}{N} \sum_\sigma \sum_{\tilde{\mathbf{k}},m_1,m_1',m_2,m_2'}  \\ \nonumber
 && \times 
 v^\eta_{\tilde{\mathbf{k}},m_1,m_1',\sigma}
 v^{\eta'}_{\tilde{\mathbf{k}},m_2,m_2',\sigma} \\ \nonumber
 && \times \frac{1}{\beta^2}\sum_{i\omega,i\omega'}G_{\tilde{\mathbf{k}},m_2',m_1\sigma}(i\omega)
 G_{\tilde{\mathbf{k}},m_1',m_2,\sigma}(i\omega')\\ \nonumber
 &&\times \frac{1}{2\hbar}
 \int_{-\beta\hbar}^{\beta\hbar}\mathrm{d}\tau e^{i(\nu+\omega-\omega')\tau},
\end{eqnarray}
where $\int_{-\beta\hbar}^{\beta\hbar}\mathrm{d}\tau e^{i(\nu+\omega-\omega')\tau} = 2\beta\hbar\delta_{\nu+\omega-\omega'}$, which immediately yields Eq.\ref{eq:bubble_freq2}.

\section{Proof that DMFT self-energy is diagonal in the non-interacting eigenbasis} \label{app:dmft_Sigma_is_diagonal}

The following shows that a local self-energy is also diagonal in the basis of $|\tilde{\mathbf{k}},m,\sigma\rangle$ states. We have
\begin{eqnarray}
 [\Sigma_{\tilde{\mathbf{k}},\sigma}(z)]_{ll'} 
 &=& \langle \tilde{\mathbf{k}},l,\sigma| \mathbf{\Sigma}_\sigma |\tilde{\mathbf{k}},l',\sigma\rangle \\ \nonumber
 &=& \sum_i [w_\sigma]^*_{(\tilde{\mathbf{k}},l),i}[w_{\sigma}]_{(\tilde{\mathbf{k}},l'),i} \Sigma_{ii,\sigma}(z) \\ \nonumber
 &=& \Sigma_{\sigma}(z) \sum_i [w_\sigma]^*_{(\tilde{\mathbf{k}},l),i}[w_{\sigma}]_{(\tilde{\mathbf{k}},l'),i}\\ \nonumber
 &=& \Sigma_{\sigma}(z) \delta_{ll'},
\end{eqnarray}
where $w_\sigma$ is the basis change matrix, for the transformation from site-space to $\tilde{\mathbf{k}},l$-space.
We can, therefore, write
\begin{eqnarray}
 [\Sigma_{\tilde{\mathbf{k}},\sigma}(z)]_{mm'} 
 &=& \langle \tilde{\mathbf{k}},m,\sigma| \mathbf{\Sigma}_\sigma |\tilde{\mathbf{k}},m',\sigma\rangle \\ \nonumber
 &=& \sum_l [\alpha^{-1}_{\tilde{\mathbf{k}},\sigma}]^*_{m,l}[\alpha^{-1}_{\tilde{\mathbf{k}},\sigma}]_{m',l} \Sigma_{ll,\sigma}(z) \\ \nonumber
 &=& \Sigma_{\sigma}(z) \delta_{mm'},
\end{eqnarray}
which immediately yields Eq.\ref{eq:dmft_Gk_diagonal_in_m_space}.

\section{Calculation of local Green's function in DMFT using the non-interacting density of states} \label{app:dmft_gloc}

Here we prove Eq.\ref{eq:dmft_Gloc}.
\begin{eqnarray}
G_{ii,\sigma}(z)&=& \langle i,\sigma| \mathbf{G}_\sigma (z) | i,\sigma \rangle \\ \nonumber
&=& \frac{1}{N}
 \sum_{\tilde{\mathbf{k}},m}
  \sum_{l,l'} \\ \nonumber
&&\times
   e^{-i\big(\tilde{\mathbf{k}}+l\frac{2\pi}{q}\mathbf{e}_x\big)\cdot \mathbf{r}_i}
   e^{i\big(\tilde{\mathbf{k}}+l'\frac{2\pi}{q}\mathbf{e}_x\big)\cdot \mathbf{r}_i}
   [\alpha_{\tilde{\mathbf{k}},\sigma}]_{l,m}
   [\alpha_{\tilde{\mathbf{k}},\sigma}]^*_{l',m}  \\ \nonumber
&&\times   
   \langle \tilde{\mathbf{k}},m,\sigma |       \mathbf{G}_\sigma (z) |\tilde{\mathbf{k}},m,\sigma \rangle      \\ \nonumber
&=& \frac{1}{N}
 \sum_{\tilde{\mathbf{k}},m}
  \sum_{l,l'} \\ \nonumber
&&\times
   e^{-i\big((l-l')\frac{2\pi}{q}\mathbf{e}_x\big)\cdot \mathbf{r}_i}   
   [\alpha_{\tilde{\mathbf{k}},\sigma}]_{l,m}
   [\alpha_{\tilde{\mathbf{k}},\sigma}]^*_{l',m}  \\ \nonumber
&&\times   
   \langle \tilde{\mathbf{k}},m,\sigma |   \mathbf{G}_\sigma (z) |\tilde{\mathbf{k}},m,\sigma \rangle.
\end{eqnarray}   
Because we know that $G_{ii}(z)$ must be uniform, we can define
\begin{eqnarray}
 &&G_{\mathrm{loc},\sigma}(z) = \frac{1}{N} \sum_i G_{ii,\sigma}(z) \\ \nonumber
 && = \frac{1}{N^2}               
        \sum_{\tilde{\mathbf{k}},m} \sum_{l,l'}
        [\alpha_{\tilde{\mathbf{k}},\sigma}]_{l,m}
        [\alpha_{\tilde{\mathbf{k}},\sigma}]^*_{l',m}  
        \sum_i e^{-i\big((l-l')\frac{2\pi}{q}\mathbf{e}_x\big)\cdot \mathbf{r}_i}   \\ \nonumber
&&\times   
   \langle \tilde{\mathbf{k}},m,\sigma |   \mathbf{G}_\sigma (z) |\tilde{\mathbf{k}},m,\sigma \rangle \\ \nonumber             
 && = \frac{1}{N^2}               
        \sum_{\tilde{\mathbf{k}},m} \sum_{l,l'}
        [\alpha_{\tilde{\mathbf{k}},\sigma}]_{l,m}
        [\alpha_{\tilde{\mathbf{k}},\sigma}]^*_{l',m}  
        N \delta_{ll'}  \\ \nonumber
&&\times   
   \langle \tilde{\mathbf{k}},m,\sigma |   \mathbf{G}_\sigma (z) |\tilde{\mathbf{k}},m,\sigma \rangle \\ \nonumber             
&& = \frac{1}{N}               
        \sum_{\tilde{\mathbf{k}},m}         
   \langle \tilde{\mathbf{k}},m,\sigma |   \mathbf{G}_\sigma (z) |\tilde{\mathbf{k}},m,\sigma \rangle \sum_{l} |[\alpha_{\tilde{\mathbf{k}},\sigma}]_{l,m}|^2  \\ \nonumber             
&& = \frac{1}{N} \sum_{\tilde{\mathbf{k}},m}                       
   \langle \tilde{\mathbf{k}},m,\sigma |   \mathbf{G}_\sigma (z) |\tilde{\mathbf{k}},m,\sigma \rangle.
\end{eqnarray}
Therefore, we can identify
\begin{eqnarray}
G_{ii,\sigma}(z)
&=& \frac{1}{N}               
 \sum_{\tilde{\mathbf{k}},m} \frac{1}{\hbar z-\varepsilon_{\tilde{\mathbf{k}},m,\sigma} - \Sigma_{\sigma}(z)} \\ \nonumber
&=& \int \mathrm{d}\varepsilon \frac{\rho_0(\varepsilon) }{\hbar
z-\varepsilon - \Sigma_{\sigma}(z)}.
\end{eqnarray}

\section{Conductivity expression in DMFT: Proof of Eqs.\ref{eq:sigma_xx_final} and \ref{eq:sigma_xy_final}} \label{app:analytical_continuation_lambda}

Starting from Eq.\ref{eq:dmft_bubble_imfreq}, we first perform the Hilbert transform of the Green's function
\begin{equation}
 G(i\omega) = -\frac{1}{\pi} \int \mathrm{d}\varepsilon
 \frac{\mathrm{Im}G(\varepsilon+i0^+)}{i\omega-\varepsilon},
\end{equation}
where $\varepsilon$ and $i\omega$ have the units of frequency.
We obtain
\begin{eqnarray} \nonumber
 \Lambda^{\eta\eta', \mathrm{disc}}_\mathbf{q=0}(i\nu) &=& \frac{t^2e^2}{c\hbar^2\pi^2}\sum_\sigma \frac{1}{\beta}\sum_{i\omega} 
 \int \mathrm{d}\varepsilon\int \mathrm{d}\varepsilon'
 v^{\eta\eta'}_\sigma(\varepsilon,\varepsilon') \\ \nonumber
 && \times \int \mathrm{d}\omega \int \mathrm{d}\omega'
 \frac{\mathrm{Im}G(\varepsilon,\; \omega)}{i\omega+i\nu-\omega}
 \frac{\mathrm{Im}G(\varepsilon',\; \omega')}{i\omega-\omega'}. \\
\end{eqnarray}
Now we apply the partial fraction expansion $\frac{1}{z-a}\frac{1}{z-b}=\frac{1}{a-b}(\frac{1}{z-a}-\frac{1}{z-b})$:
\begin{eqnarray}
 \Lambda^{\eta\eta', \mathrm{disc}}_\mathbf{q=0}(i\nu) &=& \frac{t^2e^2}{c\hbar^2\pi^2}\sum_\sigma \frac{1}{\beta}\sum_{i\omega} 
 \int \mathrm{d}\varepsilon\int \mathrm{d}\varepsilon'
 v^{\eta\eta'}_\sigma(\varepsilon,\varepsilon') \\ \nonumber
 && \times \int \mathrm{d}\omega \int \mathrm{d}\omega'\mathrm{Im}G(\varepsilon,\; \omega)\mathrm{Im}G(\varepsilon',\; \omega') \\ \nonumber
 && \times 
 \frac{1}{-i\nu+\omega-\omega'}
 \Bigg[ \frac{1}{i\omega+i\nu-\omega}
        - \frac{1}{i\omega-\omega'}\Bigg].
\end{eqnarray}
We apply $\frac{1}{\beta}\sum_{i\omega} \frac{1}{\hbar(i\omega-z)} = n_\mathrm{F}(z)$
\begin{eqnarray}
 \Lambda^{\eta\eta', \mathrm{disc}}_\mathbf{q=0}(i\nu) &=& \frac{t^2e^2}{c\hbar\pi^2}\sum_\sigma
 \int \mathrm{d}\varepsilon\int \mathrm{d}\varepsilon'
 v^{\eta\eta'}_\sigma(\varepsilon,\varepsilon') \\ \nonumber
 && \times 
 \int \mathrm{d}\omega \int \mathrm{d}\omega'\mathrm{Im}G(\varepsilon,\; \omega)\mathrm{Im}G(\varepsilon',\; \omega') \\ \nonumber
 && \times 
 \frac{1}{-i\nu+\omega-\omega'}\Big[n_\mathrm{F}(-i\nu+\omega)-n_\mathrm{F}(\omega')\Big].
\end{eqnarray}
The bosonic frequency does nothing in the argument of $n_\mathrm{F}$ so we can rewrite
\begin{eqnarray}
 \Lambda^{\eta\eta', \mathrm{disc}}_\mathbf{q=0}(i\nu) &=& \frac{t^2e^2}{c\hbar\pi^2}\sum_\sigma 
 \int \mathrm{d}\varepsilon\int \mathrm{d}\varepsilon'
 v^{\eta\eta'}_\sigma(\varepsilon,\varepsilon') \\ \nonumber
 && \times 
 \int \mathrm{d}\omega \int \mathrm{d}\omega'\mathrm{Im}G(\varepsilon,\; \omega)\mathrm{Im}G(\varepsilon',\; \omega') \\ \nonumber
 && \times 
 \frac{1}{-i\nu+\omega-\omega'}\Big[n_\mathrm{F}(\omega)-n_\mathrm{F}(\omega')\Big].
\end{eqnarray}
Formal continuation to the real axis is performed by replacing $i\nu\rightarrow \nu$
\begin{eqnarray}
 \Lambda^{\eta\eta', \mathrm{disc}}_\mathbf{q=0}(\nu) &=& \frac{t^2e^2}{c\hbar\pi^2}\sum_\sigma 
 \int \mathrm{d}\varepsilon\int \mathrm{d}\varepsilon'
 v^{\eta\eta'}_\sigma(\varepsilon,\varepsilon') \\ \nonumber
 && \times 
 \int \mathrm{d}\omega \int \mathrm{d}\omega'\mathrm{Im}G(\varepsilon,\; \omega)\mathrm{Im}G(\varepsilon',\; \omega') \\ \nonumber
 && \times 
 \frac{1}{-\nu+\omega-\omega'}\Big[n_\mathrm{F}(\omega)-n_\mathrm{F}(\omega')\Big].
\end{eqnarray}

As we are interested in the real part of the conductivity, and having in mind $\mathrm{Im}\Lambda^{\eta,\eta'}(\nu=0)=0$,
we get
\begin{eqnarray}
  \mathrm{Re}\sigma^{\eta\eta', \mathrm{disc}}_\mathbf{q=0}(\nu) &=& c\frac{\mathrm{Im}\Lambda_\mathbf{q=0}^{\eta\eta', \mathrm{disc}}(\nu)}{\nu} \\ \nonumber
  &=& \frac{t^2e^2}{\hbar\pi^2}\sum_\sigma \mathrm{Im}
 \int \mathrm{d}\varepsilon\int \mathrm{d}\varepsilon'
 v^{\eta\eta'}_\sigma(\varepsilon,\varepsilon') \\ \nonumber
 && \times 
 \int \mathrm{d}\omega \int \mathrm{d}\omega'\mathrm{Im}G(\varepsilon,\; \omega)\mathrm{Im}G(\varepsilon',\; \omega') \\ \nonumber
 && \times 
 \frac{1}{-\nu+\omega-\omega'}\frac{\Big[n_\mathrm{F}(\omega)-n_\mathrm{F}(\omega')\Big]}{\nu}.
\end{eqnarray}
For the longitudinal conductivity specifically, $v^{\eta\eta}$ is purely real, so the imaginary part comes from the delta-peak part of the $\omega,\omega'$ integrals through
$\int \mathrm{d}x \frac{1}{x-y+i0^-} = {\cal P}\int \mathrm{d}x \frac{1}{x-y} +i\pi \delta(x-y)$. In the limit $\nu\rightarrow 0$ we get Eq.\ref{eq:sigma_xx_final}.

For Hall conductivity the imaginary part comes from the principal part of the integral, and one can estimate it through\cite{Markov2019}
$ {\cal P}\int \mathrm{d}\omega\int \mathrm{d}\omega' \frac{1}{\nu} \frac{1}{-\nu+\omega-\omega'} = {\cal P}\int \mathrm{d}\omega\int \mathrm{d}\omega' \frac{1}{\nu} \frac{-\nu+\omega-\omega'}{(-\nu+\omega-\omega')^2} = -\int \mathrm{d}\omega\int \mathrm{d}\omega' \frac{1}{(-\nu+\omega-\omega')^2}+\int \mathrm{d}\omega\int \mathrm{d}\omega' \frac{1}{\nu}\frac{\omega-\omega'}{(-\nu+\omega-\omega')^2}$. In the limit $\nu\rightarrow 0$ the second term cancels exactly due to the anti-symmetry of the integrand with respect to the exchange $\omega\leftrightarrow\omega'$, and we get Eq.\ref{eq:sigma_xy_final}.

%
%

\begin{figure}[ht!]
 \begin{center}
 \includegraphics[width=3.2in, trim=0cm 0cm 0cm 0cm, clip]{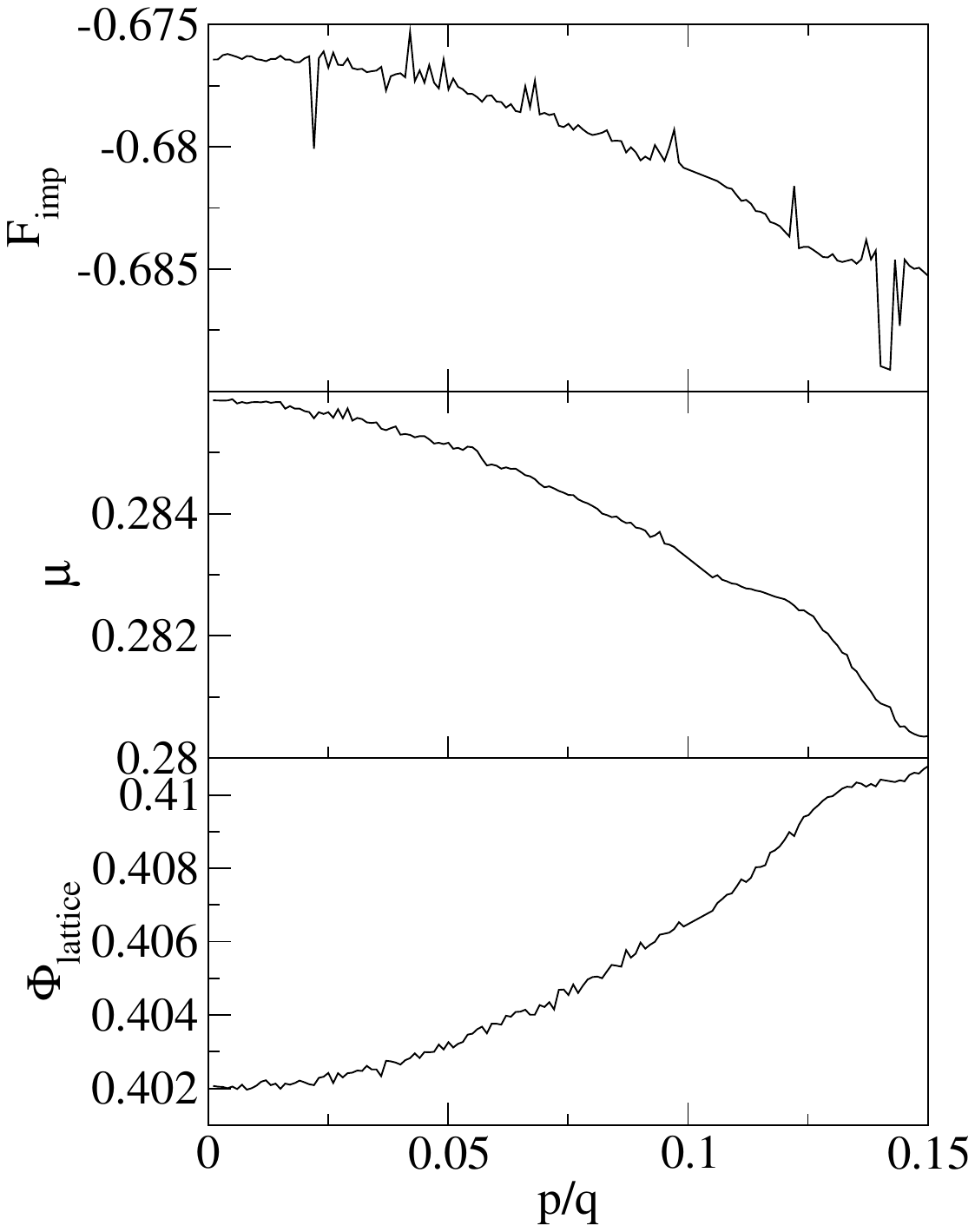}
  \end{center}
 \caption{Dependence of total free energy and its components on the magnetic field.
 }
 \label{fig:free_energy}
\end{figure}

\section{Oscillations in thermodynamic properties}

\label{app:thermodynamics}

With the NRG impurity solver, it is possible to directly calculate the
thermodynamic properties of the lattice problem (i.e., without any
integrations over parameters such as $T$ or $\mu$, which is
error-prone). This is based on Eq.~(46) from
Ref.~\onlinecite{RevModPhys.68.13}, which relates the lattice grand potential
(Landau free energy) $\Omega=F-\mu N_\mathrm{tot}=E-TS-\mu N_\mathrm{tot}$ 
($N_\mathrm{tot}$ is total number of particles)
and the impurity grand potential $\Phi_\mathrm{imp}=F_\mathrm{imp}-\mu n_\mathrm{imp}$:
\begin{equation}
\begin{split}
\frac{\Omega}{N}
=& \Omega_\mathrm{imp} - k_\mathrm{B}T \sum_{i\omega,\sigma}
\Big(
\int_{-\infty}^{+\infty} \mathrm{d}\epsilon\,\rho_0(\epsilon) \\
& \times \ln[ (i\omega\hbar+\mu-\Sigma_\sigma(i\omega)-\epsilon)
 G_\sigma(i\omega)]\Big),
\end{split}
\end{equation}
where $G_\sigma$ is the local Green's function. This may be
analytically continued to the real axis to give \cite{brandt1991,shvaika2003}
\begin{equation}
\begin{split}
\frac{\Omega}{N}
=& \Omega_\mathrm{imp} + \frac{1}{\pi} \sum_\sigma
\int_{-\infty}^{+\infty} \mathrm{d}\epsilon\,\rho_0(\epsilon)
\int_{-\infty}^{+\infty} \mathrm{d}\omega \\
&\times \mathrm{Im}
\ln \left[ (\hbar\omega+\mu-\Sigma_\sigma(\omega)-\epsilon) G_\sigma(\omega)
\right]
n_\mathrm{F}(\omega),
\end{split}
\end{equation}
with the Fermi-Dirac distribution
$n_\mathrm{F}(\omega)=1/(1+\exp[\hbar\omega/Tk_\mathrm{B}])$. The impurity free energy can
be directly calculated in the NRG using the full-density-matrix
approach.

We consider the case shown in Fig.~\ref{fig:fixedSigma_check} which exhibited significant
transport oscillations at the high-frequency, while the self-energy
and the Green's function showed instead oscillations at the SdH
frequency. In fact, at this temperature, the Fourier transform of the
oscillatory part of the inverse conductivity shows no component at the
SdH frequency, it is already thermally washed out. Indeed, we find no
remnants of the SdH/dHvA oscillations in the thermodynamic properties
either. In Fig.~\ref{fig:free_energy} we plot three elements that enter the full
thermodynamic potental: impurity free energy $F_\mathrm{imp}$, the
chemical potential $\mu$ (which enters as $-\mu n_\mathrm{imp}$ with
constant $n_\mathrm{imp}=n=0.85$), and the lattice contribution from
the double integration, $\Phi_\mathrm{lattice}$. None of these shows
any clear oscillations; if they exist, they are below the numerical
uncertaintly. We only observe a weak quadratic dependence on the
magnetic field in all three contributions. This confirms yet again
that the high-frequency quantum oscillations show up exclusively in
the transport properties through the vertex factors, thus they are, in
this sense, a purely kinetic effect.

\begin{figure}[ht!]
 \begin{center}
 \includegraphics[width=3.2in, trim=0cm 0cm 0cm 0cm, clip]{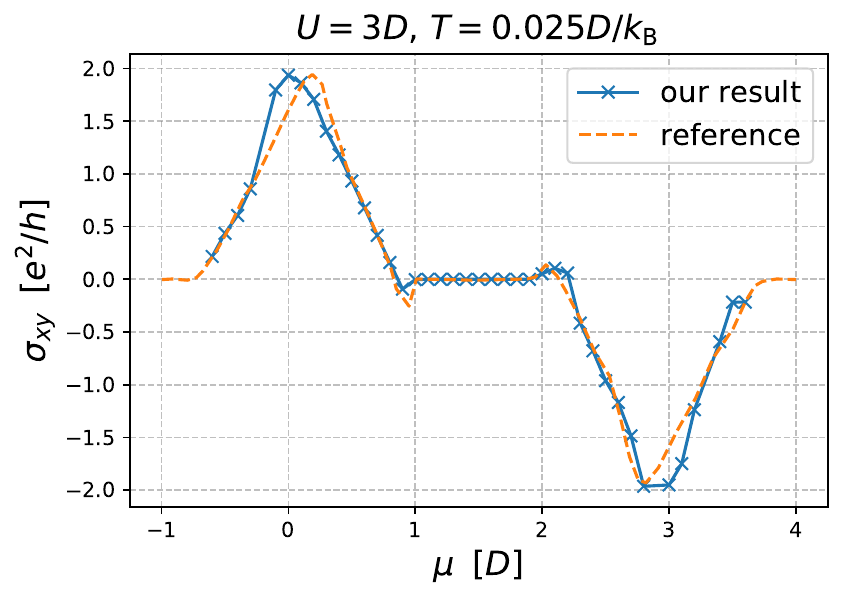}
  \end{center}
 \caption{Benchmark with the data from Markov \emph{et al.}, Ref.~\onlinecite{Markov2019}. Our data: DMFT(NRG solver). Reference data: DMFT(exact diagonalization solver with 5 bath sites)+Pad\' e analytical continuation used to obtain continuous spectra.
 }
 \label{fig:sigma_xy_benchmark}
\end{figure}

\section{Benchmark} \label{app:sigmaxy_benchmark}

To benchmark our formalism and implementation, we cross-check our $\sigma^{xy}(\nu=0)$ results with the data from Ref.~\onlinecite{Markov2019}.
We perform a chemical potential scan at a fixed $U=3D$ and $T=0.025D/k_\mathrm{B}$, which corresponds to the (doped) Mott insulator regime.
The results are shown in Fig.~\ref{fig:sigma_xy_benchmark}. The agreement is solid. Neither of the curves fully satisfy $\sigma^{xy}(\nu=0;\mu)=-\sigma^{xy}(\nu=0;-\mu)$, which reveals the extent of the systematic error bars. The biggest difference is the position of the two peaks in the curves, which can be attributed to the difference in the impurity solvers used (we have used NRG\cite{wilson1975,krishna1980a,bulla2008,resolution}, directly on the real axis; in Ref.~\onlinecite{Markov2019} they used exact diagonalization with 5 bath sites, and Pad\' e analytical continuation to obtain continuous spectra). We reproduce the change of sign of $\sigma^{xy}(\nu=0)$ as chemical potential crosses the edge of the Hubbard band into the Mott gap (at around $\mu=1$ and $\mu=2$), which appears to be a robust feature of the solution.

\begin{acknowledgments}
Computations were performed on the PARADOX supercomputing facility
(Scientific Computing Laboratory, Center for the Study of Complex
Systems, Institute of Physics Belgrade). J.~V. acknowledges funding
provided by the Institute of Physics Belgrade, through the grant by
the Ministry of Education, Science, and Technological Development of
the Republic of Serbia, as well as by the Science Fund of the Republic
of Serbia, under the Key2SM project (PROMIS program, Grant No.
6066160). R.\v Z. is supported by the Slovenian Research Agency (ARRS)
under Program P1-0044 and Projects J1-1696 and J1-2458.
\end{acknowledgments}

\bibliography{refs.bib}
\bibliographystyle{apsrev4-1}

\end{document}